\documentclass{tADP2e}
\usepackage{rotating}
\begin{document}
\bibliographystyle{tADP}           
%\listoffigures 
\title{Exactly solvable model of avalanches dynamics for
  Barkhausen crackling noise }

\author{Francesca Colaiori, 
SMC, INFM-CNR and Dipartimento di Fisica, 
``La Sapienza" Universit\`a di Roma, P.le Aldo Moro 2, I-00185
Roma, Italy}
\markboth{Francesca Colaiori}{Barkhausen noise: mean field and beyond}
\maketitle
\begin{abstract}
We review the present state of understanding of the Barkhausen effect
in soft ferromagnetic materials.  Barkhausen noise (BN) is generated
by the discontinuous motion of magnetic domains as they interact with
impurities and defects.  BN is one of the very many examples of
crackling noise, arising in a variety of contexts with remarkably
similar features, and occurring when a system responds in a jerky
manner to a smooth external forcing.  Among all crackling system, we
focus on BN, where a complete and consistent picture emerges thanks to
an exactly solvable model of avalanches dynamics, known as ABBM model,
which ultimately describes the system in terms of a Langevin equation
for the velocity of the avalanche front. Despite its simplicity the
ABBM model is able to accurately reproduce the phenomenology observed
in the experiments on a large class of magnetic materials, as long as
universal properties are involved.  To complete the picture and to
understand the long--standing discrepancy between the ABBM theory and
the experiments --- which otherwise agree exceptionally well ---
consisting in the puzzling asymmetric shape of the noise pulses,
microscopic details must be taken into account, namely the effects of
eddy currents retardation. These effects can be incorporated in the
model, and result, to a first order approximation, in a negative
effective mass associated with the wall. The progress made in
understanding BN is potentially relevant for other crackling systems:
on one hand, the ABBM model turns out to be a paradigmatic model for
the universal behavior of avalanches dynamics; on the other hand, the
microscopic explanation of the asymmetry in the noise pulses suggests
that inertial effects may also be at the origin of pulses asymmetry
observed in other crackling systems.

\end{abstract}
%\pacs{75.60.Ej, 75.60.Ch, 64.60.Lx, 68.35.Ct, 05.40.$-$a, 05.45.Tp}
%\tableofcontents
%\listoffigures 
\section{Introduction}
\label{sec:INTRO}

\subsection{Crackling noise}
\label{CN}

The term ``crackling noise''
\cite{sethna_crackling_2001,sethna_course_2007} refers to the signal
that some disordered systems produce as a response to an external
driving field smoothly changing in time.  Due to the presence of
disorder, crackling signals are extremely irregular, despite the
steady increase of the external forcing. They are typically
characterized by a sequence of pulses of very different sizes and
durations, separated by quiescence intervals. Tiny events occur very
frequently, while large ones are rare, with power laws probability
distributions.

In this review we focus on one particular case of noise, the so called
Barkhausen noise (BN), which is emitted by ferromagnetic materials
during the magnetization reversal process
\cite{barkhausen_zwei_1919,bertotti_hysteresis_1998,durin_barkhausen_2006}.

Systems that ``crackle'' are found in many different situations, and,
remarkably, the corresponding signals often share some common
characteristic features. Examples of crackling signals, besides BN,
include the shear response of a granular media
\cite{baldassarri_granular_2007, baldassarri_brownian_2006,
bretz_broad_2006}, the acoustic emission during martensitic phase
transitions \cite{vives_distributions_1994}, the bursts of
dislocations activity in plastic deformation
\cite{miguel_intermittent_2001, weiss_three_2003, zaiser_scale_2006,
csikor_dislocation_2007, sethna_crackling_2007}, the dynamic of
superconductors \cite{erta_anisotropic_1994,
field_superconducting_1995, erta_anisotropic_1996} and superfluids
\cite{guyer_capillary_1996, lilly_spatially_1996}, the fluctuations in
the stock market \cite{bak_price_1997, bouchaud_power-laws_2001,
bouchaud_subtle_2005}, the dielectric polarization of ferroelectric
materials \cite{colla_barkhausen_2001}, the acoustic emission in
fractures \cite{curtin_analytic_1991, petri_experimental_1994,
zapperi_plasticity_1997, garcimartn_statistical_1997,
salminen_acoustic_2002}, and the seismic activity in earthquakes
\cite{main_statistical,mehta_universal_2006}.

Crackling noise signals are expected to encode information on the
physical process that generates them. Understanding the statistical
properties of these jerky emissions, is therefore a step towards the
understanding of the microscopic dynamics taking place in the system
that crackles. Moreover, the fact that very diverse systems behave in
a remarkably similar manner, suggests that some general basic
principle may exist in the underlying physics.  If this is the case,
then, understanding in detail one of them, may give insight to others,
and propel other fields of research.

BN can be seen as a case study, for which a satisfying picture is
finally available. A main ingredient in this picture is a model of
avalanche dynamics known as ABBM model, named after the authors of the
original papers, Bruno Alessandro, Cinzia Beatrice, Giorgio Bertotti
and Arianna Montorsi, whose companion seminal papers
\cite{alessandro_domain-wall_1990, alessandro_domain-wall_1990-4}, one
devoted to the developing the theory, the other to the comparison with
experiments, have represented a huge step forward in the understanding
of BN. The ABBM model, corresponds to a mean field description of an
elastic magnetic wall moving in a disordered ferromagnet under the
effect of an external driving field. The system is ultimately
described by a simple Langevin equation for the velocity of the center
of mass of the wall.

Despite its simplicity, the ABBM model is able to reproduce with
striking accuracy most of the phenomenology observed in BN experiments
on a large class of magnetic materials. Moreover, thanks to a mapping
onto a simple stochastic process, all the results concerning the noise
statistics can be derived analytically, and have a clear and direct
interpretation. In particular, this simple mean field description
allows to explain the origin of the power law distributions of size
and duration of pulses, both in the quasi--static limit, where the
applied field variations are extremely slow, and at finite driving
field rates. In this second case, the model is also able to predict
how these distributions depend on the driving field rate.

A more refined analysis of the experimental data against theoretical
predictions includes the comparison of the shape of the
avalanches. This is where the ABBM model unexpectedly fails, being
unable to predict the characteristic leftward asymmetric form of BN
pulses.  The origin of this asymmetry lies in the non--instantaneous
response of the eddy fields to the domain wall displacement. To
understand and evaluate the effect of such delay, one as to take into
account the dynamical effect of eddy currents in the Maxwell equations
for the eddy field. Eddy currents retardation gives rise to an
``anti--inertial'' effect, that can be accounted for, to the first
order, by associating a negative effective mass to the wall.

Once the effective mass is identified, the corresponding inertial term
can be incorporated into the ABBM equation of motion.  The modified
ABBM model that comes out from this analysis gives a complete picture
of the Barkhausen effect, consistent with the phenomenological
observations: it is still successful in reproducing the distribution
of sizes and durations of pulses, and also correctly reproduces the
asymmetry in their shape.

It is interesting to underline the role of universality versus non--
universality in the theory of Barkhausen effect.  The standard
statistical mechanics approach usually focuses on universal
quantities. Definitely, universality is a key and extremely powerful
concept at the basis of this approach, as it allows to predict the
essential behavior of self--similar systems by means of very simple
models.  This consideration indeed applies to the case of Barkhausen
noise, where, in analogy to critical phenomena, most of the
statistical properties of the signals only depend on general
properties of the system, while they are independent of the
microscopic details: different magnetic samples respond to the
forcing by an external field by producing events of magnetization
reversal characterized by the same power law distributions, regardless
of the specific microscopic structure of the material.

However, some interesting features of BN turn out to be of microscopic
origin.  Although, consistently with universality, microscopic details
would become negligible on extremely large scales, they have an
unusually large effect in Barkhausen signals, an effect that is still
significant at the experimental scale. On one hand this can be seen as
an inconvenient and a limitation of the methods of statistical
mechanics, since it means that not all the relevant aspects of the
phenomenon can be captured by simple models.  On the other hand, the
identification of some macroscopic effect of the microscopic details,
provides a mean to extract information on microscopic quantities by
Barkhausen measurements. The asymmetry of the pulses gives indeed a
measure of this type. As an example, the skewness of the pulses
measured as a function of the pulse duration shows a peak, that allows
to identify a characteristic timescale for relaxation, which
corresponds to the ratio between mass and damping constant.

In this review we will deal with both these aspects of the theory, in
the attempt to put together a number of results achieved in recent
years. We will also try to underline which points are still unclear
and would need further analysis.

\subsection{Outline}
\label{outline}

The paper is organized as follows. Section \ref{sec:BN} reviews some
general facts about Barkhausen noise. We start with a short historical
note (\ref{sec:history}). Then we discuss how magnetic materials are
classified, in order to clarify to which ones the following theory
applies (\ref{sec:soft}). We shortly describe the typical set--up and
the problems arising in BN experiments in subsection \ref{sec:exp}.
BN phenomenology, as it comes out from experiments, is described in
subsection \ref{sec:PH}. In subsection \ref{sec:approach} we discuss
the two main theoretical approaches to BN that have been developed in
recent times: one based on a microscopic description in terms of a
random field Ising model, the other describing the magnetization
process in terms of the dynamic of magnetic interfaces, that will be
further developed in the next section.

In section \ref{sec:ABBM} we focus on the second approach: first we
introduce the ABBM as a phenomenological model (\ref{sec:ABBM_abbm})
and underline its main limitations (\ref{sec:limitations}). To put
the model in a more general framework we then briefly discuss the
problem of the dynamic of a generic elastic interface in a random
media (\ref{sec:elastic_interface}), and then we specialize to the
case of a magnetic interface (\ref{sec:mesoscopic}).  The purpose of
this digression is to recover the ABBM model as a mean field for a
model of interface dynamic (\ref{sec:MF}), therefore corroborating its
validity.

Section \ref{sec:RW} is the core part of the paper: we show how the
ABBM equation is mapped onto a Langevin equation describing a biased
random walk in logarithmic potential with absorbing boundary
conditions at the origin (\ref{sec:RW_rw}). This approach is amenable
of analytical treatment and allows to work out exactly a number of
properties, and to give a direct interpretation of the phenomenology
observed in the experiments in terms of properties of the
corresponding stochastic process (\ref{sec:LOG}). The power law
distributions in avalanches' sizes and durations can be exactly
calculated in terms of distributions of return times to the origin
(\ref{sec:PER}). The continuous dependence of the exponents in the
power law distributions on the driving field rate can also be
calculated, and turns out to be related to the marginality of the
logarithmic perturbation to the free random walk (\ref{sec:MAR}). The
existence of a threshold in the driving field rate can be derived from
recurrence properties of a free random walk (\ref{sec:REC}), and the
existence of a cut--off in the power law distributions can be
associated to the bias term (\ref{sec:BIAS}). The average shape of the
Barkhausen pulse can be derived in terms of the excursion of the
process (\ref{sec:EXC}), and finally the power spectra can be derived
from the process correlations (\ref{sec:PS2}). In subsection
\ref{sec:sum} we summarize the ABBM predictions and compare them with
the phenomenological observations. In the last subsection
(\ref{sec:cutoff}) we discuss the scaling of the cut--offs with the
demagnetizing factor. This scaling is not entirely captured by the
mean field approximation, although it is predicted by the full elastic
interface model in $d=3$.

Section \ref{sec:RW} collects a number of results, most of which, but
not all, are spread in literature, but to which has never been given a
comprehensive exposition in the BN framework before. Some of these
results have been indeed developed in an unrelated context, to deal
with a completely different problem \cite{bray_random_2000}. For this
reason in this part we made an effort to give a self--contained
exposition.

The issue of the asymmetry of Barkhausen pulses (that the ABBM model
is not able to reproduce) is treated in section \ref{sec:ASYM}: a
characteristic leftward asymmetry in the average pulse shape is
observed in experiments (\ref{sec:ASYM_exp}), which is due to the
non--instantaneous response of the eddy currents field to the wall
displacement (\ref{sec:ASYM_eddy}). The eddy current retardation gives
rise to inertial effects, that are accounted for in a generalized ABBM
model by associating a negative effective mass to the domain wall
(\ref{sec:ASYM_ABBM}). In subsection \ref{sec:ASYM_geo} we discuss how
this non--universal effect is affected by the sample geometry, and,
finally, in subsection \ref{sec:ASYM_comp} we compare theory
predictions and experiments.

In section \ref{sec:OTHER} we discuss some other crackling systems
that may be tackled with an approach similar to the one used for the
Barkhausen effect: the case of the dynamic of a granular media under
shear (\ref{sec:granular}), and the case of seismic activity during
earthquakes (\ref{sec:earth}).

Section \ref{sec:CONCLUSIONS} and \ref{sec:ak} are devoted to
conclusions and acknowledgments respectively.

\section{The Barkhausen effect: history, experiments and theoretical approaches} 
\label{sec:BN}

\subsection{Short historical note}
\label{sec:history}

Barkhausen noise is probably one of the first crackling signals ever
recorded. Indeed the Barkhausen effect has been known for almost a
century: its first observation dates back in 1919, when Heinrich
Barkhausen noticed that ``iron produces a noise when magnetized: as
the magnetomotive force is smoothly varied [...] it generates
irregular induction pulses in a coil wound around the sample that can
be heard as a noise in a telephone" \cite{barkhausen_zwei_1919}.

\begin{figure}
\centerline{\epsfxsize=9.5cm \epsfbox{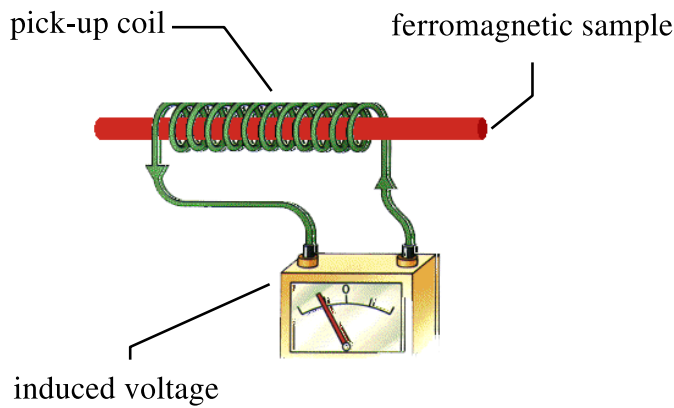}}
\caption{Sketch of experimental set--up used in the first BN
  experiment. The bar is the ferromagnetic sample, which is inserted
  in the pick--up coil. As the external field is changed, an
  electromotive force is induced in the coil. The voltage can
  eventually be transformed in real noise through an amplifier.}
\label{setup}
\end{figure}

Barkhausen's first experiment consisted in observing the variations of
the magnetization in a ferromagnetic sample subject to a slowly and
smoothly changing magnetic field. A pickup coil winded around the
sample was used to detect the variations of the magnetization as
schematically shown in figure \ref{setup}.  As the magnetization
reverses, the variations of the magnetic flux induce a voltage in the
coil, that indirectly measures the changes in the magnetization of the
sample, and that may eventually be translated in real noise through an
amplifier. The signal recorded in correspondence to jumps in the
magnetization, appears to be very irregular, no matter how smooth is
the variation of the external field.

Nowadays, magneto--optical methods are also available, that allows a
direct observations of moving domains, however, the most reliable
experiments in terms of signal statistics are still based on inductive
measurements. In order to be capable to record clean signals with
reduced background noise and high statistic, modern experimental
settings make use of sophisticated equipments, however, the basic
set--up for inductive BN experiments is conceptually still the same as
the one shown in figure \ref{setup}.

Barkhausen wondered about what could be the source of the signal
emitted during the magnetization of the iron sample, and he mistakenly
concluded that the sequence of pulses was generated by a corresponding
sequence of sudden and complete reversals of entire magnetic domains.
At that time, his experiment was indeed taken as the first indirect
observation of magnetic domains, whose existence had been postulated a
few years before (1907) by Weiss \cite{weiss_lhypothse_1907}.  Almost
20 years later (1938) Elmore was able to perform an experiment on a
cobalt crystal where the motion of domain boundaries was directly
observed and measured for the first time
\cite{elmore_magnetic_1938}. However Elmore did not recognize this
motion as the source of BN.  Only more then 10 years later (1949),
after another experiment by Williams and Shockley on a $FeSi$ crystal
\cite{williams_simple_1949}, the origin of BN was finally ascribed to
the correct cause, namely to the irregular fluctuations of magnetic
domain boundaries, rather then to the sudden inversion of
domains. This fact was definitely clarified in the same year by Kittle
in his fundamental review on the physical theory of ferromagnetic
domains \cite{kittel_charles_physical_1949}. A nice exposition of the
first steps in the understanding and misunderstandings of Barkhausen
effect is given in a recent review by G. Durin and S. Zapperi
\cite{durin_barkhausen_2006}, which also includes in appendix a
translation from German to English of the original paper by
H. Barkhausen.

Since its first observation, a lot of work has been done in order to
``decode'' the Barkhausen signal, both on the experimental side and on
the theoretical one. For a long time, theoretical studies have been
based on a phenomenological approach, which described the signal as a
superposition of random elementary jumps. Then, it became clear that
BN could represent a powerful tool to investigate the magnetization
process and the hysteretic properties on a microscopic scale, and much
effort has been devoted in developing physically grounded models that
could allow to relate the phenomenology to its microscopic origin. The
theoretical studies have followed two kind of approaches: one based on
a microscopic description in terms of spins in a random magnetic field
(see section \ref{sec:approach}); the other one, which is the one on
which we will focus in this review, describes the system in terms of a
fluctuating magnetic interface that moves under the action of the
external field.

The research on BN has been in part motivated, especially at the
beginning, by the applications of Barkhausen effect to material
testing.  Barkhausen emissions are indeed commonly used to check in a
non--destructive way the integrity of magnetic samples: the signal
intensity is sensitive to the changes in material microstructure, and
to the presence of residual stresses (in magnetostrictive positive
materials compressive stresses will decrease the intensity of
Barkhausen noise while tensile stresses will increase it).  These
properties make BN an efficient tool for the detection of
micro--imperfections and for the evaluation and mapping of the local
distribution of residual stresses \cite{tiitto_use_1989,
govindaraju_evaluation_1993, sipahi_overview_1994, lo_monitoring_1999,
baldev_r_characterisation_2001, raj_b_assessment_2003,
sagar_magnetic_2005, moorthy_magnetic_2005}.

Most of the recent research on the statistical properties of BN is
however theoretically oriented, and primarily aimed to gain
understanding on the hysteretic properties of ferromagnetic materials,
and to investigate the magnetization reversal process on a microscopic
scale.

Recently, further interest has been raised by the fact that many other
driven dissipative systems have been found to respond to a smooth
forcing in a disordered manner, with very similar and
reproducible statistical features. To this kind of response has been
given the general denomination of ``crackling noise''. Clearly, the
irregular response is due to the presence of disorder in the system,
and the challenge is to get information on the specific system under
study by decoding the noise that it generates. As the recent
encouraging advances in the understanding of BN compose a rather
satisfying picture of the phenomenon, BN emerges as an attractive case
study to understand crackling noise in general.

\subsection{Soft magnets versus hard magnets}
\label{sec:soft}

The understanding of hysteretic properties of magnets has an
interesting overlap with material science, and the theory that we
describe has been built always keeping close contact with the
experimental side.  Therefore, although the aim of this review is
mainly theoretical, we believe that, in order to put in a correct
framework the theory, understand its limitations, and to keep close
the connection with the experimental counterpart, it is worth to
discuss briefly how magnetic materials are classified, and to clarify
to which ones this theory applies, and to which ones it does not.

\begin{figure}
\centerline{\epsfxsize=10cm \epsfbox{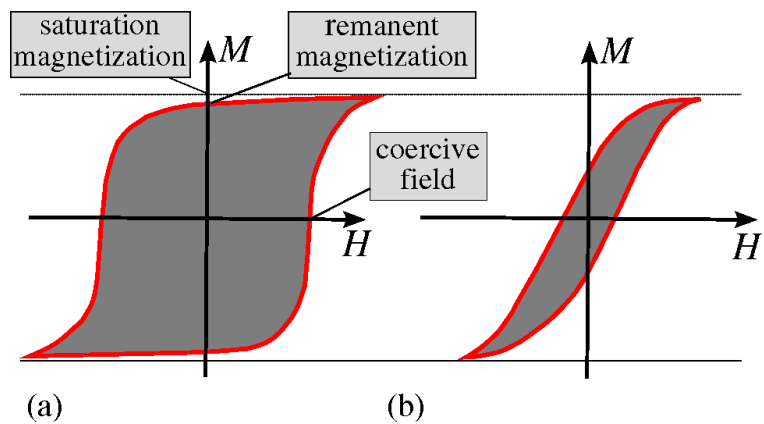}}
\caption{Sketch of a typical hysteresis loop for a hard magnet (a)
  and a soft magnet (b).}
\label{hardsoft}
\end{figure}

Magnetic materials are roughly classified in soft and hard. Soft
magnets are those characterized by very tiny hysteresis loop,
reflecting the fact that they are easily magnetized and demagnetized
by applying relatively small fields. Hard magnets are the exact
opposite: they have very wide and squared hysteresis loops, and they
need a big effort in terms of applied field in order to be magnetized
and demagnetized. The two classes of magnets are quite well separated,
and there is not much in between.  A good quantitative measure of
``magnetic hardness'', is given by the coercive field, which is the
field needed to bring back to zero the magnetization of a saturated
sample. This quantity differs, between soft and hard magnets, by
several orders of magnitudes (see figure \ref{hardsoft})
\cite{ohandley_modern_1999}.

The markedly different hysteretic behavior observed in the two classes
of materials is due to the fact that the magnetization reversal
proceeds by different mechanisms in the two cases. In soft magnets it
is dominated, except than close to saturation, by the nucleation and
displacement of magnetic walls. This is a process with low energetic
cost, that translates into a small loop area, which measures the
dissipated energy. On the other side, in hard magnets the wall
nucleation and/or displacement is strongly impeded, or totally
suppressed, and magnetization reversal by coherent spin
rotation enters into play.

Hard magnets have highly anisotropic magnetic properties. They are
usually composed from alloys containing rare earth elements (typical
ones are $Co_5Sm$ and $Ne_{2}Fe_{14}B$).  In single domain particle
magnets the nucleation of domains is prevented by building samples
made out of very small magnetic grains, whose size is comparable with
the domain wall thickness, and each one with a random easy
magnetization direction. Due to the high anisotropy, the magnetization
can only reverse by coherent spin rotation when the applied field is
large enough to force the spins out of the easy magnetization
direction. Due to the high coercive field, hard magnets cannot be
accidentally demagnetized when subjected to small fields. Also, they
have a remanent magnetization very close to the saturation value,
meaning that they are able to retain their magnetization when the
applied field is removed. These properties are extremely desirable in
the design of magnetic storage devices.

Examples of soft materials are Iron--Silicon alloys, crystalline
Iron--Nickel alloys (permalloys), and amorphous metallic alloys
(typical is the $FeCoB$). Applications of soft magnetic materials
exploit the large flux changes obtained with a small change in the
applied field, and the small energy loss. Soft magnets are used for
example in transformers, motors, and inductors, and also as field
sensors in magnetic recording devices.  The magnetization reversal in
these materials is dominated by the low--cost process of magnetic
domain displacement. Domain walls, and the domain walls movement can
be directly observed in experiments based on the magneto--optical Kerr
effect (MOKE). Reflection of a beam of linearly polarized light from
the surface of a magnetized sample causes the polarization to become
elliptical, with the principal axis rotated with respect to the
incoming light, depending on the magnetization direction. An analyzing
filter in the reflected beam is then used to generate magnetic domain
contrast.

The theory described in this review is based on a mesoscopic
description in terms of dynamics of domain walls, therefore it only
applies to soft magnets. The magnetization reversal process in hard
magnets is indeed worth of attention, however, in this case, a
microscopic description in terms of spin models would be more
appropriate.  As it will be discussed in subsection \ref{sec:PH}, soft
materials have different statistical properties depending on the range
of the interactions ruling the dynamics. The mean field theory which
is the focus of this review applies to those with long range
interactions.

\subsection{Barkhausen noise experiments}
\label{sec:exp}

In inductive experiments BN is detected by a pick--up coil wound around
the ferromagnetic sample during the magnetization of the material,
which takes place under the action of a varying external field
$H(t)$. As the magnetization reverses in the sample, the variations of
the magnetic flux $\dot\Phi$, induce an electromotive force $V$ in the
pick--up coil, which has a contribution from the applied field across
the pick--up coil, and another one given by the magnetic field inside
the material:
\begin{equation}
V=-N\dot{\Phi}=- N \mu_0 \left( A_{c} \dot{H}+A \dot{M}\right) \,,
\label{picup}
\end{equation}
where $\mu_0$ is the vacuum magnetic permeability, N is the number of
coil turns $A_{c}$ is the coil cross section, $A$ is the sample cross
section, and $M$ the magnetization.  

In high susceptibility materials the contribution coming from the
first term in equation (\ref{picup}) can be neglected, since, as long
as the susceptibility $\chi=(\mu/\mu_0-1)$ is much larger then $1$
(where $\mu$ is the magnetic permeability of the ferromagnetic
material), the contribution from the magnetization rate $\dot{M}$ is
much larger then the one due to the variation of the applied field
$\dot{H}$.  To neglect the first term in equation (\ref{picup}), it is
also necessary that the sample cross section $A$ is not too small with
respect to the pick--up coil section $A_c$.

In low susceptibility materials, or in cases where the condition
$A\simeq A_c$ is not easy to satisfy, as for example in the case of
thin films, it is possible to compensate the induced flux in air by
using another pick--up coil with the same number of turns and cross
section, but wound in the opposite direction.  

With the eventual addition of the compensation coil, it can therefore
always be assumed that the variation of the magnetic flux $\dot\Phi$ is
approximately proportional to the rate of change of the magnetization:
\begin{equation}
V\simeq- N \mu_0 A \dot{M} \,.
\label{picup2}
\end{equation}

\begin{figure}
\centerline{\epsfxsize=8cm \epsfbox{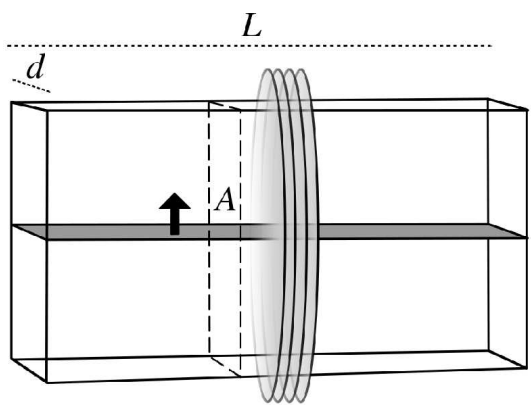}}
\caption{Sample geometry for a typical inductive BN experiment.}
\label{measure}
\end{figure}

In the simplest experimental conditions, where there is just one
domain wall separating two regions of opposite magnetization, the rate
of change of $M$ is twice the magnetization of the region spanned per
unit time. The volume of this region is given by $v d L$ in the
geometry and notations of figure \ref{measure}, and where $v$ is the
average wall displacement velocity. This volume, times the
magnetization per unit volume $M_s/A L$, gives the magnetization rate
\begin{equation}
\dot{M}=2 M_s v d/A \,,
\label{mag_rate} 
\end{equation}
where $M_s$ is the saturation magnetization.

Therefore, the induced voltage $V$ measured by the pick--up coil is
just proportional to the wall velocity $v$:
\begin{equation}
V=- 2 N \mu_0 d M_s  v \,.
\label{V}
\end{equation} 

Note that the sample thickness $d$ in the direction perpendicular to
the wall movement and to the coil axis is the only geometrical
parameter that enters in the intensity of the BN signal. 

The number of turns in the pick--up coil amplifies the signal by a
factor $N$, thus one could be tempted to use wide detection coils.
However, we stress that this is not a good choice: the magnetostatic
field produces a counterfield which has a crucial role in the domain
wall dynamics. This counterfield depends on sample geometry and domain
structure, and it is homogeneous only in special geometries. However
it is convenient to assume it to be constant also in more general
cases. This is a good approximation only as long as one limits the
region of the pick--up coil.

In the more complex case where several domain walls are present, the
signal encodes the complex collective effect of many interacting
domains. In this case the quantity $v$ in equation (\ref{V})
approximately measure the velocity of the active walls.

\begin{figure}
\centerline{\epsfxsize=15cm \epsfbox{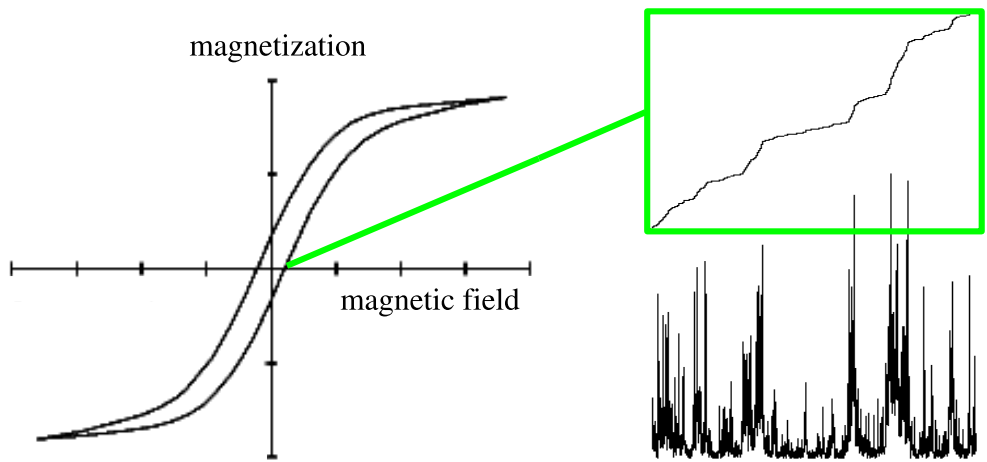}}
\caption{Sketch of a typical hysteresis loop for a soft magnet.  The
loop is not smooth, rather the magnetization increases in jumps as
shown in the zoom.  The inset on the bottom right side is the
derivative of the magnetization, proportional to the typical
Barkhausen signal, detected from the induced voltage.  The voltage
directly measures the velocity of the domain wall: pulses correspond
to jumps of the magnetic interface that gets repeatedly stuck by the
various kind of impurities and imperfection, acting as pinning
centers.  }
\label{barkhausen}
\end{figure}

Due to the presence of various types of disorder, like non--magnetic
inclusions and dislocations, the wall movement is discontinuous, since
the disorder locations act as pinning points.  As the external field
is smoothly increased, the magnetization changes in steps, in
correspondence to jumps of the magnetic interface, or, in microscopic
terms, in correspondence to avalanches of spin flips. The hysteresis
loops is also discontinuous, and the signal, which is proportional to
the derivative of the magnetization, plotted versus time, looks like a
disordered series of pulses (see figure \ref{barkhausen}).

Note that, in identifying the Barkhausen signal with the derivative of
the hysteresis curve, we are assuming that the applied field $H(t)$ is
increased linearly in time: indeed, in BN measurements, often --- but
not always --- a driving field with a triangular time profile is
applied, instead of one with the sinusoidal shape commonly used in
hysteresis experiments. In this case, the variations of magnetization
with respect to the applied field are directly translated in
derivatives with respect to time.

There exists a vast literature reporting results and data about BN
experiments in soft materials. However, not all the data collected are
reliable when one is interested in statistically analyzing and
modeling the self--similar properties of the noise, as the details of
the experimental procedure may have a serious impact on the quality of
the signal. Unfortunately, the need to establish a standard in the
experimental set--up that could allow a rigorous comparison of the
data, has been recognized as an important issue only in recent times.

The most relevant factor to take into account in order to ensure a
reliable noise signal is to ascertain the stationarity during the
dynamics. This question is crucial, since non--stationarity strongly
biases the scaling of the avalanches distributions. However, it is not
always easy to check whether the system is in a stationary state or
not.

In the context of BN, the statistical properties of the signal
strongly depend on the region of the hysteresis loop where the measure
is recorded. This question was first raised by Bertotti and Fiorillo
\cite{bertotti_barkhausen_1981}, who suggested to restrict the
measures to the part of the loop corresponding to constant average
permeability, where the condition of stationarity is guaranteed.  This
important point has however been disregarded in many experimental
studies.

\begin{figure}
\centerline{\epsfxsize=12.5cm \epsfbox{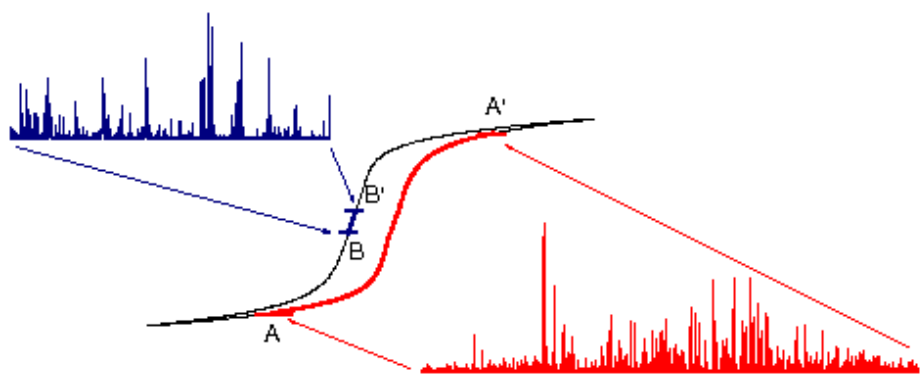}}
\caption{Hysteresis loop of a $Fe_{85}B_{15}$ amorphous alloy under
  moderate tensile stress ($10 MPa$). The signal measured along the
  whole half cycle (from $A$ to $A'$) is highly non--stationary. The
  interval $BB'$ represent the interval where the permeability can be
  assumed to be constant, and where the Barkhausen signal is
  stationary. (Reprinted with permission from \cite{durin_role_2006}). 
\cite{durin_role_2006} Copyright 2006 by Springer.}
\label{stationar}
\end{figure}

\begin{figure}
\centerline{\epsfxsize=12.5cm \epsfbox{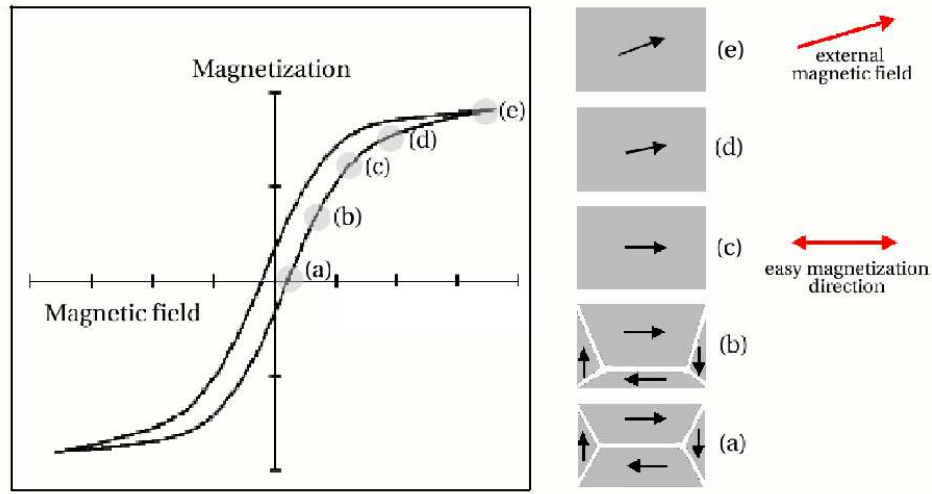}}
\caption{Typical magnetization loop for a soft magnet.  On the right
hand side the magnetization domain structures corresponding to five
different regions of the hysteresis loop are sketched, assuming that
only a single domain wall is present. The domain structure in (a)
corresponds to a zero magnetization configuration. The domain wall
divides the sample in two regions of opposite magnetization, oriented
along the material easy magnetization directions. The two small
triangular domains are the so called {\it closure domains}. Their
presence is neglected in the following treatment.  Increasing the
applied field, the domain wall moves as in (b) in order to increase
the size of the domain magnetized closer to the direction of the
applied field. In this region the hysteresis loop is linear. This
motion proceeds as the external field is increased, until the wall has
spanned the whole sample as in configuration (c). At this point the
sample is all magnetized in the same direction, which however does not
coincides in general with the direction of the external applied
field. From now on, further increasing the field, the magnetization
proceeds by coherent rotation of the spins (d), until the saturation
magnetization is reached (e), where all the spins are all aligned
parallel to the direction of the applied field.}
\label{magnetiz_reversal}
\end{figure}

When collecting BN data, it is indeed crucial to take into account
that the Barkhausen signal is stationary only in the linear region
around the hysteresis loop, where the intensity of the applied field
is close to the coercive field value (see figure \ref{stationar}). This is
precisely the region where the domain wall displacement is the
dominating ---if not the unique--- mechanism involved in the
magnetization reversal process. Getting closer to saturation, other
mechanisms enter into play, as the spins, which during the previous
phase of the magnetization process stay parallel to the easy
magnetization axis, start to coherently rotate in order to align to
the external field (see figure \ref{magnetiz_reversal}). 

Experimental studies that do not take this point into account, and
average Barkhausen signals over the whole hysteresis loop, will mix
the effect of the different magnetization mechanisms and of different
dynamical conditions, making very difficult to attribute to the
corresponding data a clear physical interpretation.

The role of stationarity has been recently further analyzed in great
detail by Durin and Zapperi \cite{durin_role_2006}. In their work they
show both with simulations and experiments that in non--stationary
conditions the distributions of avalanches, being integrated over
different values of the control parameters, are characterized by
larger effective exponents.  This might explain some experimental
results which give unusually large values for the exponents in the
power law distributions of sizes and durations of avalanches
\cite{spasojevi_barkhausen_1996}. Evaluating the effects of
non--stationary conditions on the statistical behavior of a system is
an issue that arises in many driven non--equilibrium situations, and
is analyzed in a more general context in
\cite{sornette_sweeping_1994}.

A detailed analysis of the data from many Barkhausen experiments
reported in literature is given by Durin and Zapperi in their
review \cite{durin_barkhausen_2006}, where the reliability of each
experiment considered is also discussed.

\subsection{Barkhausen noise phenomenology}
\label{sec:PH}

BN has some remarkable statistical properties, that are nowadays
measured with great precision in experiments. We briefly report the
main experimental results here. 

\subsubsection{Signal amplitude distribution}
\begin{figure}
\centerline{\epsfxsize=13cm \epsfbox{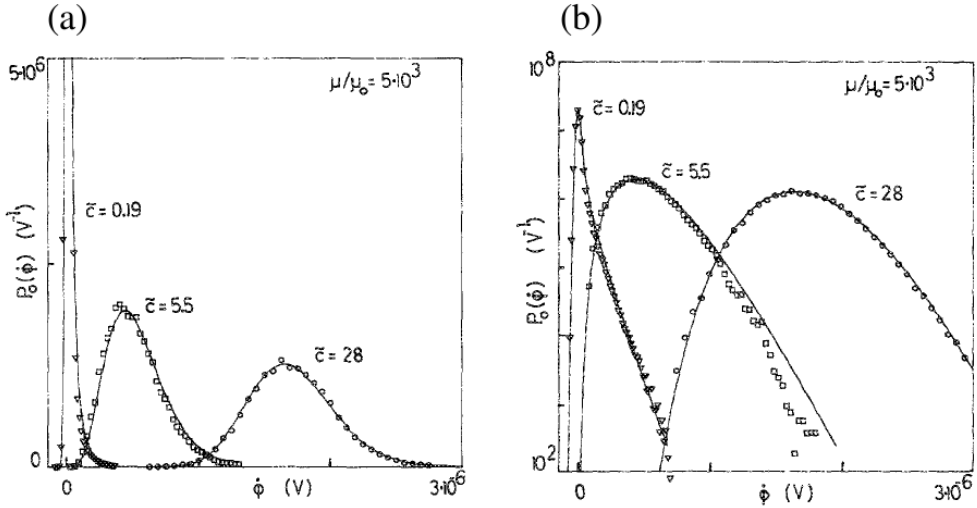}}
\caption{(a) Amplitude probability distribution $P_0(\dot{\Phi})$
  (which corresponds to $P(V)$ in our notations) at three different
  values of the driving field rate. $\dot{\Phi}$ is the variation of
  magnetic flux assumed to be proportional to the wall
  velocity. $\tilde{c}$ is proportional to $c$. The experimental data
  on a polycrystalline $3\%$ $FeSi$ sample are fitted with equation
  \ref{V_dist}. (b) Same as (a), in semilogarithmic scale. (Reprinted
  with permission from
  \cite{alessandro_domain-wall_1990-4}). \cite{alessandro_domain-wall_1990-4}
  Copyright 1990 by the American Institute of Physics.}
\label{Pv}
\end{figure}

The amplitude $V$ of the BN signal, i. e. the instantaneous value of
the voltage, is found to follow a distribution well fitted by
\begin{equation}
P(V) \propto V^{-(1-c)} \exp{(-V/V_0)} \,,
\label{V_dist}
\end{equation}
where $c$ is proportional to the driving field rate, and $V_0$ is some
characteristic cut--off value (see figure \ref{Pv}).

\begin{figure}
\centerline{\epsfxsize=10cm \epsfbox{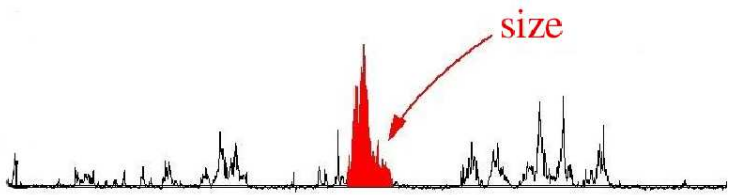}}
\caption{Size of an avalanche.}
\label{size}
\end{figure}

\begin{figure}
\centerline{\epsfxsize=10cm \epsfbox{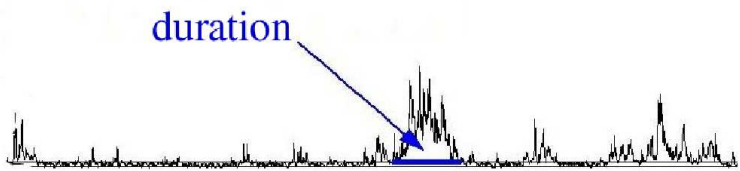}}
\caption{Duration of an avalanche.}
\label{duration}
\end{figure}

The amplitude distribution of the signal is an average quantity over
avalanches. To go more into detail, and investigate the statistical
properties of the single avalanches, it is necessary to define
precisely where a jump starts and ends. This is commonly done by
fixing a threshold value, which cannot be strictly zero due to the
unavoidable presence of background noise (see figures \ref{size},
\ref{duration}). That of signal thresholding is however a very
complicated issue, which arises in many cases when one has to analyze
temporal series, and that would need further attention.  The choice of
the threshold is crucial, since it defines the objects whose
statistical properties one is looking at, and the absence of a clear
rule makes the identification of single pulses rather ambiguous.  On
the other hand, signal thresholding is inescapable if one wants to
analyze directly experimental time series, it would therefore be
desirable to be able at least to control its effects on the measured
quantities. An alternative approach to the statistical analysis of the
signals, which bypasses completely ambiguous thresholding methods, is
to focus on the spectral properties, which indicate temporal
correlations in the system, and then try to infer from there the
statistical properties of the avalanches. This is indeed possible in
some cases thanks to the fact that the power spectra exponents are
related to the exponents in the size and time distribution of pulses
as it was shown by Kuntz and Sethna \cite{kuntz_noise_2000} for the
random field Ising model and some of its variants, and by Laurson,
Alava, and Zapperi \cite{laurson_power_2005} for the case of sand
piles.

In the case of BN, the experimental data have been shown to be
relatively insensitive to the exact choice of the threshold
\cite{durin_fractals_1995}. Background noise limits the detection of
small avalanches. Choosing the threshold below its typical amplitude
would result in over--estimating very small avalanches, but it is
expected not to affect sensibly the rest of their distribution. The
characteristic signal amplitude $V_0$ in equation (\ref{V_dist}) gives
a good reference value for the threshold, which is usually taken
between $5$ and $15\%$ of $V_0$.


\subsubsection{Distribution of sizes and durations of Barkhausen pulses}
Once the pulses are well defined one can look at the statistics of
their sizes and durations. By drawing an histogram it appears that they
are also both characterized by power law distributions, limited by an
exponential cut--off: the probability of having
a pulse of a given size has the form
\begin{equation}
p(S)=S^{-\tau}F_S(S/S_0)\,,
\label{size_distrib}
\end{equation}
where $F_S$ is some scaling function that accounts for the
cut--off, and $S_0$ is some characteristic avalanche size. 

The distribution of durations has a similar behavior: the probability
of having a pulse of duration $T$ is given by
\begin{equation}
p(T)=T^{-\alpha}F_T(T/T_0)
\label{time_distrib}
\end{equation}
where again $T_0$ is some characteristic time, and $F_T$ accounts for
the exponential cut--off.

\subsubsection{Universality classes}
\begin{figure}
\centerline{\epsfxsize=9cm \epsfbox{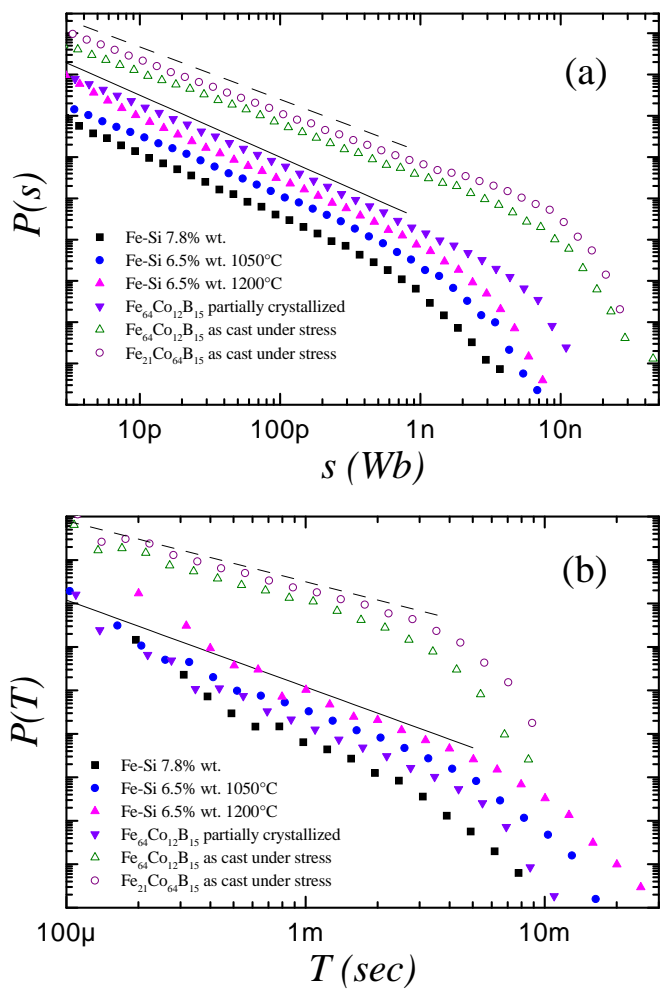}}
\caption{The data refer to Barkhausen measurements in different
  materials at the lowest available driving frequency. (a)
  Distribution of size of avalanches for different long range (filled
  symbols) and short range (empty symbols) materials. A solid line
  corresponding to slope $1.5$ (long range) and a dashed line
  corresponding to slope $1.27$ (short range) are shown for
  comparison. (b) Distribution of duration of avalanches for different
  long range (filled symbols) and short range (filled symbols)
  materials. A solid line corresponding to slope $2.0$ (long range)
  and a dashed line corresponding to slope $1.5$ (short range) are
  shown for comparison. (Reprinted with permission from
  \cite{durin_scaling_2000}).  \cite{durin_scaling_2000} Copyright
  2000 by the American Physical Society.}
\label{pspt_lrsr}
\end{figure}

By repeating BN experiments on different samples of different
materials and in different external conditions, it has been observed
that the exponents characterizing the power law distributions are
universal, and tend to cluster around two sets of values (see figure
\ref{pspt_lrsr}) \cite{durin_scaling_2000}.  In soft magnets, this
allows to identify two distinct universality classes, which are named
{\it long range} and {\it short range}, according to the kind of
elastic interaction which dominates the behavior of the magnetic
walls.

The two classes are characterized by exponents that, in the zero
frequency limit of the applied field, take the values $\tau \simeq
1.50$, $\alpha\simeq 2.0$ for materials in the long range class, and
$\tau \simeq1.27$, $\alpha \simeq 1.5$ for those in the short range
class, respectively for the avalanches' sizes and durations
distributions (see figure \ref{pspt_lrsr}).

Typical materials belonging to the long range class are
polycrystalline $FeSi$ with high $Si$ content, and partially
crystallized $Fe_{64}Co_{21}B_{15}$ amorphous alloys. Short range
behavior is observed for instance on Perminvar
($Fe_{30}Ni_{45}Co_{25}$) and amorphous alloys with composition
$Fe_xCo_{85-x}B_{15}$ under tensile stress.

We will restrict here to materials belonging to the long range class,
since for these systems the mean field description given by the ABBM
model, which is the focus of this review, turns out to be exact for
universal properties at long length and time scales (see
section \ref{sec:MF}).

\subsubsection{Distributions of pulses at finite driving rates}
The values $\tau\simeq1.5$ and $\alpha\simeq2.0$ for the long range
elasticity class, refer to experiments performed in the quasi--static
limit, namely, such that the applied field is increased very slowly
with respect to the typical time scale of the system, in such a way
that the system has time to rearrange after every infinitesimal change
of the field before the field is increased again.

\begin{figure}
\centerline{\epsfxsize=8cm \epsfbox{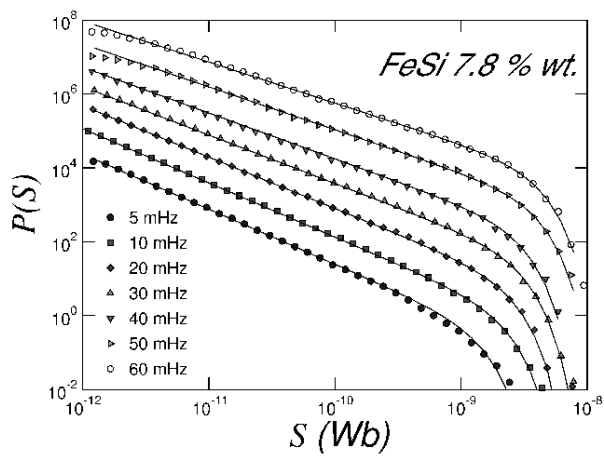}}
\caption{Cumulated distribution of sizes $P(S)=\int_{S}^{\infty} 
  p(s) ds$ for a polycrystalline $FeSi$ material at different applied
  field frequencies.  The solid lines are fits with equation
  \ref{size_distrib}, and with $\tau-1=(1-c)/2$. The exponent $\tau$
  varies linearly with frequency between $1.5$ and $1.15$. (Reprinted
  with permission from \cite{durin_barkhausen_2006}).
  \cite{durin_barkhausen_2006} Copyright 2005 by Academic Press. }
\label{ps_c}
\end{figure}

In experiments where the frequency of the field is finite but still
small, one can observe a linear dependence of the exponents $\alpha$
and $\tau$ on the driving field rate $c$, well fitted by the
expressions $\alpha=2-c$, and $\tau=(3-c)/2$ (see figure \ref{ps_c},
where the cumulated distribution of sizes of BN pulse for a
polycrystalline $FeSi$ sample is shown for seven values of the field
sweep rate). No similar dependence has been observed in materials
belonging to the short range class, where the exponents keep their
quasistatic limit values, independently of the driving field rate.

\subsubsection{Threshold in the field rate, transition from intermittent to
continuous regime}
\begin{figure}
\centerline{\epsfxsize=8cm \epsfbox{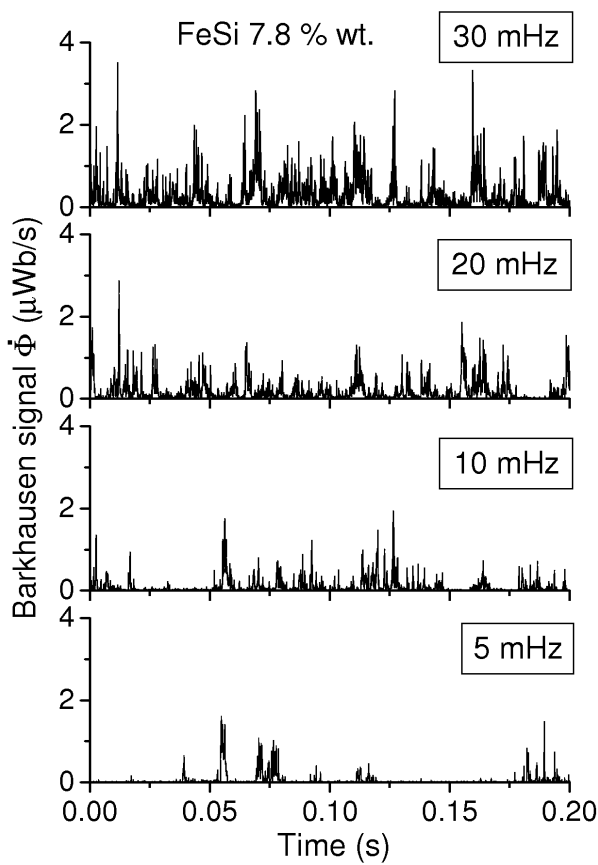}}
\caption{Barkhausen signal recorded at increasing sweep
  rates. (Reprinted with permission from
  \cite{durin_barkhausen_2006}).  \cite{durin_barkhausen_2006}
  Copyright 2005 by Academic Press. }
\label{BN_c}
\end{figure}

The variation of the exponents $\alpha$ and $\tau$ with the driving
field rate $c$ reflects the fact that, increasing the sweep rate,
large events become more frequent. Indeed, even without a statistical
analysis, it is clear by eye inspection that the Barkhausen signals
become more and more pronounced as the frequency of the applied field
is increased (see figure \ref{BN_c}). 

Further increasing the driving field rate, it is observed that, above
some well defined threshold, the typical intermittent behavior is
lost, single pulses disappear, and the magnetization reverses in a
unique avalanche. This is a sort of meta--phase transition, which
separates two completely different regimes of the dynamics. It would
be interesting to see how the system approaches the critical driving
rate, by monitoring for example the quiescence times separating the
avalanches as $c$ approaches the threshold value. This, to our
knowledge, has never been done. A similar situation arises for example
in sheared granular matter, where a threshold driving separates the
stick--slip form the sliding regime. This effect is due to avalanches
overlapping aver time: since at finite driving rates the external
field increases during the avalanche, it can keep active an avalanche
that would have stopped otherwise (see subsection \ref{sec:MAR} and
the paper by White and Dahmen \cite{white_driving_2003}).

\subsubsection{Cut--off in the distribution of sizes and durations}

\begin{figure}
\centerline{\epsfxsize=9cm \epsfbox{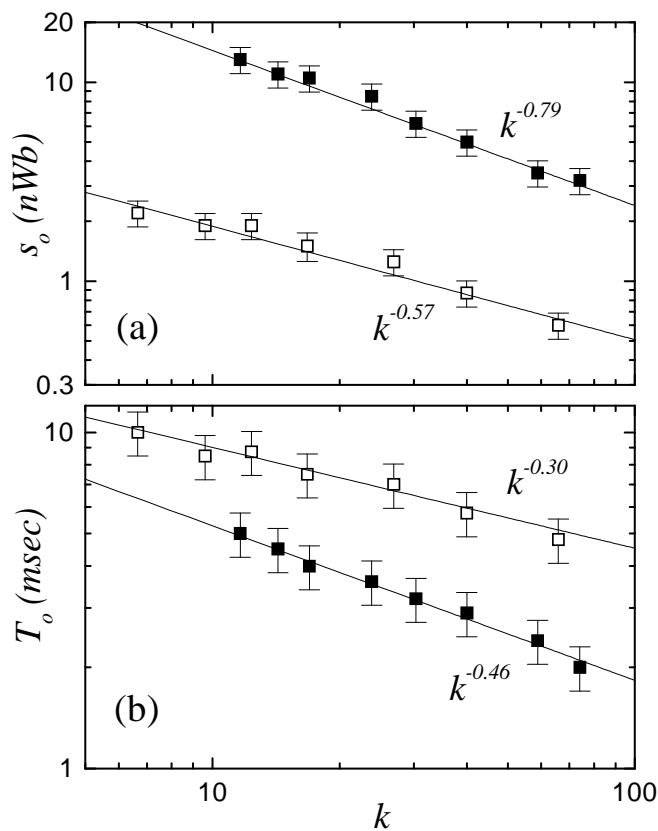}}
\caption{Top graph: experimental estimations of the cut--off $S_0$ of
  Barkhausen pulse size distribution as a function of the
  demagnetizing factor $k$ in a $FeSi$ 6.5 wt\% alloy (empty symbols),
  and an amorphous alloy $Fe_{21}Co_{64}B_{15}$ under stress (filled
  symbols). The empty symbols correspond to the long range class
  considered here. Bottom graph: same as the top for the cut--off
  $T_0$ of the Barkhausen pulse duration distribution.  (Reprinted
  with permission from \cite{durin_scaling_2000}).
  \cite{durin_scaling_2000} Copyright 2000 by the American Physical
  Society.  }
\label{So_To}
\end{figure}

The power law distributions of sizes and durations of avalanches are
limited by an exponential cut--off, which turns out to be related to
the demagnetizing field (see subsection \ref{sec:BIAS}), which in its
turn depends on the sample geometry.  Experiments repeated on samples
with different geometries, indeed confirm that the position of the
exponential cut--offs $S_0$ and $T_0$ in the distributions
(\ref{size_distrib}) varies with the shape of the sample. However, a
precise experimental evaluation of the cut--off dependence on the
parameter $k$ measuring the strength of the demagnetizing field is
very difficult, also due to the further dependence of the distribution
on the driving rate.

Earlier studies \cite{zapperi_dynamics_1998} report the scaling
$S_0\sim {k^{-1}}$, and $T_0\sim k^{-1/2}$. The most reliable
experiments are probably those reported in \cite{durin_scaling_2000,
durin_universality_2000}, done by taking measurements on the same
ribbon repeatedly cut along one direction, in order to modify the
effect of the demagnetizing field, while stresses and internal
disorder are kept constant. For a $FeSi$ sample, the scalings reported
are $S_0\sim {k^{-0.57}}$, and $T_0\sim k^{-0.30}$ (see figure
\ref{So_To}).

\subsubsection{Average pulse shape}
Averaging avalanches of the same duration $T$, one obtains, for every
given $T$ a characteristic pulse shape. 

\begin{equation}
\langle V(t \mid T) \rangle
\label{vav}
\end{equation}

\begin{figure}
\centerline{\epsfxsize=9cm \epsfbox{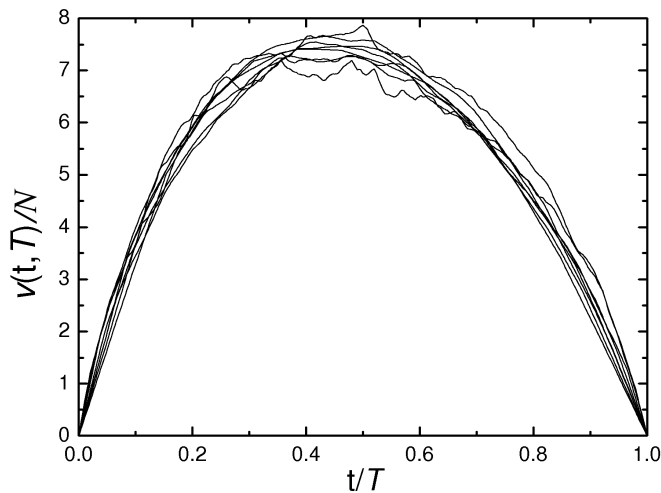}}
\caption{Attempt to collapse average pulses of different durations
  onto the same function. The data are from a BN experiment on a
  polycrystalline $FeSi$ sample. Pulses are normalized with
  $N\propto\int_0^1 d(t/T) \langle v(t \mid T) \rangle$. The curves
  are averages over several pulses for each of eight different values
  of $T$, ranging from $0,16$ to $2,03$ ms. (Reprinted with permission
  from \cite{durin_barkhausen_2006}).  \cite{durin_barkhausen_2006}
  Copyright 2005 by Academic Press.}
\label{shape}
\end{figure}

In analogy to conventional critical phenomena one might expect that
average pulses of different durations, properly rescaled by some power
of their duration, will depend on $t$ only through the scaling
variable $t/T$, and therefore collapse onto the same universal
function.  This quantity was recently proposed in
\cite{sethna_crackling_2001} to test theory against experiments, and
indeed it provides a much stringent tool than the simple comparison of
scalar quantities, such as critical exponents.

As shown in figure \ref{shape}, which reports data from a BN
experiment on a polycrystalline $FeSi$ sample, the rescaling of
experimental pulses can approximately be performed.  As we will see,
the shape of the scaling function is influenced by some non--universal
effects that however disappear, as expected, on very long time scales.

\subsubsection{Power spectra}
In most of earlier literature, experimental papers focus on the
detection of the power spectra of the noise. The most accurate report
of power spectra are reported in the seminal paper by Alessandro {\it
et al.} \cite{alessandro_domain-wall_1990-4} in experiments on a $3\%$
$FeSi$ sample. For experiments on single crystals and other
polycrystalline $FeSi$ materials see also
\cite{durin_barkhausen_1997}.

\begin{figure}
\centerline{\epsfxsize=9cm \epsfbox{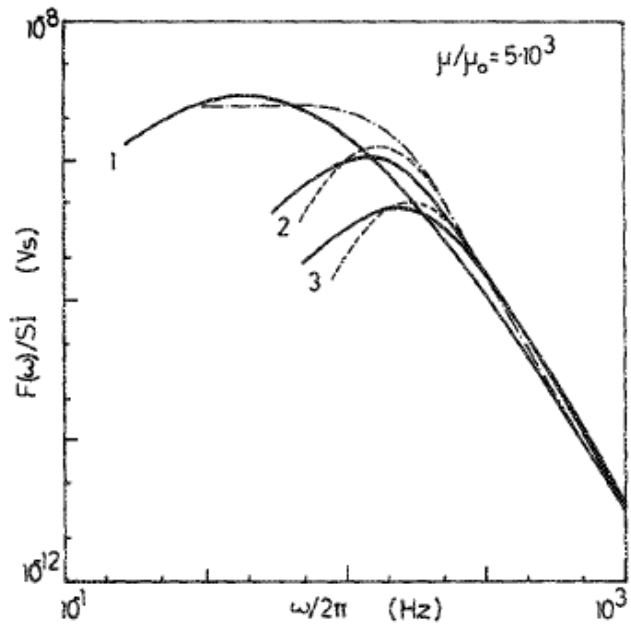}}
\caption{Log--log plot of the power spectra divided by the driving
  field rate, versus frequency measured in a polycrystalline $3\%$
  $FeSi$ sample. The three solid lines are the experimental results for
  three different driving field rates, the dotted lines are predictions
  from the ABBM model. (Reprinted with permission from
  \cite{alessandro_domain-wall_1990-4}). \cite{alessandro_domain-wall_1990-4}
  Copyright 1990 by the American Institute of Physics.}
\label{pws_b}
\end{figure}

\begin{figure}
\centerline{\epsfxsize=12cm \epsfbox{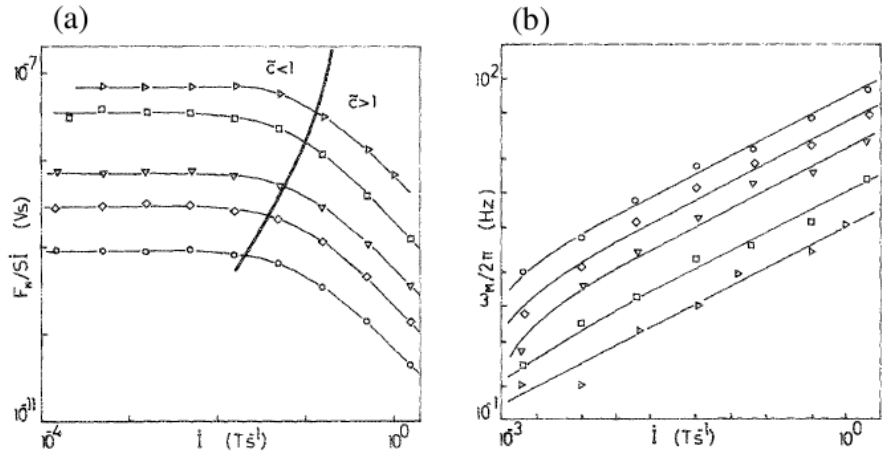}}
\caption{ (a) Amplitude of the maximum $F_M$ in the frequency spectrum
  divided by driving field rate versus driving field rate in log--log
  scale. Different data--sets refer to different values of the
  permeability. The bold line separates the two regimes intermittent
  and continuous. $\tilde{c}$ is equivalent to $c$ in our
  notations. (b) Same as (a), for the frequency $\omega_M$
  corresponding to the maximum of the power spectrum. (Reprinted with
  permission from
  \cite{alessandro_domain-wall_1990-4}). \cite{alessandro_domain-wall_1990-4}
  Copyright 1990 by the American Institute of Physics.}
\label{pws_b_c}
\end{figure}

For materials in the long range universality class the power spectra
is found to decay at large frequencies $\omega$ as $\omega^{-2}$ (see
figure \ref{pws_b}), although in same cases a deviation from this
scaling is observed at intermediate frequencies. At small frequencies
the spectrum is characterized by the presence of a maximum. The
position of the peak depends on the driving field rate $c$: it occurs
at a frequency $\omega_M$ roughly proportional to $c^{1/2}$, and with
an amplitude $F_M$ that scales as $c$ in the intermittent regime
(slow driving), while it is roughly constant in the continuous regime
(fast driving)(see figure \ref{pws_b_c} where $F_M/c$ is plotted
versus $c$).

The behavior of the power spectra reported in
\cite{alessandro_domain-wall_1990-4} and shown in figure \ref{pws_b}
is indeed common to many ferromagnetic materials. Similar spectra are
found for example in $NiFe$ alloys with vanishing anisotropies
\cite{coudercon_magnetization_1989}, grain oriented $FeSi$ alloys
\cite{bertotti_barkhausen_1981}, $FeSi$ single crystals
\cite{bertotti_microscopic_1984}.

At frequencies below the peak the power spectrum has been claimed to
behave like some power $\omega^{\psi}$, with approximately $\psi\simeq
0.6$. However, these claims are based on fits of data that are spread
on less than one decade, and therefore are not conclusive.

Although the large $\omega$ tail and the existence of the peak are
reproduced by the theory, the correct interpretation of all the
details in the complex behavior of the frequency spectrum still
represents a challenging problem. In principle the contributions from
the correlations among different avalanches, and that from the
correlations within avalanches are mixed together in the power
spectrum, which therefore is not easily amenable of physical
interpretation.

Some authors have also considered higher order spectra to further
characterize the BN signal \cite{obrien_statistical_1994,
petta_barkhausen_1998}. In principle this analysis could indicate
whether events on a certain scale would systematically precede events
on another scale.  While no precursor events have been found, time
asymmetries of the signal emerge from the analysis of the imaginary
part of the moment of order $1.5$ consistently with that observed in
the skewed pulse shape (see below).

\subsubsection{Summary of experimental observations}
\label{sec:sum_p}
It is useful at this point to summarize the main phenomenological
observations collected from different BN experiments, that we will
want to compare to the theoretical predictions in the next sections.

For soft materials with long range interactions: 

\begin{enumerate}
\item{The distribution of the intensity of the BN signal follows a power law
  decay with an exponent $1-c$ where $c$ is proportional to the
  driving field rate. }
\item{The sizes and durations of avalanches are power law
  distributed. In experiments performed in the quasi--static limit the
  exponents are $\alpha\simeq 1.5$ and $\tau\simeq 2.0$, respectively.}
\item{The distributions of sizes and durations of avalanches flatten as
  the sweep rate is increased. The corresponding exponents vary
  linearly with the field rate $c$, and their values are well fitted
  by the relations $\alpha=2-c$, and $\tau=(3-c)/2$.}
\item{The previous statement hold as long as the sweep rate is not too
  high: there exist a well defined threshold in the driving field rate
  $c$ above which the Barkhausen pulses become indistinguishable and
  magnetization reverses in a unique avalanche.}
\item{The exponential cut--offs $S_0$, and $T_0$ in the distributions
  of sizes and durations of avalanches are related to the sample
  geometry through the demagnetizing field, and scale as
  powers of $k$. }
\item{Average pulses properly rescaled by their duration can
  approximately be collapsed onto the same function. }
\item{The power spectrum decays as $\omega^{-2}$ at large
  frequencies. At small frequencies is characterized by a peak that
  occurs at $\omega_M \simeq c^{1/2}$, with an amplitude
  that in the slow driving regime scales roughly as $F_M\simeq
  c$. }
\end{enumerate} 

\subsection{Modern theoretical approaches to Barkhausen noise}
\label{sec:approach}

Recent theoretical studies on BN have moved from a pure
phenomenological description, where BN is considered as a
superposition of elementary events, without any connection to the
microscopic dynamic, to more physical and detailed descriptions which
take into account the relevant interactions governing the dynamics of
the magnetization reversal process.

The theory has then developed on two different but related avenues:
one follows a microscopic description and models the magnetic system
through interacting spins. In this framework, the criticality observed
in BN experiment is explained in terms of the proximity of a disorder
driven non--equilibrium phase transition, which is observed in the
zero temperature random field Ising model (RFIM). The other approach
describes the magnetization process in mesoscopic terms, through the
dynamics of the domain walls. In this case, the criticality is related
to the depinning transition occurring when the applied field overcomes
a critical threshold value. We will briefly discuss here the first
approach, and refer to the next sections for the second one, which is
the one taken in this is review.

The attempt to model the Barkhausen phenomenon in terms of spin models
is indeed appealing. Thanks to universality on expects the statistical
properties of BN to be independent of the microscopic details of the
system, provided that the relevant symmetries of the problem are
properly taken into account. This allows in principle to derive
macroscopic properties without going through the analysis of
complicated micromagnetic equations, which are very difficult to treat
analytically. Instead, one models the avalanche dynamics with
simplified rules for the evolution of integer valued spins. This
approach allows both to perform large scale simulations, and to derive
some theoretical result.

In particular, among all the spin models, the non--equilibrium random
field Ising model (RFIM), proposed by Sethna {\it et
al.} \cite{sethna_hysteresis_1993} as a model for avalanche dynamics in
BN, has been widely studied both analytically and numerically
\cite{dahmen_hysteresis_1993, perkovi_avalanches_1995,
dahmen_hysteresis_1996, perkovi_disorder-induced_1999,
sethna_crackling_2001, travesset_crackling_2002,
carpenter_barkhausen_2003}. In this model integer valued spins
$s_i=\pm1$ are assigned to each site of a $d-$dimensional lattice, and
interact with the nearest neighbors through a ferromagnetic coupling
$J$. The spins are coupled to the external field $H$, and to a
quenched random field $h_i$ extracted from a Gaussian distribution
with variance $R$, which is associated to each site of the lattice,
and is supposed to mimic the presence of disorder. The Hamiltonian
reads
\begin{equation}
{\cal{H}}=-\sum_{\langle i,j\rangle}Js_i s_j
-\sum_i(H(t)+h_i)s_i
\label{RFIM}
\end{equation}
where the first sum is restricted to the nearest neighbors.  In the
non--equilibrium zero temperature dynamics proposed by Sethna and
co--workers \cite{sethna_hysteresis_1993}, at each time step the spins
align with the local field:
\begin{equation}
s_i=\mbox{sign} \left( J\sum_j s_j + H(t) +h_i \right) \,,
\label{RFIMd}
\end{equation}
where, again, the first sum is restricted to the nearest neighbors of
the spin $i$. Starting from a configuration with all the spins down,
the field $H$ is slowly ramped from $-\infty$ to $+\infty$. A single
spin flip can cause nearest neighboring spins to flip, and trigger an
avalanche. In the quasistatic limit the field $H(t)$ is kept constant
during the avalanches.

For small values of $R$, the random field is not strong enough to stop
the avalanches, and the hysteresis loop has a squared shape, with a
large jump at a critical field $H_c$ at which the big avalanche
starts. When $R$ is large, only small avalanches are observed:
essentially every spin flips independently when $H(t)$ is large enough
to overcome the local pinning field $h_i$, plus the comparably small
energetic contribution from the coupling $J$ with the neighboring
spins. The two regimes are separated by a critical value $R_c$ of the
variance of the random field, for which the avalanches are power--law
distributed.

This model predicts that for the critical value of the disorder
$R=R_c$, at the critical point, in $d=3$, and in the quasi--static
limit, the exponents in the power--law distributions of size and
duration of the avalanches are given by $\tau\simeq 1.6$, and
$\alpha\simeq2.05$.  These values have been obtained both by large
scale numerical simulations \cite{perkovi_disorder-induced_1999}, and
with an epsilon expansion around the mean field solution, that has
been be worked out analytically \cite{dahmen_hysteresis_1996}.

The main problem with this approach is that RFIM predicts scaling only
at a critical value of the disorder, while there is not reason to
expect this condition to be satisfied in experiments. Another problem
is that the results for the critical exponents in the non--equilibrium
RFIM are only marginally compatible with the values observed in
experiments, where the exponents are around $\tau \simeq 1.50$,
$\alpha\simeq 2.0$, or $\tau \simeq1.27$, $\alpha \simeq 1.5$
depending on the range of dipolar interactions.

\begin{figure}
\centerline{\epsfxsize=8cm \epsfbox{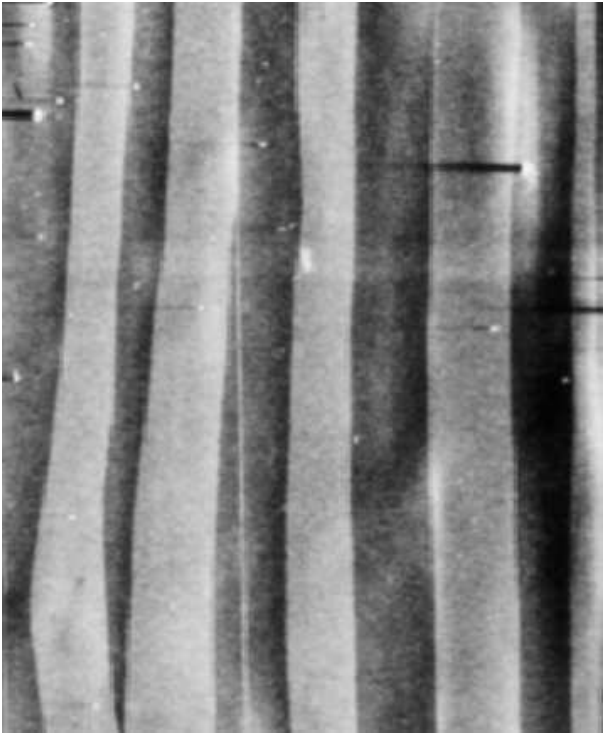}}
\caption{Domain structure of $Fe_{21}Co_{64}B_{15}$ amorphous alloy
  observed by scanning electron microscopy. This is the typical
  structure observed in soft ferromagnetic materials.  (Reprinted with
  permission from \cite{zapperi_dynamics_1998}).
  \cite{zapperi_dynamics_1998} Copyright 1998 by the American Physical
  Society. }
\label{domains}
\end{figure}

More importantly, even a rough investigation of the qualitative
behavior of the non--equilibrium RFIM shows that the domain structure
is very different from the one observed in real magnets (see figure
\ref{domains}). While in the RFIM avalanches nucleate in different
points of the system and grow with a fractal spatial structure, in
real soft magnets domain walls are rather flat
\cite{sethna_course_2007}.

In the RFIM the magnetization proceeds simultaneously by two different
mechanisms, nucleation and wall displacement, while in real soft
magnets (and in the central part of the loop, where BN is measured)
the wall movement is the only relevant mechanism of magnetization
reversal: the dynamic evolves by the motion of existing boundaries,
rather then by nucleating new domains. As a consequence, the RFIM, as
it is, is inadequate as a model for BN in soft magnets, while it could
be an interesting starting point to describe magnetization dynamics in
hard magnetic materials.

The non--equilibrium RFIM neglects dipolar interactions, which
originates, both from free dipoles which are present on the sample
surface, and from those on the magnetic interface. Magnetostatic
dipolar interactions due to the presence of ``free magnetic charges''
on the sample boundaries, generate a long range demagnetizing field,
which opposes the external applied field.  The most relevant
ingredient missing in the RFIM is indeed the demagnetizing field,
which reflects in the flat domain domain structure observed in soft
materials (see figure \ref{domains}), very different from the
structure found in the RFIM, where demagnetizing forces are not taken
into account. The other crucial effect of the demagnetizing field, as
we discuss below, is that it prevents the formation of large
avalanches.

Several variants of the RFIM have been proposed to overcome these
problems.  A particularly successful one, the front propagation model
(FPM), originally introduced in \cite{ji_percolative_1992}, has been
extensively analyzed in \cite{kuntz_noise_2000, mehta_universal_2002,
carpenter_barkhausen_2003} as a model for BN. In this variant of the
RFIM, the dynamics is such that only spins close to an existing domin
boundary take part to the dynamics, while spins not adjacent to an
avalanche front are prohibited to flip, even when it would be
energetically favorable, so that the nucleation of new domains is
totally suppressed.  This can also be implemented by introducing in the
RFIM Hamiltonian a long range term mimicking the demagnetizing field.
The FPM gives exponents $\tau\simeq 1.27$, and $\alpha\simeq 1.72$ in
$d=3$, which are compatible with BN experiments for soft magnets in
the short range universality class.

It has to be stressed that, unlike the RFIM, the FPM exhibits critical
behavior in the whole low disorder regime $R<R_c$ and not just at the
critical value $R_c$.  Indeed, the demagnetizing field opposes the
reversal process and forbids the magnetization to change abruptly, as
it is the case, below $R_c$, in the original non--equilibrium RFIM:
the demagnetizing field reduces the effective field until it stops the
domain growth, which will start again with a new avalanche only by the
increase of the external field.  This mechanism will be treated
extensively in the next section. The criticality observed is therefore
not related to the disorder induced transition of the non--equilibrium
RFIM, but to a depinning transition. 

Dipolar interactions between free dipoles on the magnetic wall produce
an effective elasticity between different parts of the domain wall,
that can be short range or long range, depending on the material. The
long range elastic interaction has been claimed to lower the upper
critical dimension to d=3 \cite{zapperi_dynamics_1998}. In such
materials, a mean field approximation is therefore expected to apply. 
In mean field, the front propagation model gives $\tau \simeq 1.5$, in
good agreement with experiment on materials in the long range
universality class. 

Another point that should be considered carefully in any RFIM approach
to BN regards the kind of disorder it represents, since random fields
are not present in real magnets. It is therefore crucial to understand
whether the RFIM non--equilibrium phase transition represents a broad
enough universality class that is preserved when other kinds of
disorder are introduced. This question is still controversial. Several
studies have attempted to understand by numerical simulations whether
other disordered spin models belong to the same universality class as
the RFIM. Examples include the non--equilibrium random bond Ising
model (RBIM), with Hamiltonian
\begin{equation}
{\cal{H}}=-\sum_{\langle i,j\rangle}J_{i,j}s_i s_j
-\sum_i H(t)s_i \,,
\label{RBIM}
\end{equation}
where the disorder enters into the exchange couplings $J_{i,j}$, and
the non--equilibrium site diluted Ising model (SDIM), described by
\begin{equation}
{\cal{H}}=-\sum_{i,j}J c_i c_j s_i s_j
-\sum_i H(t)c_i s_i \,,
\label{SDIM}
\end{equation}
where the quenched variables $c_i$, taking the values $[0,1]$,
indicate whether a given site represents a magnetic spin or not, thus
modeling the presence of non--magnetic inclusions in the material.

Simulations show that both the non-equilibrium RBIM
\cite{vives_avalanches_1994} and the non--equilibrium SDIM
\cite{vives_hysteresis_2000} exhibit a disorder induced phase
transition, with a behavior qualitatively similar to that arising in
the RFIM. The critical exponents for the avalanche dynamics differ
slightly, but are compatible, with those of the RFIM, suggesting that
RFIM may indeed represent a broad universality class.  The robustness
of RFIM with respect to different kinds of disorder is also discussed
by Dahmen and Sethna in \cite{dahmen_hysteresis_1996}, where it is
argued that universality is expected to hold on the base of
renormalization group calculations. The authors suggest that although
RBIM and the random anisotropy model have higher symmetries then the
RFIM in equilibrium, they have the same symmetries in non--equilibrium
conditions, since the external field at the critical point breaks the
up--down symmetry.
 
To conclude our short discussion of spin models, we can summarize that
the simple RFIM is inadequate to describe the Barkhausen effect in
soft magnets, rather it could be a good starting point to model
magnetization reversal process in hard magnets. The FPM, obtained from
the RFIM by incorporating the demagnetizing field, and suppressing
domains nucleation, is instead a more adequate spin model for soft
magnets, and give results in good agreement with experiments on
materials in the short range universality class for $d=3$, and for the
long range in mean field.

It is however very important to stress that in the FPM, the observed
critical behavior is not related to the proximity of the RFIM
non--equilibrium critical point (indeed criticality is observed for
all $R<R_c$, i.e. in the whole low disorder regime), but rather to the
depinning transition of the magnetic interface that will be discussed
extensively in the next sections. 

In some sense spin models can be more general then interface models,
as can easily deal for example with interface overhangs, however, in the
case of soft magnets, given the nature of the transition, a simplified
approach in terms of domain wall dynamics, seems more convenient, and
has the advantage to allow to treat separately domain wall propagation
and nucleation. The FPM naturally provides a discretized way to
implement the front dynamics, very suitable for simulations. An
alternative approach, which we follow in this review, is to formulate
directly an equation of motion for the propagating wall.

\section{The ABBM model}
\label{sec:ABBM}

The first important theoretical approach to BN was done in 1990 by
Alessandro, Beatrice, Bertotti, and Montorsi, who proposed a
phenomenological model, which was then named after the authors of the
two companion papers \cite{alessandro_domain-wall_1990,
alessandro_domain-wall_1990-4}, one dealing with theory, the other
with experiments.  By--passing all the difficulties that appear when
trying to approach the problem from basic micromagnetic equations,
Alessandro {\it et al.} managed to capture the essential ingredients at
the origin of BN, by formulating a very simplified equation for the
domain wall dynamics. Their model turned then out to be extremely
successful in reproducing the experimental observations, and played a
fundamental role in most of the successive theoretical progress, and
in the development of a consistent theory for BN.

The ABBM model was inspired by the work of Ne\'el
\cite{nel_theory_1942,nel_1943}, who was the first to introduce a
random energy model in the study of hysteresis: in Ne\'el's model, the
complex domain structure usually present in a magnetized sample is
replaced by a single domain wall, moving in a random energy landscape,
which is supposed to account for all the interactions among walls, and
of the walls with the medium. The energy profile is constructed with a
sequence of equispaced parabolic arcs, whose curvatures are extracted
from a Gaussian distribution, in such a way that, on small scales, the
motion is reversible, while, on large scales, the wall moves in random
irreversible jumps. This model was introduced to explain in simple
terms the Rayleigh law in hysteresis.

\begin{figure}
\centerline{\epsfxsize=8cm \epsfbox{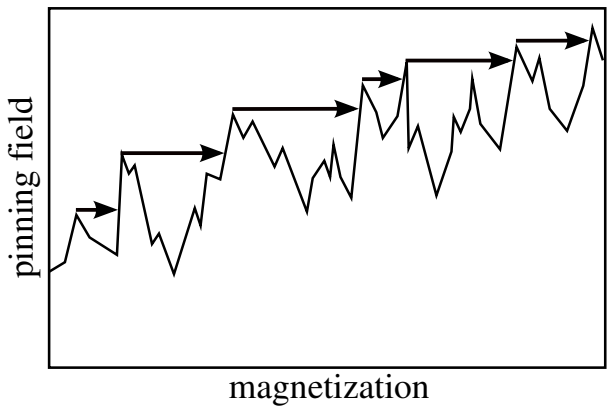}}
\caption{Random energy landscape representing the pinning field, the
  arrows represent the Barkhausen jumps. In random energy models, when
  the system has enough energy to overcome a barrier, it jumps
  irreversibly to the next valley, with a corresponding random jump in
  the magnetization. Then the system is stuck until the effective
  field has grown enough to overcome the new energy barrier.}
\label{pinning}
\end{figure}

In the Alessandro {\it et al.} approach, the idea of a random energy
landscape is generalized, and used to construct a stochastic equation
for the domain wall dynamics, in order to describe the Barkhausen
effect. The equation is derived by following a previous work by
Williams {\it et al.}  \cite{williams_studies_1950}, who studied the
dynamics of a single domain wall under an external field in a $FeSi$
single crystal sample. In this work, the motion of the wall, limited
by eddy currents damping, was described by a linear relation between
the average velocity and the net field acting on the interface, given
by the difference between the applied field $H(t)$ and some threshold
field $H_0$, which includes all kinds of random (like interaction with
pinning centers) and non--random (like dipolar forces and
demagnetizing fields) interactions (see figure \ref{pinning}).

The ABBM model is derived by making a more stringent claim, namely
by assuming that the linear relation between velocity and effective
fields holds not only on average, but also on the instantaneous
velocity. By separating in $H_0$ the random contribution of the
pinning field $H_p$ from the demagnetizing component $H_d$, one could
then write
\begin{equation}
v(t) \propto H(t)-(H_d+H_p)\,.
\label{ABBM_0}
\end{equation}

The further assumption is that the pinning field is Brownian
correlated (see Eq. \ref{brownian_correlation} below), as suggested by
some experimental evidence in special system containing just one
active wall \cite{porteseil__new_1979, vergne_statistical_1981}. Let
us point out that this assumption is quite crucial, since an
uncorrelated pinning field, as the one considered in the original work
by Ne\'el, would not give rise to power law distributed magnetization
events.

In the following section we will first describe the ABBM equation, as
it was formulated in
\cite{alessandro_domain-wall_1990,alessandro_domain-wall_1990-4}.
Then, in order to put this model in a more general framework, we turn
to the description of the dynamics of a magnetic elastic interface in
a random media. Following \cite{cizeau_dynamics_1997}, we describe a
depinning interface model for BN, that in the mean field approximation
gives back the ABBM model \cite{zapperi_dynamics_1998}. As we will
see, this will also clarify the origin of the correlated effective
pinning field.

\subsection{ABBM as a phenomenological model}
\label{sec:ABBM_abbm}

The ABBM theory assumes that there is just one single rigid wall
separating two domains of opposite magnetization. The system has
therefore just one degree of freedom, and the dynamics is described in
terms of a one--dimensional equation of motion for the center of mass
of the wall.  The wall moves under the action of an applied driving
field in an effective Brownian correlated pinning field, mimicking the
presence of disorder in the sample, as discussed above.

Without loss of generality, we choose the origin of the coordinate
indicating the position of the wall at the center of the sample,
i.e. at the zero magnetization domain configuration, so that, in what
follows, the magnetization is proportional to the position of the
wall. The equation of motion for the center of mass of the wall is
then an equation for the magnetization.

The motion is assumed to be overdamped, so that the velocity of the
wall, which is proportional to the derivative of the magnetization, is
equal to the sum of the forces acting on the interface, with a damping
coefficient $\Gamma$, characterizing the amount of dissipation:
\begin{equation}
\Gamma \frac{\partial m}{\partial t}=H(t)-km+W(m)\,.
\label{ABBM}
\end{equation}

As mentioned before, the total force acting on the wall, on the
r.h.s. of the equation, includes two terms, besides the
time--dependent applied magnetic field $H(t)$, as in equation
(\ref{ABBM_0}). One is the demagnetizing field $H_d$ generated by free
magnetic charges on the boundary of the sample, which opposes the
external field and here is assumed to be proportional to the
magnetization $H_d=-km$, through a constant $k$ ({\it{demagnetizing
factor}}) that takes into account the geometry of the sample. The
other one is the random pinning field $H_p=W(m)$, which accounts for
defects and non--magnetic impurities present in the sample, and it is
assumed to be Brownian correlated in the magnetization:
\begin{equation}
\langle\mid W(m)-W(m')\mid^2\rangle=2D\mid m-m'\mid \,.
\label{brownian_correlation}
\end{equation}

\subsection{Limitations of the ABBM model}
\label{sec:limitations}

The ABBM model has the advantage to be exactly solvable, and to
reproduce with striking accuracy most of the statistical properties
observed in the experiments for soft magnetic materials in the long
range universality class, described in section \ref{sec:PH}. However,
it is worth to underline its main limitations.  One, that we will deal
with in the last part of the review (see section \ref{sec:ASYM}), is
that it is unable to reproduce some features of non--universal origin,
in particular, the asymmetric shape of Barkhausen pulses. 

Another important limitation of the ABBM model is that it deals with a
single domain wall, while in common experiments more interacting
domains are present in the specimen and participate in the
magnetization process. The ABBM predictions on the signal power
spectra are for example cleanly observed in experiment where a single
domain wall is present \cite{vergne_statistical_1981}, while other
experiments suggest that the power spectra changes when more domains
are present \cite{durin_microstructural_1994,
durin_measurements_1996}. To fully understand these experiments a
model which includes many interacting domain walls would be needed.
This point was indeed raised in the original paper
\cite{alessandro_domain-wall_1990}, where the authors argue that in
the case of many interacting domains, the same equation of motion
approximately holds, once the correlation associated to the pinning
field (see the discussion in section \ref{sec:PS2}) is multiplied by
the number of active walls.

Another feature of BN for which the description in terms of the ABBM
model is unsatisfactory is the partial signal reproducibility observed
in some experiments \cite{petta_multiple_1996, petta_dependence_1997},
that could be due to small variations in the initial conditions or in
the driving field rate or in the peak field, or to thermal effects. A
model able to describe this behavior should treat many interacting
degrees of freedom explicitly, and not in an effective manner as in
the ABBM model.

A comparison point to point between ABBM model predictions and the
experimental evidences, which are summarized in subsection
\ref{sec:sum_p}, is postponed to subsection \ref{sec:sum}.

\subsection{Elastic interfaces in a random media}
\label{sec:elastic_interface}

An analysis of the Barkhausen phenomenon at a mesoscopic level is done
by describing the system as an elastic interface in a random media
driven by an external field, and relates the criticality observed in
the experiments to the well known {\it depinning transition} of the
interface.

The response of an elastic manifold in a random media to an external
driving force is a well studied problem in statistical physics, that
arises in many contexts, for example in the study of fluids in porous
media \cite{stokes_interfacial_1986, cieplak_dynamical_1988,
cieplak_influence_1990, narayan_threshold_1993,
alava_imbibition_2004}, flux lines in type II superconductors
\cite{blatter_vortices_1994}, charge density waves
\cite{fisher_sliding_1985, grner_dynamics_1988,
narayan_critical_1992}, and crack fronts in solids
\cite{petri_experimental_1994, fisher_statistics_1997,
garcimartn_statistical_1997, salminen_acoustic_2002,
alava_statistical_2006, koivisto_creep_2007}.  In the following we
briefly review the main facts, for the case of a $d-1$ interface in
$d$ dimensions, and refer to the extended literature for more details
\cite{kardar_nonequilibrium_1998, fisher_d._s_collective_1998}.

Assuming that the motion is overdamped, so that inertial effects may
be neglected, the general equation describing the dynamics of an
elastic interface in a quenched random media, is obtained by equating
the damping term to the total force acting on the interface:
\begin{equation}
\Gamma \frac{\partial z}{\partial t}(\vec{r},t)=
-\frac{\delta E}{\delta z}\,,
\label{overdamped}
\end{equation}
where, assuming that there are no overhangs, $z$ defines the position
of the interface as a function of the $d-1$ dimensional coordinate
$\vec{r}$ on the substrate plane, and of time $t$. The coefficient
$\Gamma$ is the damping coefficient, ruling the rate of
dissipation. 

The simplest equation of motion for an interface in an isotropic
medium is the quenched Edward Wilkinson equation in a driving field,
given by
\begin{equation}
\Gamma \frac{\partial z}{\partial t} (\vec{r},t)=\vec{F}(t)+\nu
\nabla^2{z}(\vec{r},t)+\eta(z(\vec{r},t),\vec{r})\,,
\label{depinning}
\end{equation}
where $\vec{F}$ is a uniform, eventually time dependent, applied
force, $\nabla^2z$ is the elastic term, whose coefficient $\nu$ is the
surface tension, and $\eta(z,\vec{r})$ is the noise term, that mimics
the presence of quenched disorder in the system. We consider here the
zero temperature limit, where thermal noise is negligible with respect
to quenched noise.  A more general equation should include a
kinematically generated nonlinearity of the Kardar--Parisi--Zhang
type, which changes the asymptotic scaling. However this term vanishes
at the critical point and therefore Eq. \ref{depinning} is expected to
have the correct critical behavior at the depinning transition (see
below).
 
\begin{figure}
\centerline{\epsfxsize=13cm \epsfbox{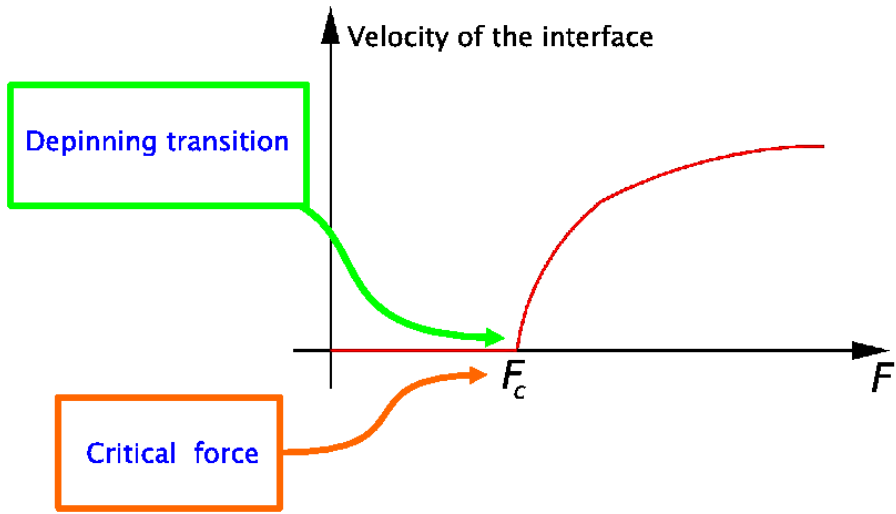}}
\caption{Sketch of the behavior of the interface velocity as an order
  parameter for the second order non--equilibrium depinning
  transition. }
\label{Fdepinning}
\end{figure}

The quenched disorder is described by a random potential
$V(z,\vec{r})$, whose derivatives gives the local pinning field
$\eta(z,\vec{r})$ in equation (\ref{depinning}).  In $d=3$, if
$P_i=(x_i,y_i,z_i)$ are the coordinates of the point-like defects, and
$\vec{r}_i=(x_i,y_i)$, then the random force is
\begin{equation}
\eta(z,\vec{r})=-R\sum_i\delta(z-z_i)\delta^2(\vec{r}-\vec{r}_i)\,,
\label{random_noise}
\end{equation}
where the sum runs over the pinning points. Here we assume the random
noise to be uncorrelated and Gaussian distributed, so that 
\begin{equation}
\langle \eta(z,\vec{r})\eta(z',\vec{r}')\rangle=2
\delta(z-z')\delta^2(\vec{r}-\vec{r}')
\label{correlation}
\end{equation}

Systems described by equation (\ref{depinning}) exhibit a second order
dynamic phase transition, called the depinning transition. The order
parameter is the velocity of the interface, and the control parameter
is the applied force, that we assume uniform and perpendicular to the
substrate plane.

The behavior of the system is sketched in figure \ref{Fdepinning}. At
small applied forces the interface is locked by the pinning potential
and $v=0$. At some threshold value $F_c$ the force overcomes the
pinning. For $F>F_c$ the interface moves with constant velocity. Just
above the threshold value $F_c$ the interface moves in power law
distributed jumps.

\subsection{Dynamics of a magnetic elastic interface}
\label{sec:mesoscopic}

\begin{figure}
\centerline{\epsfxsize=10cm \epsfbox{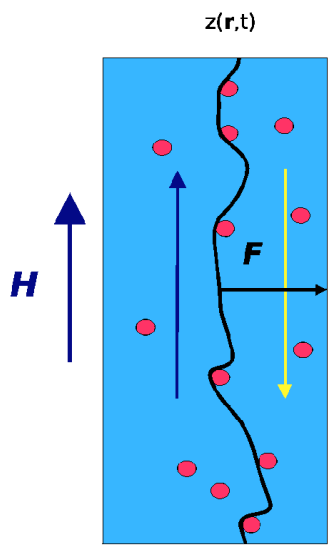}}
\caption{Magnetic wall in an external field $\vec{H}$. The system
  tends to increase the size of the domain magnetized in the direction
  of the external field, which therefore acts as an effective force
  perpendicular to the interface. }
\label{maginterf}
\end{figure}

In the case of a magnetic wall, the interface separates two domains
with opposite magnetization.  When an external field $\vec{H}$ is
applied, the system tends to increase the domain with the largest
component of the magnetization in the direction of the field, by
domain wall displacement. Thus the field $\vec{H}$ acts as an
effective force perpendicular to the wall, as shown schematically in
figure \ref{maginterf}, and we can replace $\vec{F}$ with $\vec{H}$ in
equation (\ref{depinning}). 

Unlike many other cases where an elastic interface is driven in a
disordered media, magnetic systems do not generally experience a
depinning transition of the kind predicted by equation
(\ref{depinning}) and described in the previous section. This is due
to the presence of other important interactions peculiar to the
magnetic case, that modify the behavior of the interface in an
essential way.

\begin{figure}
\centerline{\epsfxsize=10cm \epsfbox{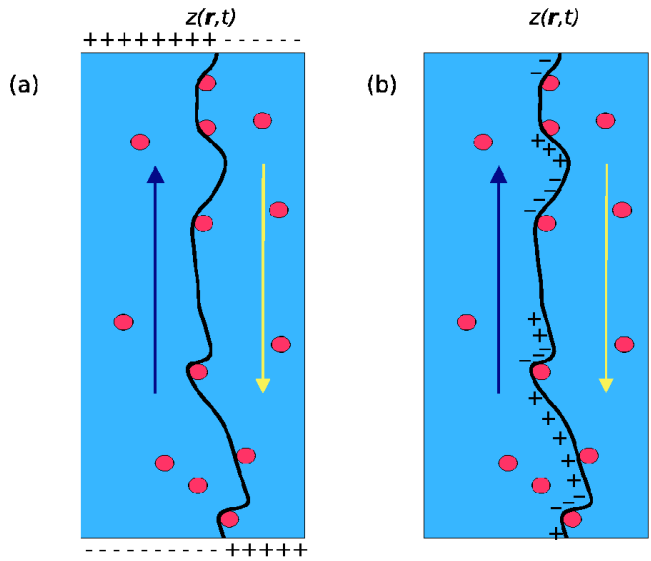}}
\caption{(a) Free magnetic charges on the boundary of the sample.(b)
Free magnetic charges on the interface.}
\label{freecharges}
\end{figure}

A first important step to adapt the general theory of elastic
interfaces in random media in order to take into account the magnetic
nature of the interactions was done by Urbach, Madison, and Markert
\cite{urbach_interface_1995} who modified equation (\ref{depinning})
to include the effect of the demagnetizing field (see also
\cite{narayan_self-similar_1996}).  The demagnetizing field is the
field generated by the presence of free magnetic charges on the
boundaries of the sample (see figure \ref{freecharges}(a)). It opposes
the applied field, and it is roughly proportional to the magnetization
of the sample.  The demagnetizing field is assumed, as in the ABBM
model, to be of the form $-km$, uniform and proportional to the
magnetization, where the {\it demagnetizing factor} $k$ takes into
account the geometry of the sample. From micro-magnetic calculations
it can be shown that the assumption of a uniform demagnetizing field
is strictly correct only in ellipsoidal geometries, however, it turns
out to be a good approximation also in more general cases.

The addition of the demagnetizing field to the equation of motion has
a crucial effect on the dynamics, since it provides the restoring
force needed to keep the interface constantly at criticality. Without
this term the system would be critical just at a threshold value of
the applied magnetic field, and experience a true depinning
transition, while, in BN experiments, criticality is observed in an
entire range of values of $H(t)$ around the coercive field. These
range of values correspond to the linear region of the hysteresis
loop, which is precisely where the magnetization reversal is dominated
by domain wall motion.

The drastic effect of the demagnetizing field on the behavior of the
system is confirmed by performing experiments in particular
geometries, as Barkhausen measurements on toroidal samples. In this
case closed interfaces can be forced so that, due to the absence of
free magnetic charges on the boundaries, the demagnetizing factor is
zero, and indeed, a true depinning transition is observed.  This
behavior has first been observed on a $FeSi$ single crystal in a
frame geometry \cite{williams_studies_1950}, and later on several
other materials \cite{galt_motion_1954,bean_influence_1955,rodbell_some_1956}. 
In any other geometry, where $k \neq 0$, the
equation of motion for the interface has to be modified with respect
to equation (\ref{depinning}) by replacing the applied force $F(t)$
not just with the applied field $H(t)$, but with an effective field

\begin{equation}
H_{eff}(t)=H(t)-km(t)\,.
\label{effective_field}
\end{equation}

The demagnetizing field produces the feedback mechanism that
guarantees the presence of criticality without any tuning of the
driving field. The mechanism is the following: the interface is stuck
as long as $H(t)$ is such that $H_{eff}(t)$ is below the critical
value $H_c$. As soon as $H_{eff}$ exceeds the critical value for the
depinning transition, the interface starts to move in a series of
discontinuous jumps. The displacement of the interface, however,
causes an increase in the magnetization $m(t)$, and therefore in the
demagnetizing field, so that at some point the effective field is back
below the threshold value, and the avalanche stops.

The work by Urbach {\it et al.} \cite{urbach_interface_1995}
represents a fundamental step towards the realistic description of an
elastic magnetic interface, since the introduction of the
demagnetizing field changes dramatically the behavior of the model,
from one that shows criticality at the pinning threshold, to one that
is critical in a finite range of values of the applied field.

However, the equation used in \cite{urbach_interface_1995} to describe
the domain wall dynamic still does not take into account all magnetic
interactions, namely, it neglects the possible presence of free
magnetic dipoles on the wall (see figure \ref{freecharges} (b)). Indeed,
the local curvature of the interface may give rise to discontinuities
in the normal component of the magnetization, and generate dipolar
forces originating in the bulk of the material.

This complicated interaction has been worked out and included in the
model by Cizeau, Zapperi, Durin, and Stanley
\cite{cizeau_dynamics_1997,zapperi_dynamics_1998}.  Cizeau {\it et
al.}  considered the effect of dipolar forces originating from free
magnetic charges on the wall, and evaluated this bulk effect, showing
that it gives rise to a further contribution to the total applied
force, acting as a long range elasticity, which works to flatten the
interface.

The long range elasticity term turns out to be of the form:
\begin{equation}
\int d{\vec r_1} K(\vec{r}-\vec{r_1})\left(z({\vec r_1})-z({\vec
  r})\right) \,,
\label{long_range_el_term}
\end{equation}
with a kernel $K$ given by
\begin{equation}
K(\vec{r})=\frac{K_0}{\mid\vec{r}\mid^3}
\left(1-\frac{3 x^2}{\mid{\vec r}\mid^2}\right)\,.  
\label{kernel}
\end{equation}
This interaction has been calculated under the mild assumption that
the surface charges are not strong enough to deviate the magnetization
from the easy axis. The coefficient $K_0$ is given by $K_0=\mu_0 m_s^2
/2\pi$, where $\mu_0$ is the vacuum permeability, and $m_s$ is the
saturation magnetization per unit volume.

The full equation of motion, including the contributions from surface
and bulk dipolar interactions finally results to be:
\begin{equation}
\Gamma \dot{z}(\vec{r},t)=\nu\nabla^2{z}(\vec{r},t)+\int d^2r_1
K(\vec{r}-\vec{r_1})
\left(z(\vec{r_1})-z(\vec{r})\right) 
+H(t)-k \overline{z}(t)+\eta(z,\vec{r})\,,
\label{cizeau}
\end{equation}
where $ \overline{z}(t)$ denotes the average position of the interface
and is proportional to the magnetization. 

The long range elasticity term is relevant in a renormalization group
sense, since it is dominating the surface tension elastic term, as
follows by power counting: rescaling $x \rightarrow x'=bx$ and
assuming that $h \rightarrow h'=b^{\alpha} h$ with some positive
$\alpha$, the elastic term from surface tension is rescaled by a
factor $b^{\alpha-2}$, while the long range elasticity term from
dipolar interactions is rescaled by $b^{d+\alpha-3}$, where $d-1$ is
the dimension of the substrate plane. This point is treated in a more
rigorous way in \cite{zapperi_dynamics_1998} where the authors,
following the functional renormalization group developed in
\cite{narayan_dynamics_1992, narayan_critical_1992,
narayan_threshold_1993}, obtain the value $d_c=2+1$ for the upper
critical dimension. The lowering of the upper critical dimension of
the system from $4+1$ (which is the one characterizing a generic
elastic interface in a random media i.e. without the long range
kernel) to $2+1$ as a consequence of the presence of the long range
dipolar interactions ensures that the mean field approximation holds
down to $d=3$, and ultimately allows to describe most experimental
situations in terms of a mean field theory. 

Corrections to scaling may however be present in $d=2+1$: indeed the
mean field power law behaviors are usually modified at the upper
critical dimension by multiplicative logarithmic factors, as it is for
example in the simplest case of the equilibrium Ising model in $d=4$
\cite{larkin_phase_1969}. As an example closer to the system described
by Eq. \ref{cizeau}, let us mention that logarithmic corrections at
the upper critical dimension for the depinning transition have
recently been calculated for the quenched Edward Wilkinson equation
with a driving force (\ref{depinning}) in
\cite{fedorenko_depinning_2003}.

Equation (\ref{cizeau}) contains all the energetic contributions
involved in the domain wall dynamics. In principle, one should add an
additional noise term to the equation, to account for thermal
fluctuations.  However, it has been observed experimentally
\cite{urbach_interface_1995} that temperature does not affect
Barkhausen measurements in bulk materials. At this regard see also
\cite{zapperi_dynamics_1998}, where a rough estimate of the signal
produced by thermal effects is given, which results to be well below
the typical instrumental background noise. This could be different in
thin films, where thermal activated motion has been observed. Recent
experiments by Kim, Choe, and Shin \cite{kim_2003,kim_2003-1} report
measures by magneto optical imaging on $Co$ thin films, where the
applied field is raised just below the coercive field and then kept
constant, so that the observed Barkhausen jumps are only triggered by
thermal noise (note that the Barkhausen jumps going outside the
imaging window makes unreliable the estimates of the critical
exponents obtained in \cite{kim_2003,kim_2003-1}, as already observed
in \cite{durin_barkhausen_2006}). therefore, in the case of thin
films, an extra time dependent random term may be required in the
equation of motion to include thermal noise.

The elastic interface model described by equation (\ref{cizeau}) has
been extensively studied by numerical simulations, by direct
integration and by simulating an automaton version of the model, and
both in finite dimensions and in the mean field limit.  Simulations in
the infinite range limit confirm the equivalence with the ABBM model
(see below), by reproducing the same amplitude distribution, and
pulses statistics.  The simulations of the interface model in $d=3$,
recover the mean field results, in agreement with $d_c=3$ being the
upper critical dimension. However, the scaling with $k$ of the
cut--off in the distributions of pulses sizes and durations does not
agree with the predictions of the mean field. This point will be
discussed more in detail in subsection \ref{sec:cutoff}.

Although in this review we are dealing only with soft magnetic
materials in the long range universality class, it is worth to note
that the model by Cizeau {\it et al.} \cite{cizeau_dynamics_1997} has
a more general validity. The results discussed above change when the
long range kernel in equation (\ref{cizeau}) is absent, and the
surface term is the dominating term. In this case the model is
equivalent to the one previously proposed in
\cite{urbach_interface_1995}. The upper critical dimension in this
case is $d_c=4+1$, and simulations in $d=3$ give $\tau\simeq 1.3$ and
$\alpha \simeq 1.45$ for the critical exponents characterizing the
pulses' distributions \cite{leschhorn_driven_1997,
durin_barkhausen_1999, bahiana_domain_1999}. Simulations results are
confirmed by renormalization group calculations
\cite{nattermann_dynamics_1992, narayan_threshold_1993,
leschhorn_driven_1997} which give $\tau \simeq 1.25$ and $\alpha
\simeq 1.43$.  In this case, no dependence on the driving field rate
$c$ is observed.  These results are in very good agreement with
experiments on materials in the short range universality class
\cite{zapperi_dynamics_1998, durin_complex_2002}, indicating that the
magnetic elastic interface model may indeed have a very general
validity, providing a description for the magnetization reversal
process suitable to deal with both universality classes of soft
ferromagnets.

\subsection{The ABBM model as a mean field theory}
\label{sec:MF}

A mean field version of the model by Cizeau {\it et
al.} \cite{cizeau_dynamics_1997} has been worked out by the same
authors in \cite{zapperi_dynamics_1998}, where they show that in the
mean field approximation, their interface model coincides with the
phenomenological ABBM model. This fact might indeed explain why the
ABBM model describes the experiments so accurately.

Mean field theory is derived by taking the limit of long range
interactions in the equation of motion. This is done by coupling all
the points on the interface uniformly with the average wall position.
Equation (\ref{cizeau}) become
\begin{equation}
\Gamma \dot{z}(\vec{r},t)=J (\overline{z}(t)-z(\vec{r},t)) 
+H(t)-k \overline{z}(t)+\eta(z,\vec{r})\,,
\label{mf}
\end{equation}
where the coefficient $J$ includes the contributions from both short
and long term elasticity. By averaging this equation over $\vec{r}$,
one gets an equation for the average position of the interface:
\begin{equation}
\Gamma \dot{\overline{z}}(t)=
H(t)-k \overline{z}(t)+
\frac{1}{S}\int d^2r \eta(z(\vec{r},t),\vec{r})\,,  
\label{mf1}
\end{equation}
where $S$ is the surface of the substrate plane.  The integral over
the noise gives rise to an effective pinning $W(\overline{z})$, which
depends only on the average position of the interface.  

Indeed from equation (\ref{random_noise})
\begin{equation}
\eta(z(\vec{r},t),\vec{r})=\int dz \delta(z-z(\vec{r},t))
\eta(z,\vec{r})=
-R\sum_i\delta^2(\vec{r}-\vec{r}_i)\delta(z(\vec{r},t)-z_i)\,,
\label{noise}
\end{equation}
which is proportional to the number of defects points that are pinning
the interface to the average position $\overline{z}$. The effective
noise $W$ is then given by
\begin{equation}
W(\overline{z})=
\frac{1}{S}\int d^2r \eta(z(\vec{r},t),\vec{r})=
-\frac{R}{S}\sum_i\delta(z(\vec{r},t)-z_i)\,.  
\label{brownian_noise}
\end{equation}

To calculate the correlations, note that, in the difference
$W(\overline{z})-W(\overline{z}')$, the only terms that contribute to
the sum are those relative to the pinning points that have actually
moved, whose number is proportional to the average wall displacement
$\overline{z}-\overline{z}'$, so that, $\eta$ being random and
uncorrelated, the effective noise is a Brownian noise in
$\overline{z}$:
\begin{equation}
\langle \left(W(\overline{z})-W(\overline{z}')\right)^2\rangle=
D (\overline{z}-\overline{z}')\,.
\label{brownian_noise2}
\end{equation}

This finally explains the nature of the effective Brownian correlated
pinning field, which was observed experimentally in
\cite{porteseil__new_1979,vergne_statistical_1981}, and used as an
assumption in \cite{alessandro_domain-wall_1990} to derive the ABBM
equation.  The long range correlations observed in the pinning field
are not a result of real long range correlations between defects in
the material. Rather, the effective correlated pinning field
experienced by the center of mass of the wall originates as a
collective effect from the interactions along the elastic wall moving
in an uncorrelated disordered medium.

The average wall position is proportional to the magnetization, so
that equation \ref{mf1} can be rewritten as
\begin{equation}
\Gamma \dot{m}(t)=H(t)-k m(t)+W(m)\,, 
\label{mf2}
\end{equation}
which coincides with the ABBM equation (\ref{ABBM}).  

The ABBM model is extremely successful in describing BN experiments,
although it was proposed on phenomenological basis, without a clear
reference to the underlying physics. The fact that the ABBM model can
be derived as a mean field theory of an interface model obtained by
taking into account all the relevant magnetic interactions is
therefore very important. We want to stress that the relation with the
elastic interface model not only validates the ABBM model, by giving
to it a more grounded microscopical derivation, but also allows to
relate the model parameters to microstructural properties of the
material, making possible a close quantitative comparison between
theory and experiments.

\section{Mapping ABBM onto a random walk process: exact results}
\label{sec:RW}

The ABBM approach allows to write a simple equation of motion for the
center of mass of the magnetic wall separating two regions of opposite
magnetization. Interestingly, the statistical properties of this
dynamics can be fully worked out analytically. 

The easiest way to do this is to transform the equation by eliminating
the correlated noise. We show below that equation (\ref{ABBM}) maps
onto an standard Langevin equation for a Brownian motion in a suitable
potential. The equation has to be solved taking care of the boundary
conditions, that ensure that the signal always stays positive.

\subsection{From the ABBM equation to the Langevin equation for a biased random walk in a logarithmic potential}
\label{sec:RW_rw}
Equation (\ref{mf2}) takes a particularly simple form in terms of
velocity versus magnetization. Taking a derivative with respect to
time, and defining $v=dm/dt$
\begin{equation}
\Gamma \frac{dv}{dm}= \frac{\dot{H}(t)}{v}-k+\frac{dW}{dm}\,. 
\label{mf3}
\end{equation}
where we used $d/dt=vd/dm$. 

Assuming that the field $H(t)$ is varied at constant rate $c$, as in
most BN experiments, and defining $w(m)=dW/dm$, we get
\begin{equation}
\Gamma \frac{dv}{dm}= \frac{c}{v}-k+w(m)\,,  
\label{mf4}
\end{equation}
where the noise term, being the derivative of a Brownian noise, is
delta correlated in magnetization, with $\langle
w(m)w(m')\rangle=2D\delta(m-m')$.  

Note that, the ABBM is defined with the assumption that just one
single and rigid wall is present. This allows the identification of
the magnetization with the coordinate defining the wall position, and
therefore the variable $v$, which is the derivative of the
magnetization with respect to time, can be interpreted both as the
average velocity of the interface, and as the induced voltage directly
measured in inductive BN experiments.

The four parameters in equation (\ref{mf2}) can be reduced to two by
rescaling the velocity $\tilde{v}=v\Gamma/D$, and defining
$\tilde{k}=k/D$, $\tilde{c}=c \Gamma/D^2$, and $\eta=w/D$, so that
$\langle \eta(m)\eta(m')\rangle=2\delta(m-m')$:
\begin{equation}
\frac{d\tilde{v}}{dm}= 
\frac{\tilde{c}}{\tilde{v}} -\tilde{k}+\eta(m)\,.   
\label{mf5}
\end{equation}
This leaves two independent parameters $\tilde{k}$, and $\tilde{c}$.
In order to avoid cumbersome notations we will avoid the tildes in the
following.

Interpreting the velocity $v$ as a coordinate $x$ and the
magnetization as time $t$, this is the Langevin equation for a random
walk in a logarithmic plus linear potential $U(x)$:
\begin{equation}
U(x)=-c \log x +k x \,,
\label{potential}
\end{equation}
where the coefficients of the two terms in the potential, that rules
the dynamics, correspond to rescaled driving field rate and
demagnetizing coefficient respectively.

The Fokker-Planck equation corresponding to the Langevin equation 
\begin{equation}
\dot{x}=\frac{c}{x}-k+\eta(t) \,.
\label{lang}
\end{equation}
is
\begin{equation}
\frac{\partial P(x,t)}{\partial t}=\frac{\partial}{\partial x}
\left(\frac{\partial P(x,t)}{\partial x} +\left(k- \frac{c}{x}\right)
P(x,t)\right)\,,
\label{FPb}
\end{equation}
where $P(x,t)$ is the probability for the walker to be at $x$ at time
$t$. 

The first thing that can easily be done is to solve equation \ref{FPb} in
the stationary $t\rightarrow \infty$ limit. This gives
\begin{equation}
P_{\infty}(x)\propto x^{c} \exp(-kx)\,.
\label{stationary}
\end{equation}

Expression \ref{stationary} gives the time independent probability
distribution for the process for all values of $c$, however, it will
become clear further on that, once the boundary conditions are taken
into account, the signal distribution can be obtained by simply taking
the asymptotic solution, as has been done here, only in the high
driving rate regime (large $c$).  We will come back to this
distribution in the next section.

\subsection{The Fokker-Planck equation for a random walk in a
  logarithmic potential confined to the semiplane}
\label{sec:LOG}

The linear term in the potential (\ref{potential}) acts as a constant
bias to the random walk, that, as we will see, has the effect to
introduce a cut--off in the return times distribution. To proceed, we
forget for now about the bias term, and focus on the behavior of an
unbiased random walk in a logarithmic potential:
\begin{equation}
\dot{x}=\frac{c}{x}+\eta(t) \,.
\label{rw-log}
\end{equation}

Interestingly, the same equation (\ref{rw-log}) in two dimensions
appears in the analysis of the annihilation dynamics of
vortex--antivortex pair in the two dimensional $XY$ model, and has
been studied in that context by Bray in \cite{bray_random_2000}, which
we follow in the next sections for what concerns the calculation of
the persistence properties of the process.

The Fokker-Planck equation associated to the Langevin equation
(\ref{rw-log}) is:
\begin{equation}
\frac{\partial P(x,t)}{\partial t}=\frac{\partial}{\partial x}
\left(\frac{\partial P(x,t)}{\partial x} - \frac{c}{x}
P(x,t)\right)\,,
\label{FP}
\end{equation}
where $P(x,t)$ is the probability for the walker to be at $x$ at time
$t$. 

Since by definition the avalanche ends when the velocity is back to
zero, we are interested to the solutions of this equation with
absorbing boundary conditions at the origin $x=0$:
\begin{equation}
P(0,t)=0\,.
\label{FP_bc}
\end{equation}

Given the absorbing condition, a walk starting at $x=0$ would never
leave the origin, thus we choose as initial condition for the walk a
small $\epsilon<<1$:
\begin{equation}
P(x,0)=\delta(x-\epsilon) \,
\label{FP_initial}
\end{equation}
and take the limit of vanishing $\epsilon$ at the end.

The solution of equation (\ref{FP}) can be searched by separation of
variables \cite{bray_random_2000} in the form
\begin{equation}
P(x,t)=r(x) \exp(-k^2t)\,,  
\label{trial}
\end{equation}
where $r(x)$ has to satisfy
\begin{equation}
x^2 r''(x)-c x r'(x)+(c-k^2x^2)r(x)=0 \,.
\label{R1}
\end{equation}
Equation (\ref{R1}) can be reduced to the Bessel equation by the
change of variable $r(x)=x^{a} R(x)$:
\begin{equation}
x^2 R''(x)+(2a-c)x R'(x)+((a-1)(a-c)+k^2x^2)R(x)=0 \,,
\label{R11}
\end{equation}
by choosing for $a$ the value $a=(1+c)/2$:
\begin{equation}
x^2 R''(x)+x R'(x)+(k^2x^2-\nu^2)R(x)=0 \,,
\label{R2}
\end{equation}
with $\nu=((1-c)/2)$.

The general solution of equation (\ref{FP}) can therefore be written as a
linear combination of Bessel functions as:
\begin{equation}
P(x,t)=x^{1-\nu}\int_0^{\infty} dk \left(
\alpha(k)J_{\nu}[kx]+\beta(k)J_{-\nu}[kx]\right)\exp(-k^2 t)\,.
\label{general_solution}
\end{equation}
The coefficients $\alpha(k)$ and $\beta(k)$ are fixed by imposing the
initial condition equation (\ref{FP_initial}), that can be enforced by using
the orthogonality properties of the Bessel functions:
\begin{equation}
\int_0^{\infty}dk k J_{\alpha}[kx_1]J_{\alpha}[kx_2]=
\delta(x_1-x_2)/x_1 \,.
\label{orto}  
\end{equation}

This gives
\begin{equation}
P(x,t\mid\epsilon,0)=\epsilon \left(\frac{x}{\epsilon}\right)^{1-\nu}
\!\!\int_0^{\infty} dk k \left( A J_{\nu}[k\epsilon] J_{\nu}[kx]+B
J_{-\nu}[k\epsilon] J_{-\nu}[kx]\right) \exp(-k^2 t)\,,
\label{solution}
\end{equation}
where $A$, and $B$ are numerical coefficients.  Finally, by performing
the integrals
\begin{equation}
P(x,t\mid \epsilon, 0)=\frac{\epsilon}{t}
\left(\frac{x}{\epsilon}\right)^{1-\nu}
\exp\left(-\frac{x^2+\epsilon^2}{4t}\right) \left( A
I_{\nu}\left[\frac{\epsilon x}{2t}\right]+B
I_{-\nu}\left[\frac{\epsilon x}{2t}\right] \right)\,,
\label{FP_fullsol}
\end{equation}
where $I_{\nu}[z]$ is the modified Bessel function.  

To simplify the notations let us rewrite equation (\ref{FP_fullsol})
as
\begin{equation}
P(x,t\mid \epsilon, 0)=
A P_{\nu}(x,t\mid \epsilon,0 )+B P_{-\nu}(x,t\mid \epsilon,0 )
\label{FP_fullsol2}
\end{equation}
and analyze the two terms separately. 

For what follows is useful to introduce the probability current
\begin{equation}
j(x,t)=-(\partial_x P-c P/x)
\label{current}
\end{equation}
so that $\partial_t P=-\partial_x j$. The absorbing boundary condition
equation (\ref{FP_bc}) corresponds to requiring a negative current at the
origin.

In order to evaluate the probability current at the origin
corresponding to the two terms in equation (\ref{FP_fullsol2}), it is
convenient to expand the Bessel functions in the limit of small
$\epsilon$, using
\begin{equation}
I_{\nu}[z]=(z/2)^{\nu}\sum_{m=0}^{\infty}(z/2)^{2m}/(m!\Gamma[m+\nu+1])\,. 
\end{equation}

The first term in equation (\ref{FP_fullsol}) simplifies to
\begin{equation}
P_{\nu}(x,t\mid \epsilon, 0)=
\frac{4}{\Gamma[1+\nu]} (4t)^{-(1+\nu)}
\epsilon^{2\nu} x \exp\left(-\frac{x^2}{4t}\right)
(1+O(\epsilon^2))\,.
\label{FP_expsol1}
\end{equation}
The corresponding probability current is
\begin{equation}
j_{\nu}(x,t\mid \epsilon, 0)=
\frac{4}{\Gamma[1+\nu]} (4t)^{-(1+\nu)}
\epsilon^{2\nu} \left((c-1)+\frac{x^2}{t}\right) 
\exp\left(-\frac{x^2}{4t}\right)
(1+O(\epsilon^2))\,,
\label{FP_PC1}
\end{equation}
and gives a non--vanishing contribution at the origin, which is
negative, as requested for an absorbing boundary, as long as $c<1$.

The second term in equation (\ref{FP_fullsol}) simplifies to
\begin{equation}
P_{-\nu}(x,t\mid \epsilon, 0)=
\frac{4}{\Gamma[1-\nu]} (4t)^{-(1-\nu)}
x^{1-2\nu} \exp\left(-\frac{x^2}{4t}\right)
(1+O(\epsilon^2))\,.
\label{FP_expsol2}
\end{equation}
The coefficient $1-2\nu-c$ of the leading term of order $x^{-2 \nu}$
in the probability current vanishes, and the next--to--leading term
gives:
\begin{equation}
j_{-\nu}(x,t\mid \epsilon, 0)=
\frac{4}{\Gamma[1-\nu]} (4t)^{-(1-\nu)}
\frac{x^{2-2\nu}}{t}
\exp\left(-\frac{x^2}{4t}\right)
(1+O(\epsilon^2))\,.
\label{FP_PC2}
\end{equation}
Since $c$ is a positive quantity, the exponent $2-2\nu=1+c$ is always
positive in the region of parameters of physical interest, and the
contribution of the term $P_{-\nu}$ to the probability current at the
origin vanishes. Therefore the physically relevant solution to our
problem, satisfying the absorbing boundary condition is $P_{\nu}$, and
we take $P=P_{\nu}$ and drop the $\nu$ sign from now on.

Finally, we can conclude that the total probability current at $x=0$
is:
\begin{equation}
j(0,t)=-
\frac{4A(1-c)}{\Gamma[(3-c)/2]}(4t)^{-(3-c)/2}\epsilon^{1-c}\,,
\label{current_0}
\end{equation}
where we have replaced $\nu$ with its expression as a function of
$c$. The current goes to zero as $c$ approaches $1$, correctly
suggesting that, as we will see more clearly in section \ref{sec:REC},
$c=1$ separates two different regimes.

It must be stressed that for $c<1$ the probability $P(x,t\mid
\epsilon, 0)$ in equation (\ref{FP_expsol1}) is not normalized to
unity, when integrated over $x$. Indeed, due to the absorbing boundary,
part of the probability density flows to the origin, so that the
conserved quantity is
\begin{equation}
\int_0^{\infty}dx P(x,t\mid\epsilon, 0)+P_{abs}(t\mid\epsilon, 0)=1
\label{conserved}
\end{equation}
where $P_{abs}(t\mid\epsilon, 0)$ is the probability that the random
walk has been absorbed at some time $t'<t$. This probability goes to
$1$ as $t \rightarrow \infty$. For this reason, the overall time
independent process amplitude distribution cannot be obtained by
taking the $t \rightarrow \infty$ in $P(x,t\mid \epsilon, 0)$, since
the asymptotic distribution is zero for any $x\neq0$.

The asymptotic distribution clearly is not the correct quantity to be
compared with the signal amplitude distribution in equation
(\ref{V_dist}).  A more meaningful probability, which indeed
corresponds to the amplitude measured in experiments is obtained by
integrating $P(x,t\mid \epsilon, 0)$ over $t$. In general, for a
stationary process, the asymptotic distribution $P_{\infty}(x)$
coincides with the average over time of the probability $P(x,t)$:
\begin{equation}
P_{\infty}(x)=\lim_{T\rightarrow \infty} \frac{1}{T}\int_0^{T} dt
P(x,t) \,.
\label{stat}
\end{equation}
However, the quantity in equation (\ref{V_dist}) that is measured in
experiments is the average over many avalanches of the probability for
the signal to have a given value. During the avalanche, the process
described by equation \ref{lang} is non--stationary, and thus the
integral over $t$ cannot be replaced by the asymptotic distribution.

This point needs a further remark. In subsection \ref{sec:exp} we
stressed the fact that the BN experiments has to be restricted to the
region where the signal is stationary, and this may generate confusion
when compared with the present argument. The BN process is described
by the stochastic equation (\ref{mf3}) only when $v>0$, that is as
long as the effective field (applied field diminished by the
demagnetizing field) overcomes the pinning field.  As long as $v>0$
the magnetization increases, causing a variation in the effective
field, which decreases until it goes somewhere below the critical
depinning threshold. At this point, the interface stays locked, until
the applied field brings the effective field again above
threshold. Therefore while $v=0$, the process is just deterministic.
Given that the wall is at rest, there is no noise term involved.  The
stationarity of the whole BN process (the random process at $v>0$ plus
the deterministic one for $v=0$) should not be confused with the
stationarity of the process described by equation \ref{lang}, which,
in the low driving field rate regime, is reached only when the random
walk is absorbed by the boundary.

Let us now go back to our problem of evaluating the signal amplitude
distribution. By taking integral over $t$ in equation
(\ref{FP_expsol1}) to the first order in $\epsilon$ we can extract the
small $x$ power law behavior:
\begin{equation}
\int_0^{\infty} dt P(x,t\mid\epsilon,0)\simeq
x^{2\nu+1} = x^{c} \,.
\label{intP}
\end{equation}
This expression is not valid for large $x$, where the small $\epsilon$
expansion does not hold.  The power law dependence on $x$ of the
process amplitude coincides with that of equation (\ref{stationary})
obtained in the previous section, apart from the exponential cut--off
depending on the bias term, and which has been neglected here. Indeed
equation (\ref{stationary}) gives an expression for the process
amplitude distribution correct for all values of $c$.

The large $t$ asymptotic has been used in literature
\cite{durin_barkhausen_1999} to explain the experimental power law in
equation (\ref{V_dist}), however, we stress that this behavior cannot
be obtained in general by simply taking the asymptotic solution.
 
To compare this result with the experimental one on the distribution
of the Barkhausen signal we must remember that here $t$ corresponds to
the physical variable $m$, and $x$ to the wall velocity, or BN
amplitude $v$.  Going back to physical time will therefore produce an
extra $v^{-1}$ factor, recovering the $c-1$ exponent found
experimentally (see equation (\ref{V_dist})).

\subsection{Distributions of sizes and durations of pulses from persistence properties}
\label{sec:PER}

Let us now turn to the analysis of the statistical properties of the
signal pulses. The distribution of the sizes and durations of Barkhausen
pulses can be derived from the persistence properties of the process
(\ref{lang}).

Let us introduce the first passage time distribution $P_0(0,t\mid
\epsilon,0)$, i.e. the probability that a random walk starting at some
threshold value $\epsilon$ at time $t=0$ goes back to $\epsilon$ for
the first time at time $t$, and the persistence probability $P_1(t\mid
\epsilon,0)$, defined as the probability that a random walk that
starts at $\epsilon$ at time $t=0$ stays positive up to time $t$.

These two quantities are related by the following relation:
\begin{equation}
P_1(t\mid \epsilon,0)=\int_{t}^{\infty} dt' P_0(t'\mid \epsilon,0) \,,
\label{P1}
\end{equation}
since the probability for a random walk to be positive up to time $t$
is equal to the probability that it comes back to $0$ for the first
time at some later time $t'>t$, which is given by the r.h.s. of
equation (\ref{P1}).

The persistence probability $P_1(t\mid \epsilon,0)$ is also related to
the solution of the Fokker Planck equation with absorbing boundary condition by
\begin{equation}
P_1(t \mid \epsilon,0)=\int_0^{\infty}dx P(x,t\mid \epsilon, 0)\,,
\label{P1_2}
\end{equation}
since, to survive up to some time $t$, a walker must be found at time
$t$ at some positive value $x$. This is equivalent to the
normalization condition (\ref{conserved}) where $P_{abs}(t \mid
\epsilon,0)=1-P_1(t \mid \epsilon,0)$.

Using the previous two equations we can calculate $P_0$ as follows:
\begin{equation}
P_0(t,\mid \epsilon,0)=-\partial_t
P_1(t\mid\epsilon,0)=-\int_0^{\infty}dx \partial_t P(x,t\mid \epsilon,
0) \,,
\label{P0}
\end{equation}
where we passed the derivative under the integral. 

Rather than replacing $P(x,t\mid\epsilon, 0)$ with its explicit
expression and then dealing with the complicated integral, it is
convenient to replace the integrand on the r.h.s. with its
expression given by the Fokker-Planck equation:
\begin{equation}
P_0(t,\mid \epsilon,0)= -\int_0^{\infty}dx \partial_x \left(\partial_x
P(x,t\mid \epsilon, 0)-\frac{c}{x}P(x,t\mid\epsilon,0)\right)=j(0,t)
\label{P0_2}
\end{equation}
where, in replacing the integral with the current in zero we use the
fact that $P(x,t\mid \epsilon, 0)$ and its derivatives decay
exponentially to zero for $x\rightarrow \infty$. Eq. (\ref{P0_2}) is
consistent with the probability $P_{abs}$ for the process to having
being absorbed before time $t$ being equal to the total current flow
through the origin up to time $t$: $P_{abs}=-\int_0^{t} j(0,s) ds$.

Finally we get from equation (\ref{current_0}):
\begin{equation}
\label{P0_3} 
P_0(t,\mid \epsilon,0)\propto (1-c)t^{-(3-c)/2}\,.
\label{per}
\end{equation}
When $c \rightarrow 0$ equation \ref{per} gives back the well known
result for the persistence probability of a free random walk.

To see how this relates to the distribution of sizes and durations of BN
pulses one has to go back to the physical variables, where $t$ has to
be interpreted as magnetization. Equation (\ref{P0_3}) gives the
probability that the avalanche ends at some given value of the
magnetization. It gives therefore directly the distribution of sizes
of the avalanches, which is $p(S)\sim S^{-(3-c)/2}$. This gives the
correct exponent $\tau=3/2$ in the quasi--static limit $c=0$, and the
expected linear dependence on $c$ for finite driving field rates.

The distribution of durations of BN pulses can be also derived by this
same result, by imposing that $p(S)dS=p(T)dT$, and using the scaling
$S\sim T^2$ that we will show in subsection \ref{sec:EXC}.  This gives
$p(t)\sim T^{-(2-c)}$, with $\alpha=2$ for vanishing field rate, and a
linear dependence on $c$ that agrees with the fit of experimental
histograms. Note that the exponential cut--offs cannot be recovered
from this analysis, since they involve the demagnetizing factor that
has been put to zero here (see beginning of section \ref{sec:LOG}).

\subsection{Non--universality of the exponents and marginality of the
  logarithmic potential}
\label{sec:MAR}

The continuous dependence on a parameter in otherwise universal
exponents is usually related to the presence of some potential which
is marginal in the renormalization group sense. Indeed, also in the
present case, the dependence of the exponents $\alpha$ and $\tau$ on
the driving field rate $c$ can be attributed to some marginal
contribution. 

This becomes clear in the mapping onto the random walk process.  The
non--universality of the exponents in the persistence and first
passage time probabilities are special to the logarithmic potential,
and are related to the marginality of the $1/x$ perturbation to a free
random walk in a renormalization group sense.  Indeed, analyzing the
relevance of a power law perturbation $x^{\alpha}$ to a free random
walk, one can show by simple scaling arguments that the force is
irrelevant for $\alpha> -1$ (the random walk behaves essentially as if
it where free), while it is relevant, and dominates the noise term for
$\alpha <-1$ (the motion is in many respects deterministic, and
persistence properties are governed by rare fluctuations of the
noise).

The equation for a random walk in a generic power law potential is
\begin{equation}
\label{rw_potential}
\dot{x}=cx^{-\psi}+\eta(t) \,.
\end{equation}
Under rescaling $x\rightarrow ax$, $t\rightarrow a^z t$, the
coefficients rescale as $c\rightarrow a^{z-1-\psi}c$, $D\rightarrow
a^{z-2}D$, where $D$ is the noise correlation. The equation of motion 
therefore becomes
\begin{equation}
\label{rw_potential_rescaled}
\dot{x}=c a^{z-1-\psi} x^{-\psi}+a^{z/2-1}\eta(t) \,.
\end{equation}

This equation has two fixed points, depending on the value of $\psi$.
The potential term dominates with respect to the noise term for
$z-1-\psi>z/2-1$. In this case, the noise term renormalizes to zero,
and the value of the dynamical exponent is obtained by balancing the
potential term with the l.h.s., which gives $z=1+\psi$. Replacing the
fixed point value of $z$ in the inequality we deduce that this is the
fixed points for $\psi<1$.

The other regime is the one where the noise dominates over the
potential, for $z/2-1>z-1-\psi$. In this case, the coefficient of the
potential renormalizes to zero, and the value of the dynamical exponent
is obtained by balancing the noise with the l.h.s., which gives
$z=2$. Replacing the value of $z$ in the inequality we verify that the
system renormalize to this fixed point as long as $\alpha>1$.

Therefore, for $\psi >1 $, $z=2$, and $c$ rescales to zero: the force
is irrelevant, and the random walk behave essentially as if it were
free. For $\psi<1$, $z=1+\psi$, the noise correlation $D$ rescales to
zero: the noise term is (dangerously) irrelevant, and the process is
almost deterministic.

The case of our equation, $\psi =1$, corresponding to the logarithmic
potential, is marginal: it is the value of $\psi$ for which all the
three terms in the equation balance, with $z=2$. This is at the origin
of the non--universality of the persistence exponents and of their
continuous dependence on the coefficient $c$ of the potential.

The effect of the driving field rate on crackling events has been
recently studied by White and Dahmen in \cite{white_driving_2003}. In
this work the authors make a distinction between an avalanche
(sequence of events triggered by an initial event at constant driving
field), and a pulse (uninterrupted sequence of events measured in
experiments). In general, at non--zero sweep rate a pulse can be a
superposition of several simultaneously propagating avalanches, while
avalanches and pulses coincide only in the adiabatic limit. Under some
general assumptions that are fulfilled by many crackling systems, the
authors of \cite{white_driving_2003} show by scaling arguments that
the value of the exponent $\alpha$ in the distribution of events
durations determines whether or not the pulse statistic will be
affected by the sweep rate $c$. In particular, they claim that no
dependence on $c$ has to be expected for $\alpha<2$, while a linear
dependence of the kind $\alpha=2-c$ is predicted in the case
$\alpha=2$. It would be very interesting to make a connection between
the results obtained in \cite{white_driving_2003} in terms of
superposition of adiabatic avalanches with those reported here for the
ABBM model, but this has not been done yet.

\subsection{Existence of a threshold in the driving field rate from
  recurrence properties of the free random walk}
\label{sec:REC}

Another evidence in Barkhausen noise phenomenology is the fact that
the intermittent behavior typical of the avalanche dynamics is lost
when the driving field is cycled too fast: above a threshold value of
the driving field rate the typical intermittent signal disappears, and
the magnetization reversal proceeds in a unique avalanche.

The existence of a threshold value in the driving field rate $c$,
corresponds to the existence of a special value $c$ for the
coefficient of the logarithmic potential, that discriminates among two
qualitatively different behaviors.  The value $c=1$ already appeared
in equation (\ref{current_0}) as the threshold below which the
probability current at the origin is negative.

Essentially, this value of $c$ determines whether or not the random
walk is recurrent.  The random walk is certainly recurrent for $c<0$,
which corresponds to an attractive potential at the origin. This,
however, does not correspond to the physical case of BN. The case of
physical interest is $c>0$, corresponding to a repulsive
potential. Interestingly, the random walk may come back to the origin
even when it is subject to a logarithmic repulsion, as long as the
repulsive potential coefficient is not too strong. This gives a
condition on the coefficient $c$.

A a simple and clear way to interpret this behavior is to observe that
there exists a natural mapping between the free random walk in $d$
dimension and the confined one dimensional random walk in a
logarithmic potential.  Starting from the Fokker--Planck equation for
a free random walk in $d$ dimensions.
\begin{equation}
\partial_t P({\bf x},t)=\nabla_{x}P({\bf x},t)
\label{rw_ddim}
\end{equation}
we can write the Fokker--Planck equation for the radial part in polar
coordinates as follows
\begin{equation}
\label{radial}
\partial_t P_R(r,t)=\partial_x\left(\partial_x
P_R(r,t)-\frac{d-1}{r}P_R(r,t)\right) \,.
\end{equation}
This equation shows that the radial part of free random walk in $d$
dimensions performs a one dimensional random walk in a logarithmic
potential confined to a semiplane, with coefficient $c$ equal to
$d-1$ for the potential.

Using this mapping, the existence of a threshold value in the
coefficient of the potential is explained in terms of the well known
recurrence properties of the free random walk.  The condition $d<2$
for the random walk to be recurrent translates into $c<1$.

\subsection{Cut--off in the distributions of pulses' sizes and durations as an effect
of the bias to the random walk}
\label{sec:BIAS}

Let us now analyze the effect of to the bias term $k$. We consider a
simple biased random walk first:
\begin{equation}
\dot{x}(t)=-k+\eta(t)\,.
\label{bias}
\end{equation}

\begin{figure}
\centerline{\epsfysize=5.5cm \epsfbox{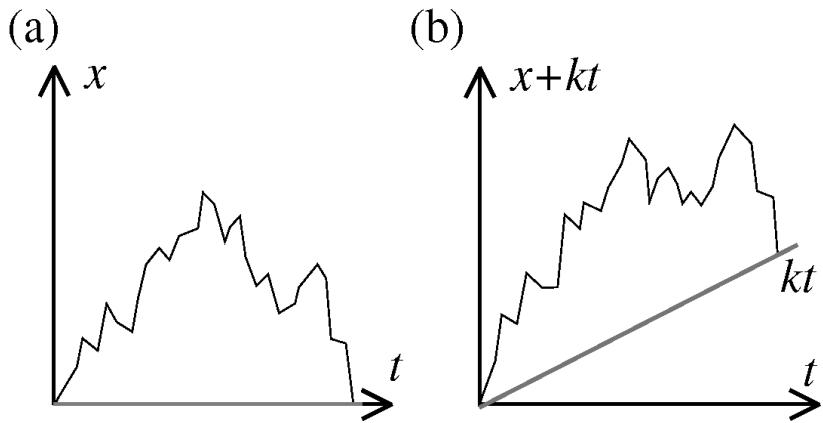}}
\caption{Sketch of the mapping of the return of the biased random walk
  to the origin (a) onto the return of the free random walk to the
  line $kt$ (b). The characteristic time $\tau$ is defined as the
  typical time at which the free random walk crosses the line $kt$. }
\label{s_bias}
\end{figure}

By rescaling $x\rightarrow x+kt$, it is clear that the probability for
a biased random walk with constant bias $k$ to return to the origin is
equivalent to the probability for a free random walk to return onto
the line $kt$. Given that the random walk is sub--linear, this
introduces a characteristic timescale $\tau$, after which the random
walk will very unlikely, namely with exponentially small probability,
go back to the line $kt$ (see figure \ref{s_bias}). 

The value of the characteristic time $\tau$ is such that $k\tau \sim
\sqrt{\tau}$, which gives $\tau \sim k^{-2}$. Since $t$ corresponds to
magnetization in the physical variables, $\tau$ corresponds to the
characteristic avalanche size $S_0$ in the cut--off of equation
(\ref{size_distrib}). As we will see further on, $T_0\sim S_0^{1/2}$,
giving $T_0 \sim k^{-1}$ for the cut--off in equation
(\ref{time_distrib}).

As we have seen in subsection \ref{sec:MAR}, the logarithmic potential
term does not alter the dynamic exponent $z$ which is still equal to $2$
as for the free random walk, thus the walk in the logarithmic
potential is still sublinear, and is affected by the bias in the same
way as the free random walk.  

The fact that the bias term $k$ generates an exponential cut--off in
the distribution of the return times, physically means that the size
of the largest magnetization event is regulated by the demagnetizing
field.

\subsection{Average shape of Barkhausen pulses as excursion of the random walk process}
\label{sec:EXC}

In analogy to conventional critical phenomena one might expect that
pulses of different durations have an average shape that, once
properly rescaled, collapses onto a universal scaling function. In a
recent paper by Sethna {\it et al.} \cite{sethna_crackling_2001} it
was suggested that this scaling function is an interesting quantity to
measure in order to make quantitative comparison between the
theoretical models and experiments. This quantity provides a much
sharper tool to test theory against experiments than the simple
comparison of critical exponents, that are scalar quantities.

Indeed, by performing this kind of analysis on experimental data,
average pulses, properly rescaled by their durations, approximately
collapse onto the same function (see figure \ref{shape})
\cite{colaiori_shape_2004}.

The average shape of the pulse is related, in terms of the
corresponding stochastic process, to what is known in the mathematical
literature as the {\em excursion}, namely the average trajectory
between two successive return of the process to some reference
value. 

The average shape for the process (\ref{rw-log}) has been calculated
in \cite{colaiori_shape_2004}.  This work was exactly motivated by
the problem, raised in \cite{sethna_crackling_2001}, of finding the
average form of a Barkhausen pulse as a function of time. 

Let us consider a generic random walk in a potential $U(x)$:
\begin{equation}
\dot{x}(t)=U(x)+\eta(t) \,.
\label{sto_proc}
\end{equation}

The average over trajectories starting at $x=0$ at time $t=0$, and
constrained to come back to $x=0$ for the first time at $t=T$, that we
will denote as $\langle x(t)\rangle_T$ is found to scale, in appropriate
time regimes depending on the potential $U$, as
\begin{equation}
\langle x(t)\rangle_T=T^{\gamma}f(t/T)
\label{exc1}
\end{equation}
where $f$ is some scaling function
\cite{baldassarri_average_2003,colaiori_average_2004}.

Let us consider processes constrained to the semiplane $x>0$, and
focus on the returns to the origin (take as reference value a small
threshold $x=\epsilon$). Once the solution with absorbing boundary
conditions of the Fokker Planck equation corresponding to the Langevin
equation (\ref{sto_proc}) is known, the excursion can be expressed by
averaging over all trajectories going through $x$ at time $t$.

In general, the probability for a Markovian process starting at
$\epsilon$ at time $0$ and going back to $\epsilon$ at time $T$ to be
found at position $x$ at time $t$, is given by $c(x,t;\epsilon,T)=
P(x,t;\epsilon,0) P(\epsilon,T;x,t)$. Using time translational
invariance $c(x,t;\epsilon,T) = P(x,t;\epsilon,0)P(\epsilon,T-t;x,0)=
P(x,t;\epsilon,0)\overline{P}(x,T-t;\epsilon,0)$,  where
$\overline{P}$ is the conditional probability for the inverse process.

The average excursion is therefore obtained by taking the
following limit for $\epsilon \rightarrow 0$:
\begin{equation}
\langle x(t)\rangle_T=\lim_{\epsilon \rightarrow 0}
\frac{\int_0^{\infty}dx x
  P(x,t;\epsilon,0)\overline{P}(x,T-t;\epsilon,0)}
{\int_0^{\infty}dx
  P(x,t;\epsilon,0)\overline{P}(x,T-t;\epsilon,0)}\,, 
\label{exc2}
\end{equation}
which counts the fraction of excursions that go through $x$ at time
$t$.

It is instructive to calculate the excursion for the simplest case of
a free random walk \cite{redner_guide_2001}. The probability
$P(x,t;\epsilon,0)$ for the walk confined to the semiplane can be
calculated by the image method as a linear combination of two
solutions of the Fokker Planck for the free process:
\begin{equation}
P(x,t;\epsilon,0)=\sqrt{\frac{2}{\pi t}} \left(
\exp^{-(x-\epsilon)^2/2t}- \exp^{-(x+\epsilon)^2/2t} \right) \,,
\label{images}
\end{equation}
which for small $\epsilon$ gives
\begin{equation}
P(x,t;\epsilon,0)=\sqrt{\frac{2}{\pi }} t^{-3/2}\epsilon x 
\exp^{-x^2/2t} \,.
\label{images2}
\end{equation}
In this case $\overline{P}=P$ due to the process time reversal
invariance, therefore the average excursion is given by 
\begin{equation}
\langle x \rangle_T=\frac{\int_0^\infty dx x^3 
 \exp^{-x^2/2\tau}}{\int_0^\infty dx x^2 
 \exp^{-x^2/2\tau}}\,,
\label{images3}
\end{equation}
with $\tau=t(T-t)/t$. Evaluating the integrals gives
\begin{equation}
\langle x
\rangle_T
=T^{1/2}\sqrt{\frac{8}{\pi}}
\sqrt{\frac{t}{T}\left(1-\frac{t}{T}\right)}\,.
\label{images4}
\end{equation}

Therefore the exponent $\gamma$ is equal to the wandering exponent of
the free process $\gamma=1/2$, and the scaling function is
\begin{equation}
f(x)=\sqrt{8/\pi}\sqrt{x(1-x)}\,,
\label{f_free}
\end{equation}
normalized to have $\int_0^1 dx f(x)=1$.

Let us now consider the case interesting to our theory of the
excursion for the random walk in a logarithmic potential. This can be
calculated in a similar way, by observing that the process is still
time translational invariant, and it is invariant under $t\rightarrow
-t$, as long as the coefficient $c$ is also inverted in the revers
process. Therefore $\overline{P}(x,T-t;\epsilon,0)$ is obtained by
replacing $c$ with $-c$ in $P(x,T-t;\epsilon,0)$.

Indeed, by repeating the previous steps with the solution of the
Fokker Plank equation (\ref{FP}), the calculation finally reduces,
also in this case, to evaluating the ratio between two successive
moments of a half Gaussian with variance $w=2t(T-t)/T$, so that the
average excursion results to be proportional to
$\sqrt{w}=\sqrt{t(T-t)/T}$, as in the case of the free random walk,
and it is still given by equation (\ref{f_free}).

\begin{figure}
\centerline{\epsfxsize=12cm \epsfbox{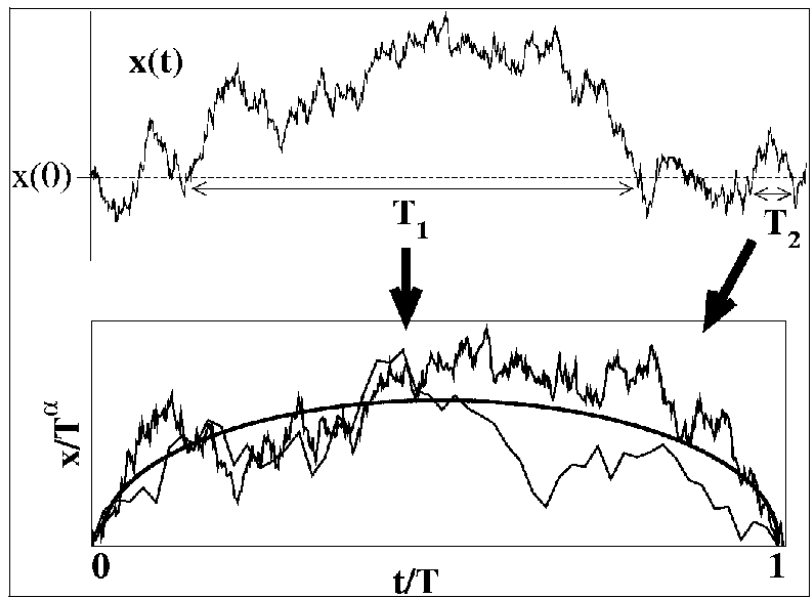}}
\caption{Schematic representation of the average excursion. Different
  realizations of the stochastic process all returning at the origin
  after the same time interval $T_1$ are averaged to give a smooth
  curve. Repeating the same process with a different return time
  interval $T_2$ gives another curve, that can eventually be rescaled
  onto the first. }
\label{excursion}
\end{figure}

To translate this result into the shape of the Barkhausen pulse, we
have to go back to our original physical variables.  The excursion
calculated above corresponds to the signal $v$ as a function of the
avalanche size $s$. The average pulse as a function of $s$ and of the
total size $S$ of the pulse is then
\begin{equation}
\langle v(s,S)\rangle=S^{1/2} f(s/S) \,.
\label{vs}
\end{equation}
However, the commonly measured experimental pulses, as those shown in
figure \ref{shape}, represent the signal $v$ as a function of duration
$t$ of the avalanche, that will have some different scaling and shape
\begin{equation}
\langle v(t,T)\rangle=T^{\beta}g(t/T) \,.
\label{vt}
\end{equation}

Following \cite{colaiori_shape_2004} we calculate the scaling function
$g$: we have to express $s$ and $S$ as functions of $t$, and $T$. By
definition, the size $s$ of the avalanche at time $t$, is the integral
of the signal $v$ from time $0$ to $t$:
\begin{equation}
\langle s(t,T) \rangle =\langle \int_0^t dt' v(t',T) \rangle=
T^{\beta+1}\int_0^{t/T}dx g(x)\,,
\label{s(t,T)}
\end{equation}
therefore, the total size $S$ scales as $S\simeq T^{\beta+1}$.
Approximating $s(t,T)$ with its average value in the
equation 
\begin{equation}
v(t,T)=v(s=s(t,T),S=S(T)) 
\label{g1}
\end{equation}
gives the following condition on $g$:  
\begin{equation}
T^{\beta} g(t/T)= T^{(\beta+1)/2}f\left(\int_0^{t/T}dx g(x)\right)\,. 
\label{g2}
\end{equation}

This enforces $\beta=1$, and gives an implicit integral equation for
$g$:
\begin{equation}
g(x)= f\left( \int_0^x g(x')dx'\right)\,,
\label{g3}
\end{equation}
that has to be solved subject to the boundary conditions
$g(0)=g(1)=0$. Equation (\ref{g3}) takes a simpler form in terms of
$h(x)= \int_0^x g(x')dx'$:
\begin{equation}
h'(x)= f\left(h(x)\right)=\pi \sqrt{h(x)(1-h(x))}\,,
\label{h_eq}
\end{equation}
which is solved by $h(x)=\sin^2(\pi x /2)$.  This gives for $g(x)$
\begin{equation}
g(x)= h'(x)=\pi/2\sin(\pi x)\,, 
\label{g4}
\end{equation}
which is normalized to unit and satisfies $g(0)=g(1)=0$. 
Finally, for the normalized average avalanche as a function of its
duration we get:
\begin{equation}
\langle v(t,T) \rangle=\frac{\pi T}{2}\sin{\pi t/T}   \,.
\label{g5}
\end{equation}

\begin{figure}
\centerline{\epsfxsize=9cm \epsfbox{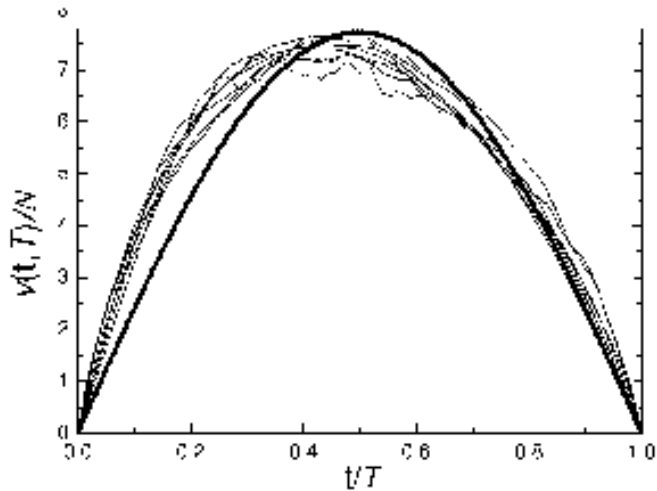}}
\caption{The approximate collapse of average pulses in figure
\ref{shape} is compared with the prediction of the ABBM model (bold
line) from equation \ref{g4}.  (Adapted from
\cite{durin_barkhausen_2006}).(Reprinted with permission from
\cite{durin_barkhausen_2006}).  \cite{durin_barkhausen_2006} Copyright
2005 by Academic Press.}
\label{asymm_abbm}
\end{figure}

Note that, while Eq. \ref{vs} with $f$ given by Eq. \ref{f_free} is an
exact result, and gives the scaling form of an avalanche as a function
of magnetization, to derive Eq. \ref{g5} one has to replace the
dependence of $s$ on time in a single realization with its average,
given by Eq. \ref{s(t,T)}.

\begin{figure}
\centerline{\epsfxsize=9cm \epsfbox{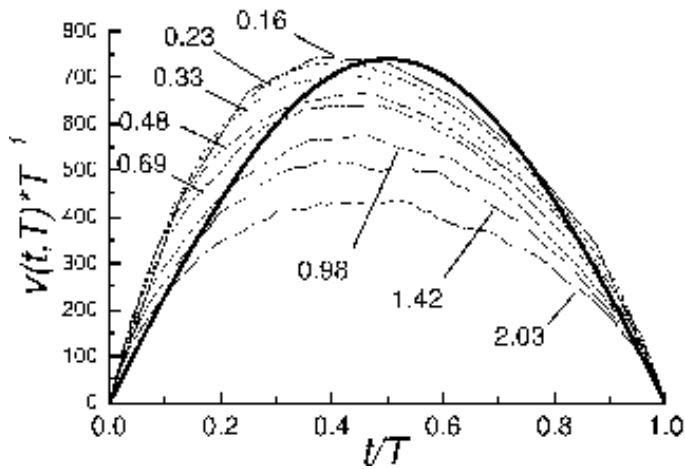}}
\caption{Average avalanche shapes for a polycrystalline $FeSi$
material. Signals as a function of time $v(t,T)$ are scaled according
to equation (\ref{g5}). The bold line is the theoretical prediction. 
Numbers in the graph denote avalanche duration in ms. (Adapted from
\cite{colaiori_shape_2004}).  (Reprinted with permission from
\cite{colaiori_shape_2004}).  \cite{colaiori_shape_2004} Copyright
2004 Elsevier.  }
\label{sin}
\end{figure}

Figure \ref{asymm_abbm} shows the comparison between rescaled
experimental pulses of figure \ref{shape}, and the scaling function
(\ref{g4}) predicted by the ABBM. The curves have similar shapes,
however, the experimental collapse is characterized by an evident
leftward asymmetry, while the ABBM predicts a symmetric shape. 
Note that normalizing the average pulses with $N$ as done in figure
\ref{asymm_abbm} should in principle be equivalent to rescaling them
by $T^{-1}$, as in equation (\ref{g5}), since
\begin{equation}
N=\int_0^{1}v(t,T)d(t/T)=S/T
\label{N}
\end{equation}
and $S\sim T^2$. However, if one directly plots $\langle v(t,T)\rangle
T^{-1}$, the curve collapse is obtained only for short avalanches,
while a clear deviation appears for longer avalanches (see figure
\ref{sin})

\begin{figure}
\centerline{\epsfxsize=9cm \epsfbox{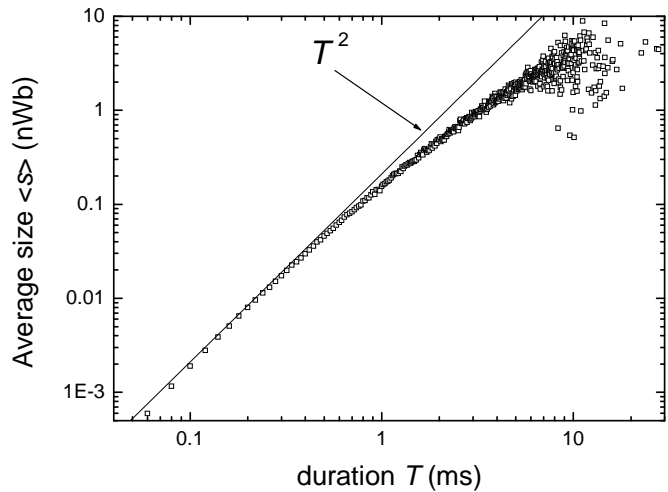}}
\caption{Average avalanche size as a function of its duration for
polycrystalline $FeSi$ material. Data are from the same experiment of
figure \ref{sin}. The solid line is the predicted scaling
$S\sim T^2$. (Reprinted with permission from
\cite{durin_barkhausen_2006}).  \cite{durin_barkhausen_2006} Copyright
2005 by Academic Press.}
\label{SvsT}
\end{figure}

This implies that the scaling $S\sim T^2$ does not hold for large
avalanches. Indeed, by plotting the average size $\langle S\rangle$
versus $T$ for the same BN signal, the curve deviates from the
predicted scaling for avalanches longer then $\sim 0.6 ms$, which is more
or less where the pulses start to deviate form the expected collapse
(see figure \ref{SvsT}). 

\subsection{Power spectra}
\label{sec:PS2}

The power spectrum for the ABBM equation has been calculated in the
original paper \cite{alessandro_domain-wall_1990}. In order to get the
correlation function it is more convenient to use directly the
physical variables, time, magnetization, and velocity (or signal
amplitude). The probability $R$ to find a signal $v$ at time $t$ is
related to the probability $P$ used in the previous sections by
\begin{equation}
R(v,t)=P(v,m)\frac{dm}{dt}\mid_{m=m(t)}=P(v,m)v\mid_{m=m(t)}\,.
\label{PtoR} 
\end{equation}
The signal--signal time correlation function is defined as 
\begin{equation}
C(t)=\langle (v(t)-c/k)(v(0)-c/k)\rangle \,. 
\label{correlat} 
\end{equation}
By deriving the probability $R$ with respect to $t$ and using the
Fokker Planck equation, it is easy to get a differential equation for
$C$ (see \cite{alessandro_domain-wall_1990} for the details), which
turns out to be
\begin{equation}
\frac{dC(t)}{dt}=-k C(t)\,. 
\label{correlat2} 
\end{equation}
Therefore $C(t)=c/k \exp{(-kt)}$, which corresponds to a Lorentzian
power spectrum
\begin{equation}
F(\omega)=\frac{2c}{\omega^2+k^2}\,. 
\label{ps0} 
\end{equation}
Equation (\ref{ps0}) captures the correct $\omega^{-2}$ large
frequency decay found in experiments, however it saturates to a
constant for small $\omega$, without showing any peak.

To get a better prediction of the power spectrum, let us go back once
again to the original formulation of the ABBM model, which includes a
more general case, that we have not considered up to now, in which a
finite correlation length $\xi_P$ is associated to the pinning field
\cite{alessandro_domain-wall_1990}.  A finite $\xi_P$ has the effect
of confining the pinning field, and physically measures the length
over which the pinning field decorrelates.

In most cases it can safely be assumed $\xi_P \rightarrow \infty$, as
done in most literature, and in the analysis reported in the previous
sections.  However, the existence of a finite correlation length in
the pinning field affects the low frequency behavior of the power
spectrum.  The power spectrum for the case of a finite pinning
correlation length is calculated in \cite{alessandro_domain-wall_1990}
in a high field velocity approximation, and results to be 
\begin{equation}
F(\omega)=\frac{2 c \omega^2}{(\omega^2+k^2)(\omega^2+\tau_P^{-2})}\,, 
\label{ps1} 
\end{equation}
where $\tau_P=\xi_P/c$ is a characteristic time associated to the
correlation length $\xi_P$. 

Equation (\ref{ps1}) reproduces the low frequency part of the
spectrum. It predicts the existence of a peak at
$\omega_M=(k\tau_P)^{1/2}\sim c^{1/2}$, with a maximum value of the
spectrum $F_M=\frac{4c}{(\tau^{-1}+\tau_P^{-1})^2}$ which scales as $c$
for small driving rates, as observed in experiments.

\begin{figure}
\centerline{\epsfxsize=9cm \epsfbox{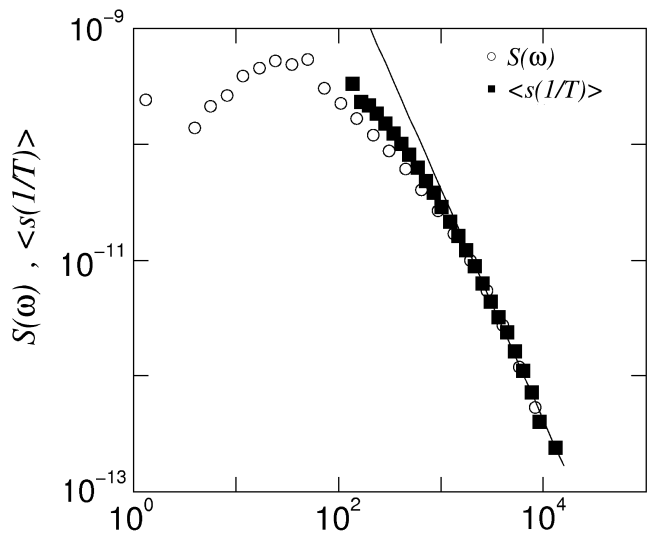}}
\caption{Comparison of the power spectrum $F(\omega)$ with the average
avalanche size $\langle S \rangle$ as a function of the inverse
duration for a polycrystalline $FeSi$ material. The solid line is the
predicted scaling $S\sim T^2$. (Adapted from
\cite{durin_power_2002}). (Reprinted with permission from
\cite{durin_power_2002}).  \cite{durin_power_2002} Copyright 2002 by
Elsevier.}
\label{ps_dev}
\end{figure}

It would be interesting to understand how the presence of a finite
$\xi_P$ affects other properties of the avalanches, as the
decorrelation of the pinning field at some finite length scale might
be responsible of some of the observed deviations from the $\xi_P
\rightarrow \infty$ ABBM theory in large events.

The deviation from the $1/\omega^2$ prediction for the tail, which is
observed at intermediate frequencies in some experiments might have
the same origin as the deviation from scaling observed in the pulse
shape scaling in the previous subsection \cite{durin_power_2002}. This
is suggested by the comparison of the power spectra with the average
size versus inverse avalanche duration. The two curves deviate for the
same value of $\omega$ (see figure \ref{ps_dev}). The origin of the
deviation in the size versus duration plot observed in some
experiments is, however, still unclear.

A possibility is that the eddy current retardation that we will
discuss in the next section is responsible for the intermediate
frequencies deviations from the asymptotic scaling, as suggested in
\cite{durin_signature_2007} based on simulation of the modified ABBM
equation. We will come back to this in the next section.

\subsection{Summary of ABBM predictions and comparison with the phenomenological observations}
\label{sec:sum}

Let us summarize the ABBM predictions and compare them with the 
phenomenological observations from BN experiments on soft
materials with long range interactions:
\begin{enumerate}
\item{The distribution of the amplitude of the BN signal follows a
  power law decay with an exponent $1-c$ as observed in experiments (see
  equation \ref{V_dist}). }
\item{The distributions of sizes and durations of avalanches can be
  calculated in terms of recurrence properties of the process
  described by equation \ref{lang}, and turn out to be power laws,
  with exponents $\alpha=2$ and $\tau=3/2$ in the quasi--static limit
  $c=0$, as found in experiments. }
\item{Equation \ref{lang} predicts a linear dependence of the
  exponents $\alpha$ and $\tau$ on the driving rate $c$, given by
  $\alpha=2-c$ and $\tau=(3-c)/2$, the same obtained in experiments.}
\item{There exist a threshold $c=1$ above which the process in
  equation \ref{lang} become transient. This corresponds to the
  threshold in the driving field rate above which the BN pulses
  collapse into a unique large avalanche of magnetization reversal.}
\item{The bias term in equation \ref{lang}, deriving from the
  demagnetizing field, is responsible for the exponential cut--offs in
  the distributions of sizes and durations of avalanches. The dependence
  on $k$ of the cut--off predicted by the ABBM equation is however
  different from that observed in experiments.}
\item{The average excursion of a trajectory for the stochastic process
  in equation \ref{lang} can be calculated and is related to the average
  pulse shape in BN. However, the predicted shape is symmetric, while
  experimental pulses show a clear leftward asymmetry. }
\item{The power spectrum decays at large frequencies as
  $\omega^{-2}$. A finite correlation length for the pinning field
  produces a peak at small frequencies at $\omega_M\sim c^{1/2}$ of
  amplitude $F_M \sim \omega$. These results are compatible with
  experimental ones, however, deviations from the predicted behavior
  are observed at intermediate frequencies in some experiments. }
\end{enumerate} 

These predictions can be compared point to point with the experimental
observations summarized in sec \ref{sec:sum_p}.  Most of the
phenomenology is captured by the ABBM model, however, two main
unexplained discrepancies between experiments and theory appear: one
is the scaling of the cut--offs with the demagnetizing factor $k$, the
other is the presence of a leftward asymmetry in the pulse shape.

The first inconsistency can be dealt with by going back to the full
elastic interface model, as we will discuss in briefly the next
subsection. The second one is more serious, as the pulses asymmetry is
related to non--universal properties of the magnetization reversal of
microscopic origin, and we will postpone its discussion to the next
section.

\subsection{Cut-off out of mean field approximation}
\label{sec:cutoff}

Let us go back for a moment to the elastic interface model discussed
in section \ref{sec:mesoscopic}. This model has been extensively
simulated both in the infinite range case, corresponding to the mean
field, and in finite dimension $d$. The case $d=3$ is supposed to give
results that, apart from logarithmic corrections, coincide with the
mean field, since $d=3$ is the upper critical dimension for this
system. This is confirmed in simulations, where one gets in both
cases, for distribution of sizes and durations of the pulses, the
exponents $\alpha$ and $\tau$ predicted by the ABBM theory, both in
the adiabatic limit $c\rightarrow 0$, and for finite driving $c$.

However, the scaling with $k$ of the cut--offs $S_0$ and $T_0$ in the
distributions is different in the case $d=3$ and in mean field. The
simulations on the infinite range model confirm the results of section
\ref{sec:BIAS}: they predict that the cut--offs scale as $S_0\sim
k^{-2}$, and $T_0\sim {k^{-1}}$, consistently with the results
obtained for the ABBM model.  However simulations in $d=3$ are not
compatible with these results.

Earlier simulations \cite{zapperi_dynamics_1998} in $d=3$ give
$S_0\sim {k^{-1}}$, and $T_0\sim k^{-1/2}$, both in the quasi--static
limit and at finite $c$. In \cite{zapperi_dynamics_1998} the full size
and duration distributions are fitted by
\begin{equation}
P(S,k)\sim S^{-(3-c)/2}f_S(S/k^{-\gamma_S})\,,
\label{culoff_s}
\end{equation}
and
\begin{equation}
P(T,k)\sim T^{-(2-c)}f_T(T/k^{-\gamma_T})\,, 
\label{culoff_t}
\end{equation}
with $\gamma_S\simeq1$, and $\gamma_T\simeq 1/2$, which are claimed to
be compatible with the scaling obtained in the experiments reported in
the same paper, where the cut--offs dependence on $k$ has been
investigated by performing experiments on samples with different
geometries. It has to be taken into account the extrapolation of a
precise cut--off value is complicated by the further dependence of the
distributions on the driving field rate $c$.

In more recent papers, different values are reported
\cite{durin_scaling_2000, durin_universality_2000}, corresponding to
$\gamma_S \simeq 2/3$, and $\gamma_T \simeq 1/3$.  These latter
results are also supported by the following argument. The maximum
interface jump is limited by a correlation length $\xi$, which is the
length at which the dipolar interaction term, of order $\xi^{-1}$
overcome the restoring force. In $d=3$ (where the interface is not
rough), this is of order $k\xi^{d-1}$, where $\xi^{d-1}h$ is the size
of the avalanche. This gives $\xi \sim k^{-{1/d}}$, and therefore $S_0
\sim \xi^{d-1} \sim k^{-(d-1)/d}=k^{2/3}$.

In the same papers \cite{durin_scaling_2000, durin_universality_2000},
results from more recent experiments are also reported, corresponding
to $\gamma_S \simeq 0.6$, and $\gamma_T \simeq 0.3$, that could be
compatible with the values $\gamma_S \simeq 2/3$, and $\gamma_T \simeq
1/3$. In these works, the experimental estimates of the dependence on
the demagnetizing field is extracted by repeating experiments on the
same sample, which is progressively cut in order to change the value
of $k$. This procedure allows to change the value of $k$ while stress
and internal disorder are kept constant in the repeated experiments
(see figure \ref{So_To}, which also reports the results of experiments
on materials belonging to the short range universality class not
discussed here).  The experimental results in
\cite{durin_scaling_2000, durin_universality_2000} are therefore
expected to be more reliable than previous ones, where different
samples where used as in \cite{zapperi_dynamics_1998}.

The scaling with $k$ of the cut--offs in the distributions of sizes
and durations of the avalanches seems to agree with that predicted by
the elastic interface model at the upper critical dimension $d=3$,
although this prediction is different from the mean field one.  The
origin of the discrepancy between mean field behavior and the behavior
at the upper critical dimension is not fully understood, and we
believe that this point would still need further analysis in order to
be definitely clarified.

\section{Beyond ABBM: Asymmetry in the pulse shape}
\label{sec:ASYM}

As we have seen in the previous section, despite its extreme
simplicity, the ABBM model is able to capture a number of
phenomenological evidences from BN experiments. However, the analysis
on the scaling functions is more subtle: although the average pulses
do approximately collapse onto each other once rescaled by their
duration, by comparing the average shape of rescaled experimental
Barkhausen pulses with the prediction from the ABBM model, one clearly
observes a discrepancy: while the ABBM model predicts a symmetric
shape, experimental pulses are systematically characterized by a
visible leftward asymmetry, as can be seen in figure \ref{asymm_abbm}.
The leftward asymmetry indicates that, unlike what would be expected
from standard inertia, avalanches start fast and end slowly.

Given the success of ABBM theory in reproducing many other aspects of
phenomenology, this inconsistency has been considered quite enigmatic,
and the issue of understanding its origin has remained an open problem
for some time.

The first attempts to identify the origin of the asymmetry have been
done in \cite{baldassarri_average_2003}, and more extensively in
\cite{colaiori_average_2004}. In these works, the authors pose the
general problem of analyzing the behavior of the average excursion in
a stochastic process, in order to understand to which extent this
quantity is universal, and what kind of information it gives on the
process. The idea was that this analysis could suggest in which
direction the ABBM equation had to be modified in order to recover a
leftward asymmetric pulse shape.

For a large class of stochastic processes, a scaling law of the form
\ref{exc1} is obeyed, in an appropriate time regime, with a scaling
function that is, to a large extent, independent of the details of the
single increment distribution, while it encodes interesting
information on the presence and form of correlations. Indeed the
introduction of correlated noise, is found to induce asymmetric
shapes. However, in the cases considered in
\cite{baldassarri_average_2003,colaiori_average_2004}, the scaling
functions are characterized by a negative skewness, corresponding to a
rightward asymmetric shape.

More recently \cite{zapperi_signature_2005,durin_signature_2007}, the
origin of the asymmetry has finally been identified as a transient
effect of eddy currents. Eddy currents are generated in conducting
magnets as a response to domain wall displacements. This response,
however, is not instantaneous and acts as an anti--inertial effect on
the domain wall dynamics.  It turns out that the effect of this
retardation can indeed be accounted for, to a first approximation, by
including a negative effective inertial term in the stochastic
equation of motion for the domain wall.

\subsection{Experimental evidence of asymmetric pulses}
\label{sec:ASYM_exp}

Figure \ref{shape} shows an attempt to collapse average pulses of
different durations, obtained from a BN experiment on a
polycrystalline $FeSi$ sample, onto the same function. The pulses do
approximately collapse, when properly rescaled, however, the supposedly
universal shape is clearly not symmetric, as would be predicted by the
ABBM model (see figure \ref{asymm_abbm}). The asymmetry appears to be
more pronounced for small avalanches.

To get a quantitative measure of the asymmetry the skewness can be
calculated as:
\begin{equation}
\Sigma(T)=\frac
{\frac{1}{T}\int_0^T dt \langle v(t,T)\rangle(t-\overline{t})^3}
{\langle \frac{1}{T}\int_0^T dt 
\langle v(t,T)\rangle (t-\overline{t})^2\rangle^{3/2}}\,,
\label{skewness}
\end{equation}
where $\overline{t}=\int_0^T dt \langle
v(t,T)\rangle/T$. 

\begin{figure}
\centerline{\epsfxsize=10cm \epsfbox{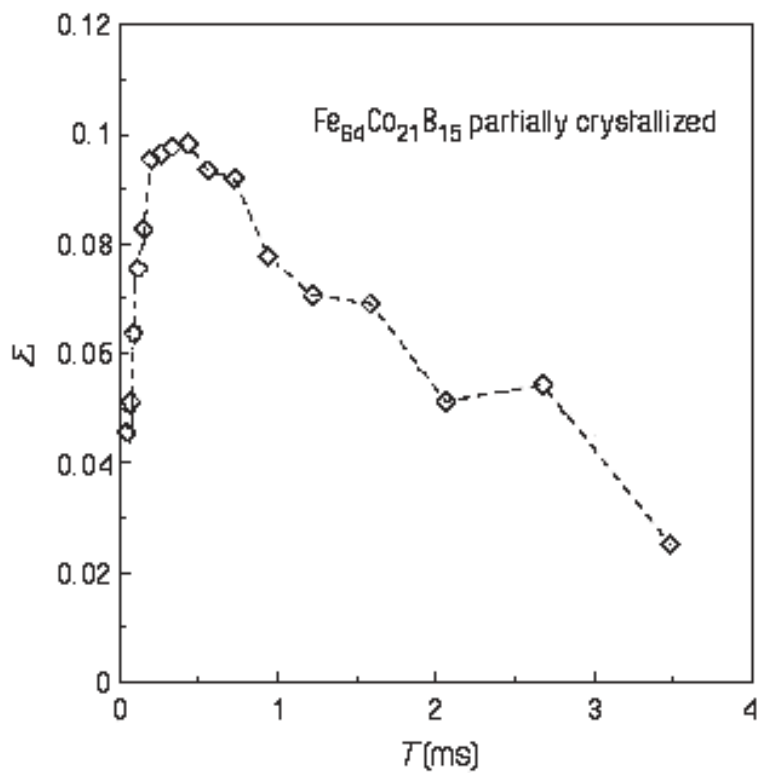}}
\caption{Skewness as a function of the avalanche duration measured in
  an experiment on a partially crystallized $Fe_{64} Co_{21} B_{15}$
  sample. (Reprinted from \cite{zapperi_signature_2005}).
}
\label{ex_skewness}
\end{figure}

Figure \ref{ex_skewness} shows the skewness as a function of the pulse
duration in an experiment on a partially crystallized $Fe_{64}Co_{21}
B_{15}$ ribbon. The skewness is always positive, corresponding to the
leftward asymmetry. It shows a peak for $T_P\simeq 400 \mu s$, and
decays to zero as $T$ goes to infinity.

The fact that, except for the initial increase at very small
durations, the asymmetry becomes smaller and smaller for larger and
larger avalanches suggests that the ingredient still missing in the
ABBM theory must involve microscopic details of the system that become
negligible on larger scales. The presence of a peak indicates the
existence of some microscopic timescale.

\subsection{Retarded eddy current pressure}
\label{sec:ASYM_eddy}

To understand why ABBM is unable to reproduce the leftward asymmetry
of the average pulse shape, we have to go back and question the
assumptions which were made in its derivation. The equation of motion
for the magnetic interface was derived by assuming that the work done
by the total effective field (the applied field corrected by the
demagnetizing and pinning fields) balances at any time the energy
dissipated by eddy currents, which, in the quasistatic approximation,
is proportional to the instantaneous velocity. The equation is
overdamped, and neglects inertial effects. Let us mention that the
existence of inertial effects in domain wall dynamics, has been
demonstrated by D\"oering \cite{d_ferromagnetic_1948}, who showed that
a moving wall differed in energy from the wall at rest by a term due
to gyromagnetic effects and proportional to the square wall speed. He
identified this as a kinetic energy term and defined the constant of
proportionality as one--half of the mass of the wall. The so called
D\"oering mass $M_D$ results to be proportional to the inverse wall
width divided by the squared gyromagnetic ratio (for a simplified
treatment see \cite{becker_dynamique_1951}, and also
\cite{kittel_magnetic_1956} for a review).  Inertial effects are
therefore always present in domain wall dynamics, however, they are
not relevant up to the GHz band, and can be safely neglected in the
equation of motion for the magnetic wall in other regimes, where eddy
currents damping largely overwhelms the contribution from D\"oering
mass.

The assumption that the work done by the net field is at any time
proportional to the instantaneous velocity is, however, only
approximately correct: eddy currents take a finite time to set up
after the corresponding wall displacement, and also, they persist for
a finite time after the magnetic reversal. Because of this delay, the
eddy pressure on the wall at a given time $t$ of the avalanche depends
on the history of the motion: is not strictly proportional to the
instantaneous velocity, but rather it is a weighted average of all the
velocities from the beginning of the avalanche up to time $t$.

To go beyond the quasistatic approximation, one has to investigate the
dynamical effects eddy current dissipation. Following the lines
indicated by \cite{bishop_contribution_1980} this has been done by
calculating the magnetic field generated by the eddy currents from the
full Maxwell equations \cite{zapperi_signature_2005,
colaiori_eddy_2007}.

\begin{figure}
\centerline{\epsfxsize=12cm \epsfbox{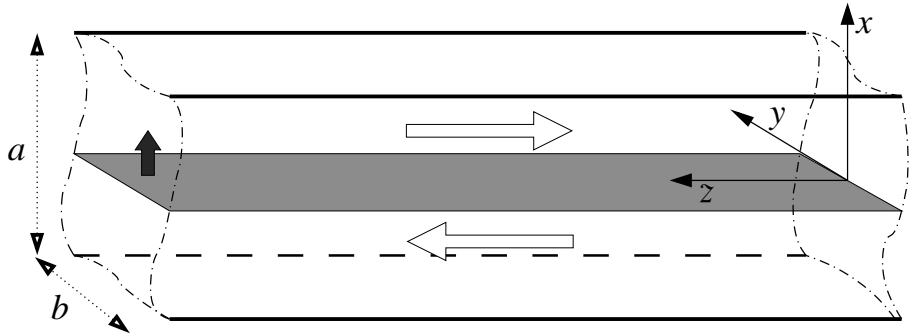}}
\caption{Horizontal arrows indicate the directions of the
  magnetization in the two domains. The black vertical arrow indicates
  the direction of motion of the wall.}
\label{eddy.eps}
\end{figure}

Consider a sample with dimension $x\in[-a/2,a/2]$, $y\in[-b/2,b/2]$,
and infinite in the $z$ direction, divided in two magnetic domains by
a rigid domain wall on the $yz$ plane, that starts to move from the
zero--magnetization position $x=0$, as in figure \ref{eddy.eps}.

The displacement of the magnetic wall in the conducting medium induces a
flow of eddy currents that generates a magnetic field $H_e$, which, in
this geometry, is parallel to the $z$ axis:
\begin{equation}
\vec{H_e}=H_e(x,y,t)\hat{z}.
\label{eddyfield}
\end{equation} 
Neglecting the displacement currents with
respect to the ohmic currents, the Maxwell equation for $H_e$ is
\begin{equation}
\nabla^2H_e(x,y,t)=\sigma \mu \partial_t H_e(x,y,t)\,,
\label{Maxwell}
\end{equation}
where $\sigma$ and $\mu$ are the electric conductivity and the
magnetic permeability of the medium respectively. 

Equation (\ref{Maxwell}) is a diffusion equation for the eddy field,
with a typical timescale for diffusion proportional to $\sigma
\mu$. In the quasi--static approximation the permeability is assumed
to be negligible within the magnetic domains, and equation
(\ref{Maxwell}) reduces to $\nabla^2H_e=0$, corresponding to an
instantaneous eddy field propagation.

This approximation is correct as long as there is a clear separation
of timescales, which happens when the observed quantities do not vary
appreciably in time on a scale of the order of the characteristic time
for diffusion. In that case, one can assume an instantaneous response
of the eddy field as given by the static Maxwell equation. However,
this is not the case for Barkhausen avalanches, especially for short
ones, and therefore the dynamic effect of eddy field has to be taken
into account.

Let us go back to equation (\ref{Maxwell}), which has to be solved
subject to the appropriate boundary condition

\begin{equation}
H_e=0
\end{equation} 
on the sample surface, and with the Faraday condition
\begin{equation}
\partial_x H_e(0^+,y,t)- H_e\partial_x H_e(0^-,y,t)=2 \sigma M_s v(t)\,,
\label{faraday}
\end{equation} 
that regulates the discontinuity of the field across the wall. In
equation \ref{faraday} $M_s$ is the saturation magnetization, and $v(t)$ is
the instantaneous velocity of the domain wall.

The solution of equation (\ref{Maxwell}) has been derived in
\cite{colaiori_eddy_2007}, by expanding $H_e$ in its Fourier
components
\begin{equation}
H_e(x,y,t)=\frac{1}{\sqrt{2\pi}}\int_{-\infty}^{\infty} d\omega
F(x,y,\omega)
e^{i\omega t}\,. 
\label{H2}
\end{equation} 
Each component has to satisfy 
\begin{equation}
\nabla F(x,y,\omega)=r^2F(x,y,\omega)\,, 
\label{F}
\end{equation}
with $r^2=i\omega \sigma \mu$. The condition for the magnetic eddy
field to be zero on the sample boundary implies that $F(x,\pm
b/2,\omega)=F(\pm a/2,y,\omega)=0$.  The solution of equation
(\ref{f}) for the Fourier components with these boundary condition is
of the form
\begin{equation}
F(x,y,\omega)=\sum_{n=0}^{\infty}A_n(x,\omega) \cos(\lambda_n y)\,,
\label{F2}
\end{equation}
where $A_n$ satisfies
\begin{equation}
\partial^2_{x}A_n(x,\omega)=\Lambda_n^2 A_n(x,\omega)\,,
\label{A}
\end{equation}
with $\Lambda_n^2=\lambda^2+r^2$, to be solved separately for $x>0$
and $x<0$. The condition on the sample boundary in the $y$ direction
implies $\cos(\pm b/2\lambda_n)=0$ which fixes
$\lambda_n=(2n+1)\pi/b$.  The condition on the other boundary in
satisfied by choosing
\begin{equation}
A_n(x,\omega)=C_n(\omega) \sinh(\Lambda_n(\mid x \mid - a/2))\,,
\label{A2}
\end{equation}
so that 
\begin{equation}
F(x,y,\omega)=\sum_{n=0}^{\infty}C_n(\omega) 
\sinh(\Lambda_n(\omega)(\mid x \mid -a/2))\cos(\lambda_n y)\,.
\label{F3}
\end{equation}

The functions $C_n(\omega)$ are fixed by the Faraday condition around
the wall, which in Fourier space reads
\begin{equation}
\partial_xF(0^+,y,\omega)=-\partial_x F(0^-,y,\omega)=
\sigma M_s \hat{v}(\omega)\,,
\label{FC}
\end{equation}
where $\hat{v}$ is the Fourier transform of the velocity of the wall
$v(t)$. Equation (\ref{FC}) implies
\begin{equation}
\sum_{n=0}^{\infty}C_n(\omega)
\Lambda_n(\omega) \cosh(\Lambda_n(\omega) a/2) 
\cos(\lambda_n y)=\sigma M_s \hat{v}(\omega)\,.
\label{}
\end{equation}
Multiplying by $\cos(\lambda_m y)$, integrating in $[-b/2,b/2]$, and
using the orthogonality relations $\int_{-b/2}^{b/2}dy \cos(\lambda_n
y) \cos(\lambda_m y)=\delta_{n,m}b/2$, and $\int_{-b/2}^{b/2}dy
\cos(\lambda_m y)=(-1)^m2/\lambda_m$ one gets
\begin{equation}
C_n(\omega)=
(-1)^n\frac{4\sigma M_s}{b} \frac{1}{\lambda_n
  \Lambda_n(\omega) \cosh(\Lambda_n(\omega) a/2)}\hat{v}(\omega)\,,
\label{C}
\end{equation}
so that the solution for the Fourier components of the field $H_e$ is
finally found to be given by 
\begin{equation}
F(x,y,\omega)=\frac{4\sigma  M_s}{b}
\sum_{n=0}^{\infty}(-1)^n
\frac{\sinh(\Lambda_n(\omega)(\mid x \mid -a/2))}
{\lambda_n  \Lambda_n(\omega) \cosh(\Lambda_n(\omega) a/2)}
\cos(\lambda_n y)\hat{v}(\omega)\,.
\label{F4}
\end{equation}

To calculate the pressure on the wall due to the presence of eddy
currents, only the value of the field at $x=0$ is needed, where $F$
simplifies to:
\begin{equation}
F(0,y,\omega)=
\sum_{n=0}^{\infty}(-1)^n
\frac{4\sigma  M_s}{b}\frac{\tanh(\Lambda_n(\omega) a/2)}
{\lambda_n  \Lambda_n(\omega) }\cos(\lambda_n y)\hat{v}(\omega)\,.
\label{F5}
\end{equation}

The average eddy current pressure on the wall is then obtained by
integrating the magnetic field over $y$ at the wall position $x=0$:
\begin{equation}
P(t)=\frac{2I}{b}\int_{-b/2}^{b/2}dy
H_e(0,y,t)=\frac{2I}{b}
\frac{1}{\sqrt{2\pi}}
\int_{-\infty}^{\infty} d\omega e^{i \omega t} 
\int_{-b/2}^{b/2}dy F(0,y,\omega)\,.
\label{Pt}
\end{equation}
In real space the pressure at time $t$ is given by a convolution of
velocities of the wall at all times prior to $t$ with some response
function $f$: 
\begin{equation}
P(t)=\frac{1}{\sqrt{2\pi}}\int_{-\infty}^{\infty} ds v(t-s) f(s),
\label{Pt22}\,.
\end{equation}
which, in Fourier space, corresponds to 
\begin{equation}
\hat{P}(\omega)=-\hat{v}(\omega)\hat{f}(\omega)\,.
\label{Po2}
\end{equation}
Replacing $F(0,y,\omega)$ in equation (\ref{Pt}) with its expression
(\ref{F5}), and performing the integral over $y$, we get, for the
Fourier transform of the response function, the expression:
\begin{equation}
\hat{f}(\omega)=\frac{16 I^2
  \sigma}{b^2}\sum_{n=0}^{\infty}\frac{\tanh(\Lambda_n(\omega)
  a/2)}{\lambda_n^2 \Lambda_n(\omega)} \,.
\label{fo}
\end{equation}
To proceed further it is convenient to simplify the $\omega$
dependence in equation (\ref{fo}), which can be done by using the
relation
\begin{equation}
\tanh(z)=8z\sum_{k=0}^{\infty}\frac{1}{\pi^2 (2k+1)^2 +4 z^2}\,,
\label{trick}
\end{equation}
with $z=\Lambda_n(\omega) a/2$, which gives
\begin{equation}
\hat{f}(\omega)=
\frac{64 a I^2 \sigma}{b^2}
\sum_{n,k=0}^{\infty}\frac{1}{\lambda_n^2 
(b^2 \lambda_k^2+a^2 \Lambda_n^2(\omega))}\,.
\label{ft}
\end{equation}
Replacing $\lambda_n$ and $\Lambda_n$ with their expressions, one gets
\begin{equation}
\hat{f}(\omega)=\frac{64 I^2 \sigma}{a b^2 \epsilon^2}
\overline{\sum_{n,k=1}^{\infty}}\frac{1}
{n^2\omega_b\left(k^2\omega_a+n^2\omega_b+i\omega\right)}\,,
\label{fo2}
\end{equation}
where $\overline{\sum}$ indicates a summation over odd numbers only,
and $\omega_a=\tau_a^{-1}=\pi^2/\epsilon a^2$,
$\omega_b=\tau_b^{-1}=\pi^2/\epsilon b^2$.  

From equation (\ref{fo2}), the expression for the response $f$ in real
space is obtained by anti--transforming term to term in the double
sum, and replacing each term of the form $1/(\omega_0+i\omega)$ with
its inverse Fourier transform $\sqrt{2\pi} \exp(-\omega_0 t)
\theta(t)$ (where $\theta(t)$ is the Heaviside theta function). This
leads to:
\begin{equation}
f(t)=\sqrt{2\pi}\frac{64 I^2 \sigma}{a b^2 \epsilon^2}
\overline{\sum_{n,k=1}^{\infty}}\frac{1}{n^2\omega_b}
e^{-\omega_{k,n}t}\theta(t)\,,
\label{f}
\end{equation}
where $\omega_{k,n}=\tau_{k,n}^{-1}=k^2\omega_a+n^2\omega_b$.  The
form (\ref{f}) of the response function gives through the convolution
(\ref{Pt}) the explicit expression of the retarded pressure on the
wall.

The response function finally results to be the sum of simple
exponential relaxations, with different relaxation times. The most
relevant relaxation time is the largest one, which is
\begin{equation}
\tau_{0,0}=\frac{\sigma\mu}{\pi^2}\left(\frac{1}{a^2}+\frac{1}{b^2}\right)^{-1}
\,.
\label{largest_tau}
\end{equation}
This is the quantity that has to be compared with the typical
timescale of the phenomena under study to decide whether or not the
dynamical effects due to eddy current relaxation are negligible.  

In a typical Barkhausen noise experiment, $\tau_{0,0}$ is of the order
of $5 \mu s$. Although this time is small with respect to overall
duration of a Barkhausen pulse, the magnetization within an avalanche
does vary appreciably on that timescale (see figure \ref{shape}).

\subsection{Generalized ABBM: negative effective mass}
\label{sec:ASYM_ABBM}

In terms of the equation of motion, the effect due to the
non--instantaneous response of the eddy field to the wall displacement
is accounted for by replacing the left hand side of equation (\ref{mf2})
with the retarded pressure $P(t)$.  This leads to a modified ABBM
equation:
\begin{equation}
\int_0^{t}ds f(t-s) v(s)=H(t)-km+W(m)\,,
\label{modified_ABBM}
\end{equation}
that includes the time nonlocal effects. 

The eddy dissipation keeps memory of the past velocities of the
interface, within a timescale of the order of the eddy field diffusion
relaxation time. 

To clarify the effects of the response function, it is useful to
consider the case of a simple exponential decay
$f(t)=\Gamma/\tau_0\exp(-t/\tau_0)$, which corresponds to the
contribution of a single term in equation (\ref{f}) to the equation of
motion. As long as $\tau_0$ is small with respect of the duration of
the avalanches considered, only velocities at small times before $t$
will contribute to the convolution, so that we can expand $v(s)$
around $t$ and perform the integrals:
\begin{equation}
\Gamma \int_0^t ds \frac{e^{-s/\tau_0}}{\tau_0} v(t-s)\simeq
\Gamma v(t)-\Gamma \tau_0v'(t) \,.
\label{exp_decay}
\end{equation} 
A simple exponential response function would therefore modify the
equation of motion by introducing an inertial term
\begin{equation}
\Gamma v(t)+Mv'(t)=H(t)-km+W(m)\,, 
\label{mod_abbm}
\end{equation}
and would lead to the identification of an effective mass $M=-\Gamma
\tau_0$, which turns out to be negative.

When the full response function is considered, a series of mass and
damping terms are generated, each coming from a single
exponential. Taking only the leading order in $\mu$, this procedure
leads in a similar way to the identification of a damping coefficient
\begin{equation}
\Gamma =\frac{64 I^2\sigma
b^2}{a \pi^4 } \Sigma_1(b/a)\,,
\label{damping} 
\end{equation}
and of a negative effective mass
\begin{equation}
M=-\frac{64 I^2 \sigma^2 \mu b^4}{a \pi^6} \Sigma_2(b/a)\,,
\label{mass}
\end{equation} 
where $\Sigma_1(\alpha)= \overline{\sum}_{n,k} \,\frac{1}{n^2}
\frac{1}{n^2+\alpha^2 k^2}$, and $\Sigma_2(\alpha)=
\overline{\sum}_{n,k} \,\frac{1}{n^2} \frac{1}{(n^2+\alpha^2 k^2)^2}$,
and $\overline{\sum}$ indicates a summation over odd numbers only. 

As in the case of the simple exponential relaxation, the ratio between
mass and damping defines a characteristic time
\begin{equation}
\tau=|M|/\Gamma=\omega_b^{-1} \frac{\Sigma_2(b/a)}{\Sigma_1(b/a)}\,.
\label{tau}
\end{equation}  

Keeping all orders in $\mu$ results in a frequency dependent
effective mass, that has been calculated in \cite{colaiori_eddy_2007}
and is given by 
\begin{equation}
M=-\frac{64 I^2 \sigma^2 \mu b^4}{a \pi^6}
\overline{\sum_{n,k=1}^{\infty}}\frac{1}
{n^2\left((k^2(a/b)^2+n^2)^2+(\omega/\omega_b)^2\right)}\,.
\label{Momega}
\end{equation}
The effective mass turns out to be negative at all frequencies. 

The leftward pulse asymmetry observed in Barkhausen experiments is
indeed consistent with a negative effective mass: the avalanches start
fast and end slowly, which is exactly the opposite of what one would
expect from standard inertia. Indeed, since the retarded pressure at
time $t$ is given by a weighted average of previous velocities of the
wall up to time $t$, this anti--inertial effect can be understood by
observing that the effective average velocity is smaller than the
instantaneous one for an accelerating wall, while the opposite is true
when the wall decelerates.

\subsection{Effective mass dependence on sample geometry}
\label{sec:ASYM_geo}

The role of sample geometry on eddy current retardation has been
studied in detail in Ref. \cite{colaiori_eddy_2007}. It turns out that
the only geometrical parameter that significantly affects the damping
coefficient, effective mass, and characteristic time, is the smallest
sample dimension.

\begin{figure}
\centerline{\epsfysize=9.5cm \epsfbox{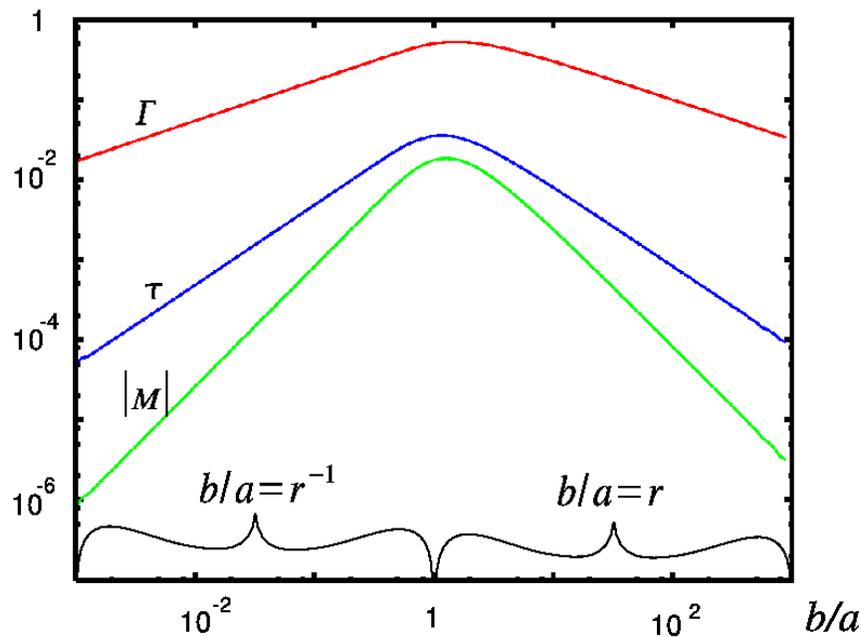}}
\caption{Damping coefficient $\Gamma$, characteristic time $\tau$, and
modulus of the effective mass $M$, as functions of the aspect ratio
$r$.}
\label{a3}
\end{figure}
\begin{figure}
\centerline{\epsfysize=9cm \epsfbox{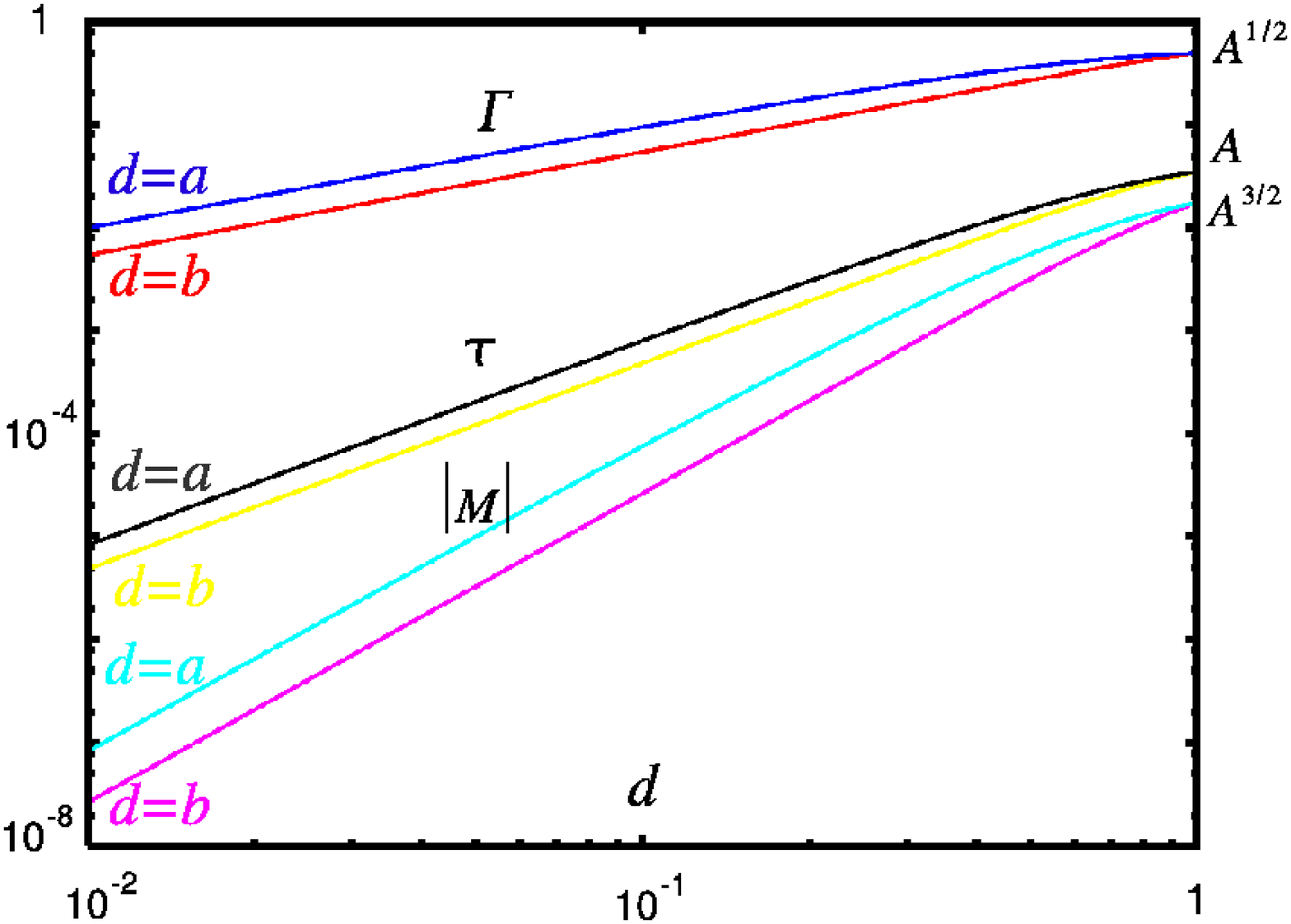}}
\caption{Damping coefficient $\Gamma$,
the characteristic time $\tau$, and the modulus of the effective mass
$M$, as functions of the smaller sample dimension $d$, both in the
two cases, when $d$ equals the dimension transverse ($a$) and parallel
($b$) to the domain wall. }
\label{b3}
\end{figure}

Figure \ref{a3} shows the variation of damping coefficient $\Gamma$,
characteristic time $\tau$, and of the modulus of the effective mass
$M$, as a function of the aspect ratio $r$ defined as
\begin{equation}
r=\max(a,b)/\min(a,b)\,,
\label{ar}
\end{equation}
and keeping constant the section area $A=ab$.  These three quantities
have a maximum in the case of a square rod. Therefore this is the
geometry where, $A$ being constant, the maximum asymmetry is expected
to be observed.

Figure \ref{b3} shows the variation of the the same three quantities
as a function of the most relevant parameter, which is the smallest
sample dimension $d=\min(a,b)$. The damping coefficient $\Gamma$,
characteristic time $\tau$, and modulus of the effective mass $M$,
increase as $d$, $d^2$, and $d^3$ respectively, and saturate to the
values $A^{1/2}$, $A$, and $A^{3/2}$, respectively.  The role of the
two dimensions $a$, and $b$, transverse and parallel to the wall is
however not symmetric, and the effect of eddy current retardation is
expected to be more severe, other things being equal, when the
transverse dimension is the smallest.

The analytical expressions of $\Gamma$, $\tau$, and $M$ used in
figures \ref{a3}, and \ref{b3} are given in \cite{colaiori_eddy_2007}.

It would be interesting to repeat BN experiments in changing sample
geometries in order to test these results.

Let us comment on the relevance of the eddy inertial term when
compared to the usual inertial term which is associated to the
D\"oring mass $M_D$. The values of $M_D$ are about four orders of
magnitude smaller then the eddy mass $M$ in typical sample geometries
used in BN experiments. Indeed the D\"oring mass originates from a
gyroscopic precessional effect, whose associated time scales are much
faster then those relative to the eddy mass.  However, the analysis
reported in \cite{colaiori_eddy_2007} on the dependence on sample
geometry predicts a fast decay $d^{-3}$ of the mass with the sample
thickness, which means that the two masses may become comparable in
thick films.

\subsection{Comparison between theory and experiments}
\label{sec:ASYM_comp}

\begin{figure}
\centerline{\epsfysize=8.5cm \epsfbox{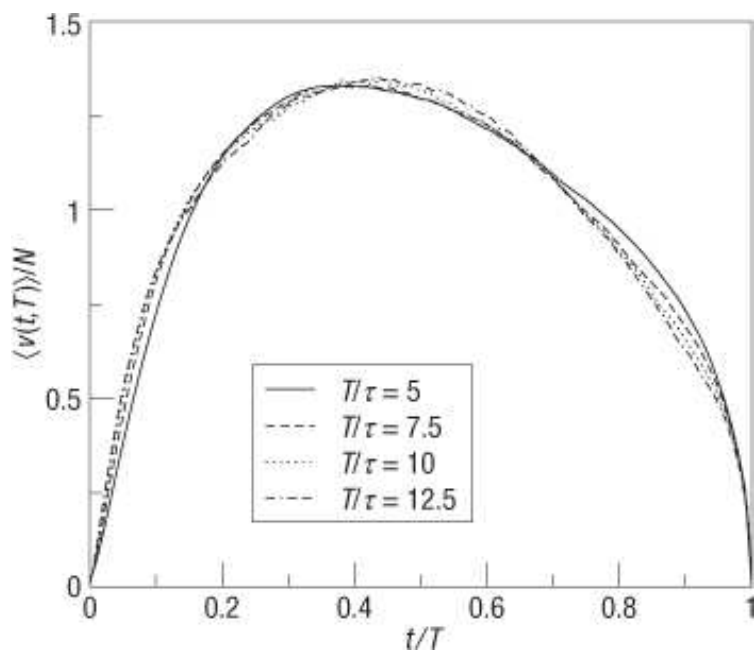}}
\caption{Asymmetry of the rescaled pulse shape: pulses obtained from
  Barkhausen noise experiments in a partially crystallized
  $Fe_{64}Co_{21}B_{15}$ ribbon. The shapes for different durations
  $T$.  (Reprinted from \cite{zapperi_signature_2005}).}
\label{nphys1}
\end{figure}

\begin{figure}
\centerline{\epsfysize=8.5cm \epsfbox{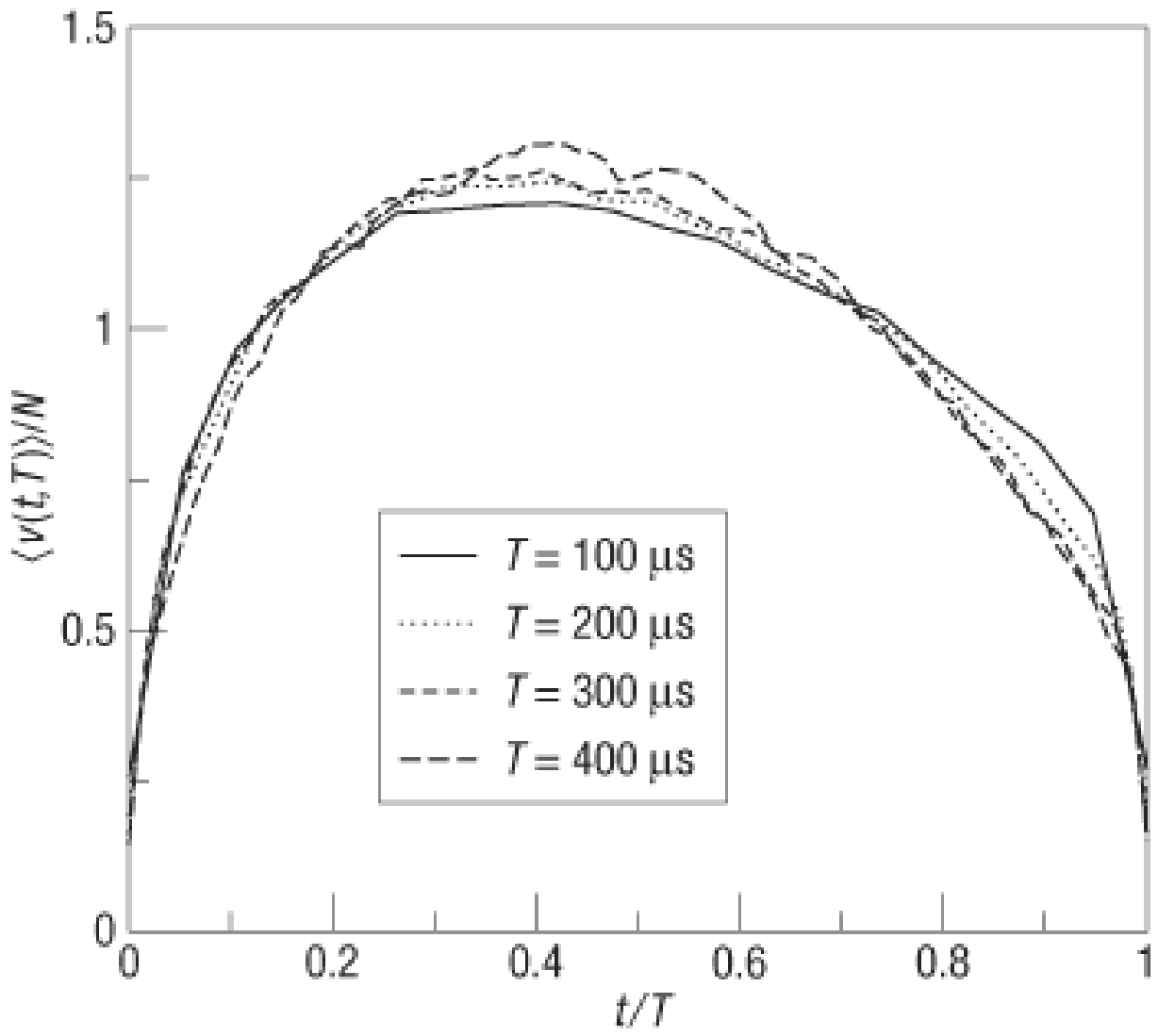}}
\caption{The shapes obtained from the modified ABBM model with the
  parameters corresponding to those of the experiment in
  figure \ref{nphys1}. (Reprinted from \cite{zapperi_signature_2005}). }
\label{nphys2}
\end{figure}

By simulating numerically the modified ABBM equation
(\ref{modified_ABBM}) one observes that, whereas the distributions of
avalanche duration and size are unaffected by the addition of the
inertial term, the pulse shapes become asymmetric and bear a
remarkable similarity with the experimental ones (see figures
\ref{nphys1}, \ref{nphys2}, where the curves are normalized with
$N=\int_0^T dt \langle v(t,T)\rangle/T$)
\cite{zapperi_signature_2005}. Note that eddy currents only change the
internal dynamics of the avalanches, so that at least the distribution
of avalanches sizes' is expected to be left unaltered, while
correction could be present in the power spectrum (see below) and in
the distribution of avalanche durations' (not observed).


\begin{figure}
\centerline{\epsfysize=10cm \epsfbox{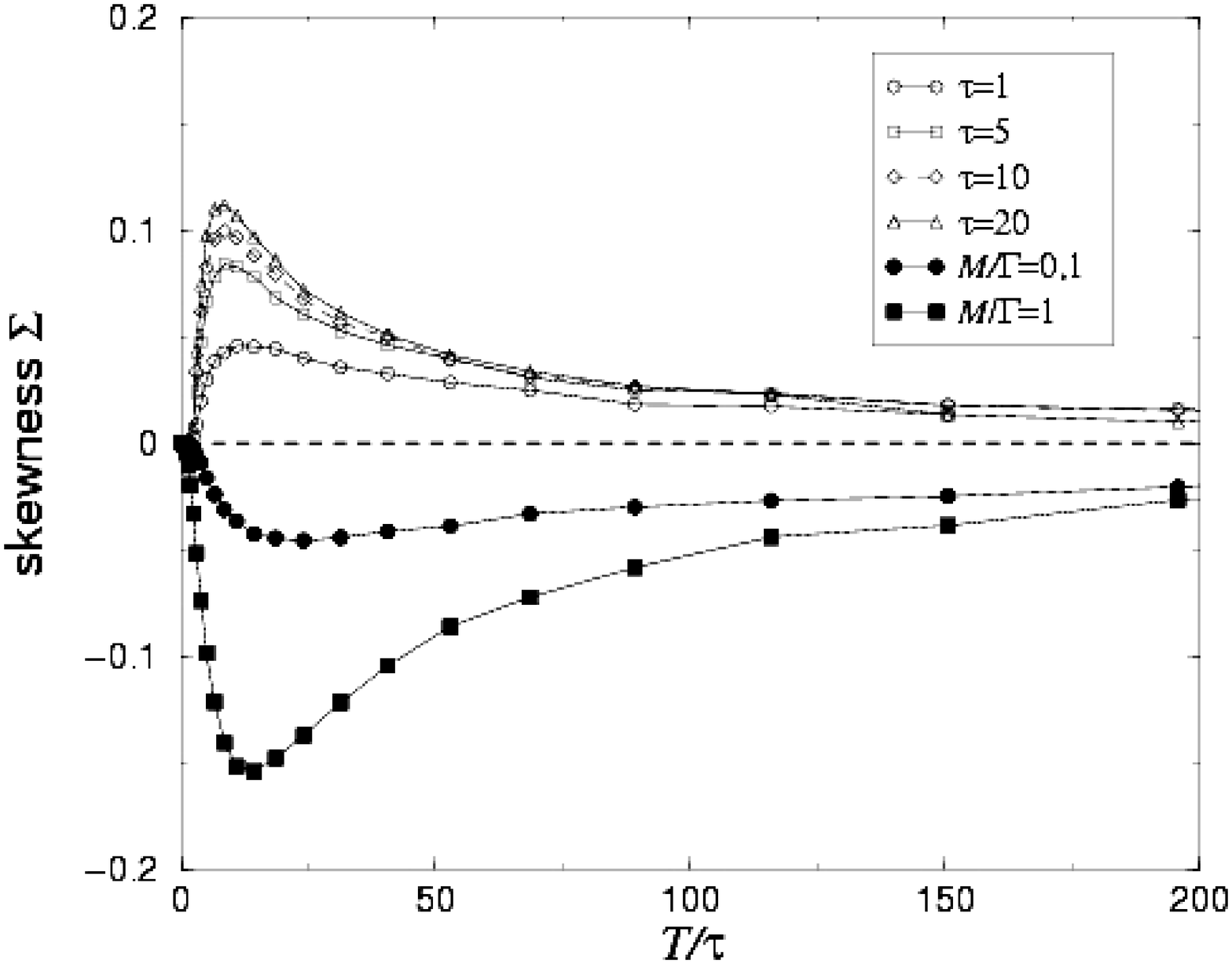}}
\caption{Skewness from simulations of the modified ABBM model. Filled
symbols correspond to positive mass (negative skewness), empty symbols
to negative mass (positive skewness).  (Reprinted from
\cite{zapperi_signature_2005}).}
\label{skewness.eps}
\end{figure}

The skewness of the simulated model results positive, indicating a
leftward asymmetry, corresponding to a negative mass, while it
would be negative, indicating rightward asymmetry, in the standard
case of a positive mass.  The qualitative behavior of the skewness as a
function of the avalanche duration in correspondence to a negative
mass (see figure \ref{skewness.eps}) is similar to the one observed in
the experiments (see figure \ref{ex_skewness}): it is always positive,
it increases from zero, shows a peak at some characteristic time, and
it decays to zero for long avalanches. The existence of a peak can be
used to extract the characteristic relaxation timescale, which
corresponds to the ratio between mass and damping constant, and to
make a quantitative comparison between the model and experiments.

A further quantitative test of the theory against experiments includes
the comparison of the location of the peak $T_P$ that characterizes
the skewness versus duration plot. Skewness extracted from
experimental pulses shows a $T_p$ at a value which is roughly twice
that obtained from simulations of the modified ABBM equation with the
microscopic parameters of the corresponding experiment. This may be
due to the fact that the model considers just one single domain wall,
while many domains may be present in general in a sample.  It is
indeed expected \cite{bishop_contribution_1980} that the relaxation
time increases with the number of domains.
  
Another interesting point to underline is the effect of the negative
inertial term on the power spectra. Power spectra obtained by
simulating the modified ABBM equation show a deviation from the large
frequency asymptotic regime at a frequency $\omega^{*}=T_P^{-1}$ very
similar to the one observed in some experiments. Moreover,
experimental data also show a correspondence between the peak in the
skewness and the frequency at which the deviation appears in the
spectrum \cite{durin_signature_2007}.

\begin{sidewaysfigure}
\centerline{\epsfxsize=18cm \epsfbox{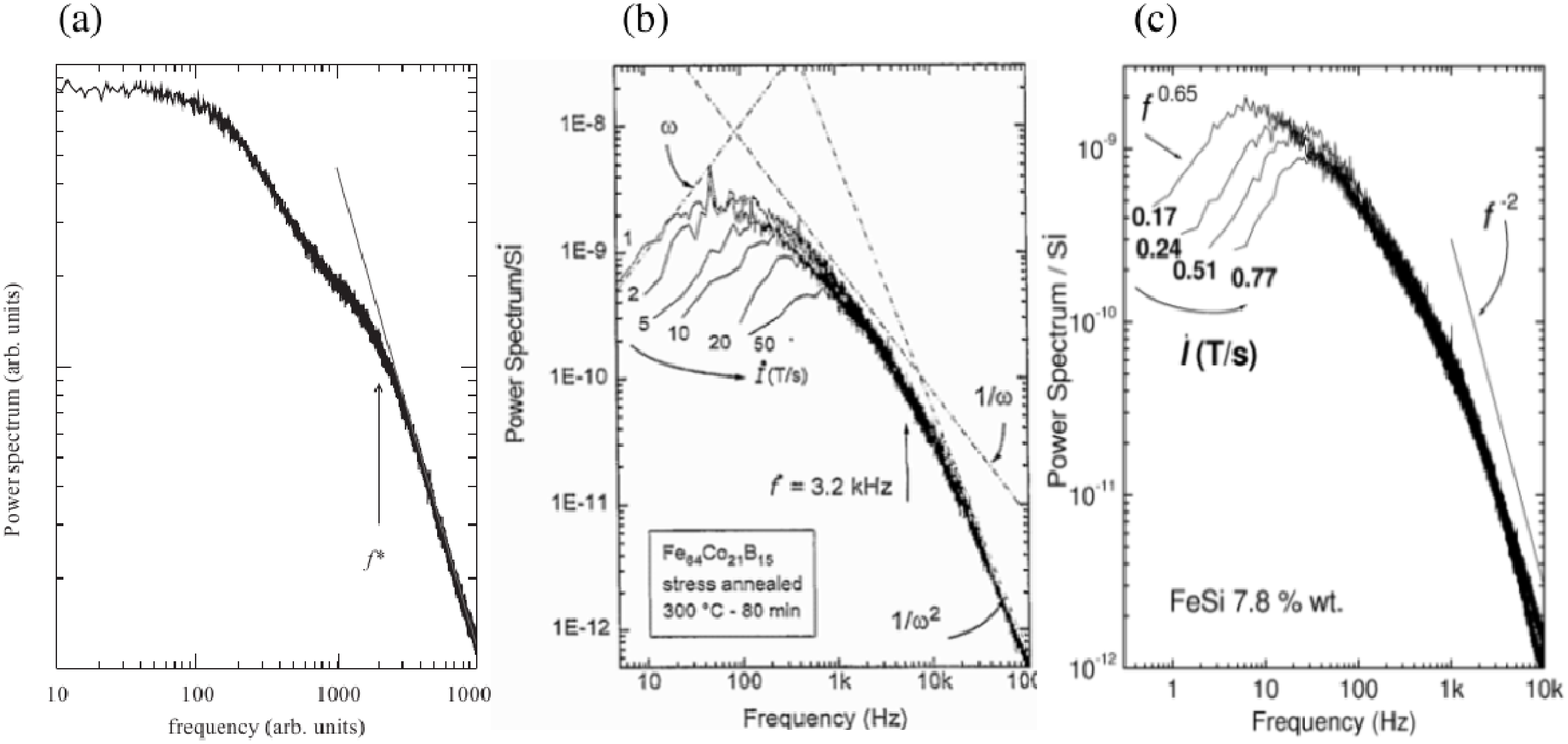}}
\caption{Comparison of power spectra where deviation from $1/\omega^2$
  is observed at intermediate frequencies with the one from
  simulations. (a) Power spectrum from a simulation of the modified
  ABBM model. (Reprinted with permission from
  \cite{durin_signature_2007}).  \cite{durin_signature_2007} Copyright
  2007 by Elsevier. (b) Power spectrum for a partially crystallized
  $Fe_{64}Co_{21}B{15}$.(Reprinted with permission from
  \cite{durin_measurements_1996}).  \cite{durin_measurements_1996}
  Copyright 1996 by Elsevier.  (c) Power spectrum for a
  polycrystalline $7.8\%$ $FeSi$. (Reprinted with permission from
  \cite{durin_barkhausen_1997}).  \cite{durin_barkhausen_1997}
  Copyright 1997 by World Scientific, Singapore. }
\label{pw_3}
\end{sidewaysfigure}

Summarizing the discussion of this section, we can conclude that the
leftward asymmetry of pulse profiles in BN is caused by the non
instantaneous response of the eddy field to the wall displacement,
that can be accounted for by associating a negative effective mass to
the moving wall.  The asymmetry is a non--universal property, that
depends on the material parameters $\sigma$, $\mu$, and $M_s$ and on
the geometrical dimensions $a$ and $b$ of the sample, and therefore it
allows to extract information about the characteristic time ruling the
microscopic dynamics.

Since the asymmetry is related to the existence of this microscopic
timescale, its magnitude depends on the avalanche duration: for
avalanches long enough with respect to the characteristic time, the
microscopic details of the material and the geometry of the sample
become irrelevant, pulse--shapes become symmetric, and universality is
expected to hold.

Let us finally comment on the physical meaning of the eddy inertial
term. It has always to be kept in mind that the negative effective
mass is just a convenient way to account, to the first order, for the
history dependent damping. The effective mass, however, is unable to
describe for example the very short time dynamics, when the domain
wall starts to move from its position at rest. Indeed, in this case,
the negative mass would cause the wall to move backward, which is
certainly not the case. Here the results of this section should be
used with more care: when the wall starts to move there is no previous
motion history, and therefore both the eddy damping and the eddy
inertial term are zero. In the very short time dynamics the D\"oring
mass, which is positive, represents the most relevant inertial term in
the equation of motion. Therefore standard inertia governs the
dynamics, until damping and eddy mass build up.

\section{Other crackling systems: granular materials, seismic activity in earthquakes}
\label{sec:OTHER}

Crackling noise emerges in very different contexts, and the similarity
in how completely different systems behave suggests the existence of
some common principle in the underlying physics.  As an example of how
ideas developed in the study of one crackling system can stimulate
advances in other fields, we report here some analogies between BN and
two other crackling system: sheared granular materials and seismic
activity in earthquakes.

\subsection{Dynamics of granular materials under shear}
\label{sec:granular}

A granular medium, subject to a slow loading rate, has an intermittent
shear response characterized by large fluctuations, typical of
crackling systems \cite{miller_stress_1996,dalton_basin_2002}. The
large fluctuations in this case originate from the rearrangement of
the network of force chains along which the stress propagate.  

Most of the past theoretical activity on granular matter has focused
on average properties, that could allow the formulation of macroscopic
laws governing the dynamics. However, more recently it became clear
that a deeper understanding of the mechanisms that regulate the
dynamics on smaller scales and cause the large fluctuations, cannot be
reached without going through a detailed analysis of the slip
statistics.

In this framework, a very interesting model has recently been proposed
to interpret a series of experiments on granular materials sheared in
a Couette geometry \cite{baldassarri_brownian_2006}.  This model, is
based on an assumption similar to the one that underlies the ABBM
model for BN, namely that the resultant from the forces acting on the
system performs a Brownian motion.

In the experiment a cylindrical cell is filled with monodisperse glass
beads with $2mm$ of diameter. The shear is created by rotating an
annular plate over the top of the circular channel, driven by a torsion
spring with stiffness $k$. After being initialized to reach the
stationary state, the system is then run at a slow driving velocity
$v$. The experimental device allows to measure the angular position
$\theta$ of the plate, and the deflection of the torsion spring. In a
given range of driving velocities, the system displays a stick--slip
regime, where repeatedly the torque on the plate accumulates as the
spring winds up, until the plate slips.

The instantaneous velocity of the plate $\dot{\theta}$ as a function of
time, shows the typical intermittent behavior of crackling signals,
with pulses of largely fluctuating amplitudes. A statistical analysis
of the signal similar to the one usually done for BN emissions reveals
the presence of power law distributions of the slip events in the
stick--slip phase, as it was already observed in a previous experiment
\cite{dalton_basin_2002}.

In this case the pulse size $s$ corresponds to the angular variation
during the slip event. The sizes and durations distributions show a more
complex shape than in the BN case, since the initial power law decay
is followed by a peak at larger scales, in correspondence to some
cut--off value. The position of the peak is independent of the driving
velocity $v$.

As in the case of BN, the intermittent stick--slip phase is observed
only as long as the driving velocity is not too large. Above a
threshold value, a transition to a steady sliding regime occurs, that
is analogous to the transition between the intermittent BN noise
regime, and the unique avalanche magnetization reversal that occurs at
high sweep field rates.

In ref. \cite{baldassarri_brownian_2006} the authors propose a simple
model that quantitatively reproduces the phenomenology, and which is
based on some minimal hypothesis. The equation of motion for the
stick--slipping plate is obtained by equating the inertial term to the
driving force on the spring diminished by the counteracting friction
of the granular media.
\begin{equation}
I \ddot{\theta}=k(vt-\theta)-F \,.
\label{granular1}
\end{equation}
Here $I$ is the moment of inertia of the system, and the whole
complexity of the granular matter dynamics that give rise to the wide
fluctuations is accounted for in the frictional torque term $F$.

Baldassarri {\it et al.} assume that, as a first approximation, the
instantaneous frictional torque is a function of $\theta$ and
$\dot{\theta}$ only (since $F$ is constant when the disk is stuck,
there is no direct dependence on time $t$ in $F$), and that $F$ can be
split into a deterministic part $F_d(\dot{\theta})$ plus a random
contribution $F_r(\theta)$. 

The deterministic part of the instantaneous friction
$F_d(\dot{\theta})$ is obtained by fitting the experimental average
velocity dependence of the frictional torque with the expression
\begin{equation}
F_d(\dot{\theta})=
F_0+\gamma\left(\dot{\theta}-2v_0\ln(1+\dot{\theta}/v_0)\right)\,,
\label{granular2}
\end{equation}
where $F_0$ is the average static friction force, $v_0$ is the minimum
average torque and $\gamma$ is the high velocity damping.  Note, that
the results obtained, however, do not depend very much on the precise
choice of the function $F_d$.

The fluctuating part of the instantaneous friction is the one that
accounts for the continuously rearranging disordered structure of the
force chain network present in the granular medium. This is assumed to
be described by a confined Brownian process
\begin{equation}
\frac{dF_r(\theta)}{d\theta}=\eta(\theta)-a F_r(\theta)\,,
\label{granular3}
\end{equation}       
where $\eta$ is a Gaussian quenched uncorrelated noise, and $a$ is an
inverse correlation length.  The idea is that the frictional torque
changes of a random amount after every small slip, but eventually
decorrelate on large slips involving a complete rearrangement of the
grains, resulting in a confinement of the Brownian process.  The
quantity $a^{-1}$ plays the same role as the correlation length of the
pinning field $\xi_P$ introduced in the ABBM model (see section
\ref{sec:PS2}).

This theory shows a number of analogies with the ABBM theory for BN.
The granular matter dynamics is more complicated than the one of
magnetic domains, due to the presence of inertial effects and of a
shear rate weakening torque. However, both theories are based on the
same assumption that the effective force acting on the system performs
a Brownian motion. Indeed, the presence of Brownian correlations in
the effective force evolution, due to collective effects (the sum over
many pinning forces over the domain wall in one case, the formation
and destruction of force chains in the granular media in the other)
may be of more general validity in driven dissipative systems.

The equation of motion is also similar in the two cases, and,
although, the one describing the granular material includes more
complex effects, it reduces to the ABBM equation in the limit $I,v_0
\rightarrow 0$ of vanishing inertia and minimal average torque.  In
this case, or in an appropriate velocity regime $v_0\ll v\ll \gamma
/I$, where the additional effects can be neglected, all the results
obtained for the ABBM model are valid, once magnetization is
interpreted as a slip, the magnetic field rate as a shear rate, the
demagnetizing factor as the stiffness of the torsional spring, the
eddy current damping as a frictional damping, and the random pinning
as a random friction.

The analogy also suggests that the transition from the stick--slip
phase at slow drivings to the sliding phase at fast drivings might
possibly be explained also in this case as a transition from transient
to recurrent for the corresponding stochastic process. 

\subsection{Seismic activity in earthquakes}
\label{sec:earth}

Another kind of crackling noise that shows analogies with BN is the
seismic activity during earthquakes. 

Earthquake phenomenology exhibits a number of power law distributions
including the Gutenberg-Richter frequency-size statistics
\cite{gutenberg_seismicity_1949, lee_international} and the Omori law
for aftershock decay rates \cite{omori_aftershocks_1894,
kisslinger_properties_1991}.  

In literature, the area of the region of the fault that slips in an
earthquake is called the {\em moment}.  The probability of occurrence
of an earthquake of a given moment $M$ decays as a power law according
to
\begin{equation}
P(M)\sim M^{-(1-\beta)} \,.
\label{gr}
\end{equation}
The exponent $\beta$ is the same, within errors, for most sets of
data, collected in different times and regions. 

As in the case of BN, scaling functions have been introduced as
additional interesting observables, which provides a powerful tool
both to check universality, and to test the theory against data. In
particular, the average moment rate profiles at fixed total earthquake
duration, have been studied in \cite{mehta_universal_2006}, and
compared with the prediction of a mean field theory for earthquakes
dynamics \cite{ben-zion_earthquake_1993}.

\begin{figure}
\centerline{\epsfysize=11cm \epsfbox{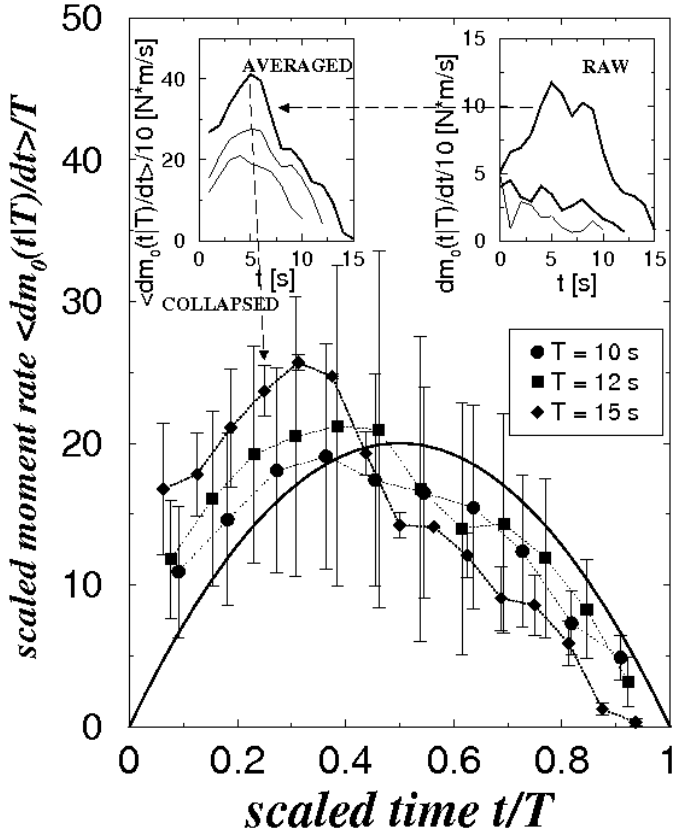}}
\caption{A collapse of averaged earthquake pulses' shapes, with a
  duration of $T$, with two to ten earthquakes averaged for each value
  of $T$. The collapse was obtained using the mean field theory for
  which the universal scaling function is claimed to be $g(x)=Ax(1-x)$
  with $x=t/T$ \cite{kuntz_noise_2000}. Note the apparent asymmetry to
  the left in the observed data while the theoretical curve is
  symmetric around its maximum. In the inset the raw data and the
  averaged data before collapse are shown.  (Reprinted with permission
  from \cite{mehta_universal_2006}).  \cite{mehta_universal_2006}
  Copyright 2006 by the American Physical Society.  }
\label{earthq}
\end{figure}

The average moment rate is supposed to behave as 
\begin{equation}
\langle dM(t,T)/dt\rangle \simeq T g(t/T) \,.
\label{moment-rate}
\end{equation}

In paper \cite{mehta_universal_2006} the universal function $g(t/T)$
is claimed to be:
\begin{equation}
g(x)=A(x(1-x)) \,. 
\label{gmf}
\end{equation}
This result was obtained by mapping the model onto the mean field
model for BN, for which the result of $g$ being an inverted parabola
was claimed in \cite{kuntz_noise_2000}. 

It is clear from figure \ref{earthq} that in seismic data the moment
rate of a given duration $T$ increases faster then it drops, as the
corresponding average moment shapes show a leftward asymmetry similar
to that observed in BN, while, also in this case, models predict a
symmetric shape.

Seismic movements are caused by the motion of fault planes in response
to the accumulated stress. The analogy with BN suggests the
possibility that the origin of the asymmetry is similar in the two
cases. The asymmetry may indeed originate from some non--local
dynamical effect, that, in the case of earthquakes could be caused by
the presence of stress overshoots \cite{dahmen_nonlinear_2005}. It
would be interesting to see how the skewness of the average pulses
changes with the duration of the seismic event (which is not clear for
the data in ref. \cite{mehta_universal_2002}), as this can help to
understand whether or not also in this case the asymmetry disappears
in large events, and whether universality is recovered oon large
time scales.

In the case of earthquakes, the analogy with domain wall dynamics is
even more stringent than in the case of sheared granular
media. Indeed, the mean field description studied in
\cite{mehta_universal_2006} exactly coincides with the ABBM model.
Long range elastic interactions lead to $d=3$ being the upper critical
dimension for earthquakes models \cite{fisher_statistics_1997,
  ben-zion_earthquake_1993, ben-zion_slip_1995, ben-zion_stress_1996}, and
playing the same role as the long range dipolar interactions in BN. 

It is interesting to mention that asymmetric avalanches are also
observed in other crackling systems. The dynamics of the
Bak--Tang--Wiesenfeld model of sand piles \cite{bak_self_1987} is
characterized by asymmetric avalanches
\cite{laurson_power_2005}. Leftward asymmetric pulses also appear in
the intermittent behavior of systems undergoing a shear deformation
that proceeds in bursts of dislocations activity
\cite{richeton_dislocation_2005}.  Recently Laurson and Alava
\cite{laurson_noise_2006} analyzed a discrete dislocation dynamics
model and observed, in agreement with the experiments, a leftward
asymmetry in the avalanche shape, corresponding to an asymmetry in the
average creation and annihilation rates of dislocations during the
avalanche. They also find that the asymmetry decreases with the signal
threshold, although the threshold value do not have visible effects on
the avalanche statistics. Finally, a strong skewness in the avalanche
shape has recently been observed in the acustic emissions during
martensitic transformations \cite{perez_training_2007}.

\section{Conclusions}
\label{sec:CONCLUSIONS}

In this review we report on the present state of understanding of the
Barkhausen effect in soft ferromagnetic materials, focusing on those
with long range dipolar interactions. In particular, we collect
several results on the mean field ABBM model for BN. This model,
originally proposed on phenomenological basis, corresponds to a mean
field description for an elastic magnetic wall in a disordered
ferromagnet under the effect of an external driving field, and is
extremely successful in reproducing the complex phenomenology observed
in the experiments, at least as long as universal properties are
involved.

The ABBM model consists of an effective equation of motion for the
velocity of the avalanche front. This equation maps onto the Langevin
equation for a biased random walk in a logarithmic potential with
appropriate boundary conditions.  This process is amenable to
analytical treatment with standard methods in the theory of stochastic
processes. Therefore this mapping has the advantage to provide several
exact results, as well as a direct interpretation of a number of
properties observed in the experiments in terms of properties of the
corresponding stochastic process. Specifically, through this mapping,
the power laws in the distributions of sizes and durations of avalanches
are exactly derived in terms of distributions of return times to the
origin, the continuous dependence of the exponents in the power law
distributions on the driving field rate can be calculated and turns
out to be related to the marginality of the logarithmic perturbation
to a free random walk, the existence of a threshold in the driving
field rate has a clear interpretation (thanks to a further mapping) in
terms of recurrence properties of a free random walk, the existence of
a cut--off in the power law distributions is ascribed to the bias
term, and therefore is due to the presence of the demagnetizing field,
and finally the average shape of the Barkhausen pulse which is related
to the excursion of the process, and the power spectra of the noise
can be computed explicitly.

Although deviations from the ABBM predictions are observed in some
experiments, the picture that emerges from this theory is quite
satisfactory. The most relevant experimental feature that is not
reproduced by ABBM is the leftward asymmetric shape systematically
observed in experiments. This asymmetry is due to a non--local damping
by the eddy currents, whose effect is neglected in the original ABBM
model formulation, as it is in most domain walls models.  This effect
can be taken into account by including in the equation of motion an
inertial term with a negative effective mass for the interface, which
leads to the definition of a generalized ABBM model.  The
asymmetry of the pulses depend on the duration and can be used to
extract important information on the characteristic time of the
underlying dynamics.

A basic point to keep in mind, is that the whole theory developed so
far models one single domain wall, while many interacting domain walls
are in general involved in the magnetization process. It would
therefore be interesting to understand how the presence of many
domains modifies the present theory.

Although the ABBM theory was derived as a model for Barkhausen noise,
it turns out to be a paradigmatic theory for avalanche dynamics, and
can give suggestions on how to model other crackling systems. We
discussed some analogies with sheared granular matter and seismic
events. These similarities point to the existence of deep connections
between the physics of avalanches in different systems.  Also the
presence of non--universal effects might be common to other crackling
system. Asymmetric avalanche shapes similar to those observed in BN
appear in other crackling systems: in martensitic transformations, in
plastic deformations, in the Bak-Tang-Wiesenfeld model of sand piles,
and in seismic events. In the last case, the leftward asymmetric
average moments may originate, from the presence of stress overshoots.

\section{Acknowledgments}
\label{sec:ak}

The author wishes to thank all those people who have contributed to
our current understanding of the problems discussed. In particular,
the personal author's understanding of these subjects is largely due
to interactions with A. Baldassarri, G. Bertotti, C. Castellano,
F. Dalton, G. Durin, A. Petri, J. P. Sethna, and S. Zapperi, to whom
the author is most grateful also for stimulating discussions and for
encouragement to write this review.

%\bibliography{REVIEW_BIBLIO}
%\begin{thebibliography}{12} 
%\end{thebibliography}
\end{document}